%% file: nyquist.camera.tex
\documentclass[a4paper,onecolumn,10pt,twoside,reqno,nonamelimits,intlimits,accepted=2023-07-06]{quantumarticle}
\pdfoutput=1
\usepackage{a4wide}
\usepackage[utf8]{inputenc}
\DeclareUnicodeCharacter{2202}{WhyTheFuckDoINeedThis}
\usepackage[T1]{fontenc}
\usepackage[colorlinks=true,urlcolor=blue,linkcolor=blue,citecolor=blue]{hyperref}

\usepackage{amsmath}
\usepackage{amsmath}
\usepackage{amssymb}
\usepackage{amsfonts}
\usepackage{amscd}
\usepackage{amsthm}
\usepackage{amsxtra}
\usepackage{wasysym}
\usepackage{mathrsfs}
\usepackage{nicefrac}
\usepackage{graphicx}%\graphicspath{ {image/} }
\usepackage{xcolor}
\usepackage{subfig}
\usepackage{tikz}
\usetikzlibrary{cd}
\usepackage[multiple]{footmisc}

\usepackage[inline]{enumitem}
\setlist[enumerate]{leftmargin=2em,itemindent=0em, labelindent=0pt,labelwidth=1.5em,labelsep=.5em, align=left, noitemsep}
\newlist{condenum}{enumerate}{1}
\setlist[condenum]{ leftmargin=4em,itemindent=0em, labelindent=0pt,labelwidth=1.5em,labelsep=.5em, align=left, noitemsep}
\newlist{txtenum}{enumerate}{1}
\setlist[txtenum]{leftmargin=0em,itemindent=1.5em, labelindent=0pt,labelwidth=1em,labelsep=.5em, align=left}
\newlist{txtitem}{itemize}{1}
\setlist[txtitem]{leftmargin=0em,itemindent=1.5em, labelindent=0pt,labelwidth=1em,labelsep=.5em, align=left}
\usepackage[boxed,noend]{algorithm2e}\DontPrintSemicolon % ,linesnumbered,
\SetKwInOut{Fixed}{Fixed}
\SetKwInOut{Input}{Input}
\SetKwInOut{Output}{Output}

\SetKwFor{SeparationRoutine}{Separation Routine}{}{end}
\SetKwRepeat{Repeat}{Repeat}{Until}
\SetKwIF{If}{ElseIf}{Else}{If}{then}{Else if}{Else}{Endif}
\SetKw{Return}{Return}
\newcommand{\myparsmall}{\par\smallskip\noindent}
\newcommand{\mypar}{\par\medskip\noindent}

\newenvironment{Claim}{\medskip\noindent\textit{Claim.}}{\par\medskip}

\input{defsN.amsart}
\input{defs.common}
\DeclareFontEncoding{FMS}{}{}
\DeclareFontSubstitution{FMS}{futm}{m}{n}
\DeclareFontEncoding{FMX}{}{}
\DeclareFontSubstitution{FMX}{futm}{m}{n}
\DeclareSymbolFont{fouriersymbols}{FMS}{futm}{m}{n}
\DeclareSymbolFont{fourierlargesymbols}{FMX}{futm}{m}{n}
\DeclareMathDelimiter{\VERT}{\mathord}{fouriersymbols}{152}{fourierlargesymbols}{147}

\numberwithin{theorem}{section}

\usepackage{xr}
\newcommand{\GenericFun}{\mathscr{G}}
\newcommand{\Fun}{\mathscr{F}}            % function space
\newcommand{\decFun}{\mathscr{F}^{\downarrow}}
\newcommand{\sumFun}{\mathscr{F}^{\oplus}}
\newcommand{\oldFun}{\mathscr{K}}
\newcommand{\oldFunOp}{\mathscr{K}^\textrm{\tiny op}}
\newcommand{\place}{{\textrm{\tiny$\,\square\,$}}}
\newcommand{\cH}{\mathcal H}      % Hilbert space
\newcommand{\Schwartz}{\mathcal{S}}
\DeclareMathOperator{\sinc}{sinc}
\DeclareMathOperator{\Li}{Li_2}
\DeclareMathOperator{\Log}{Log}
\newcommand{\tA}{\texttt{A}}
\newcommand{\tB}{\texttt{B}}
\newcommand{\tC}{\texttt{C}}

\newcommand{\tE}{\texttt{E}}

\newcommand{\tM}{\texttt{M}}
\newcommand{\tU}{\texttt{U}}
\newcommand{\tW}{\texttt{W}}

\newcommand{\tI}{\texttt{I}}  % Identity
\newcommand{\tLambda}{\mathtt{\Lambda}}
\undef\Diag
\DeclareMathOperator{\Diag}{\mathtt{D\mspace{-1.5mu}i\mspace{-2mu}a\mspace{-1mu}g}}

\DeclareMathOperator{\eig}{eig}
\newcommand{\lambdamin}{\eig_{\mathrm{min}}}
\newcommand{\lambdamax}{\eig_{\mathrm{max}}}
\DeclareMathOperator{\Diff}{Diff}
\newcommand*{\textcal}[1]{{\fontfamily{qzc}\selectfont\large#1}}
\newcommand{\opNm}[1]{\lt\VERT{#1}\rt\VERT}%
\newcommand{\bigOfrom}[1]{\mathfrak{O}^{\ge\!#1}}
\begin{document}
\title{``Proper'' Shift Rules for Derivatives of Perturbed-Parametric Quantum Evolutions}%
\author{Dirk Oliver Theis}%
\affiliation{Theoretical Computer Science, University of Tartu, Estonia}%
\email{dotheis@ut.ee}%
%\homepage{https://quantum-computing.ut.ee/?page\_id=292}%
\orcid{0000-0002-9362-0748}%
\thanks{This research was partly funded by the Estonian Center of Excellence in Computer Science (EXCITE), and by Estonian
  Research Agency (Eesti Teadusagentuur, ETIS) through grant PRG946.}
%%%%%%%%%%%%%%%%%%%%%%%%%%%%%%%%%%%%%%%%%%%%%%%%%%%%%%%%%%%%%%%%%%%%%%%%%%%%%%%%%%%%%%%%%%%%%%%%%%%%%%%%%%%%%%%%%%%%%%%%%%%%%%%%%%%%%%%%%%%%%%%%%%%%%% 
%%
%%
\date{Thu Jul~6 17:52:07 (CEST) 2023}
%%
%%
%%%%%%%%%%%%%%%%%%%%%%%%%%%%%%%%%%%%%%%%%%%%%%%%%%%%%%%%%%%%%%%%%%%%%%%%%%%%%%%%%%%%%%%%%%%%%%%%%%%%%%%%%%%%%%%%%%%%%%%%%%%%%%%%%%%%%%%%%%%%%%%%%%%%%% 
\maketitle

\begin{abstract}
  Banchi \& Crooks (\emph{Quantum,} 2021) have given methods to estimate derivatives of expectation values depending on a
  parameter that enters via what we call a ``perturbed'' quantum evolution $x\mapsto e^{i(x A + B)/\hbar}$.  Their methods
  require modifications, beyond merely changing parameters, to the unitaries that appear.  Moreover, in the case when the
  $B$-term is unavoidable, no exact method (unbiased estimator) for the derivative seems to be known: Banchi \& Crooks's method
  gives an approximation.

  In this paper, for estimating the derivatives of parameterized expectation values of this type, we present a method that only
  requires shifting parameters, no other modifications of the quantum evolutions (a ``proper'' shift rule).  Our method is exact
  (i.e., it gives analytic derivatives, unbiased estimators), and it has the same worst-case variance as Banchi-Crooks's.

  Moreover, we discuss the theory surrounding proper shift rules, based on Fourier analysis of perturbed-parametric quantum
  evolutions, resulting in a characterization of the proper shift rules in terms of their Fourier transforms, which in turn
  leads us to non-existence results of proper shift rules with exponential concentration of the shifts.  We derive truncated
  methods that exhibit approximation errors, and compare to Banchi-Crooks's based on preliminary numerical simulations.

  %%% arXiv:
  % Banchi & Crooks (Quantum, 2021) have given methods to estimate derivatives of expectation values depending on a
  % parameter that enters via what we call a "perturbed" quantum evolution $x\mapsto e^{i(x A + B)/\hbar}$.  Their methods
  % require modifications, beyond merely changing parameters, to the unitaries that appear.  Moreover, in the case when the
  % $B$-term is unavoidable, no exact method (unbiased estimator) for the derivative seems to be known: Banchi & Crooks's method
  % gives an approximation.
  %
  % In this paper, for estimating the derivatives of parameterized expectation values of this type, we present a method that only
  % requires shifting parameters, no other modifications of the quantum evolutions (a "proper" shift rule).  Our method is exact
  % (i.e., it gives analytic derivatives, unbiased estimators), and it has the same worst-case variance as Banchi-Crooks's.
  %
  % Moreover, we discuss the theory surrounding proper shift rules, based on Fourier analysis of perturbed-parametric quantum
  % evolutions, resulting in a characterization of the proper shift rules in terms of their Fourier transforms, which in turn
  % leads us to non-existence results of proper shift rules with exponential concentration of the shifts.  We derive truncated
  % methods that exhibit approximation errors, and compare to Banchi-Crooks's based on preliminary numerical simulations.

  \myparsmall%
\end{abstract}

\setcounter{tocdepth}{2}
\tableofcontents

\section{Introduction}\label{sec:intro}%-------------------------------------------------------------------------------------
One paradigm in near-term quantum computing is that of variational quantum algorithms (VQAs): Quantum algorithms which contain
real-number parameters and which must be trained, i.e., the parameters must be optimized --- similarly to classical
differentiable programming, e.g., artificial neural networks.

In fault-tolerant quantum computing, evidence has been given (see, e.g.,
\cite{McClean-Rubin-Lee-Harrigan-OBrien-Babbush-Huggins-Huang:foundQC4chem:2021} and the references therein) that the concept of
quantum programs depending on parameters that are fitted to data may turn out to be an important component in applications of
quantum computing, beyond the realm of machine learning and AI.

Pre-fault-tolerance, under the name of Parameterized Quantum Circuits or Variational Quantum Circuits, VQAs are at the heart of
proposals for quantum-computer based simulations of molecules and condensed matter materials, quantum machine learning and
quantum AI, some approaches to quantum-based combinatorial optimization, for example some uses of the Quantum Approximate
Optimization Algorithm, and even applications such as linear-system solving and factoring (e.g.,
\cite{Yuan-Suguru-Qi-Ying-Benjamin:theory-VQS:2019, Mitarai-Negoro-Kitagawa-Fujii:qcirclearn:2018,
  Benedetti-Lloyd-Sack-Fiorentini:PQCs:2019, Farhi-Goldstone-Gutmann:QAOA:2014,
  Anschuetz-Olson-AspuruGuzik-Cao:var-factoring:2018, BravoPrieto-LaRose-Cerenzo-Subasi-Cincio-Coles:VQLS:2019}).
But the concept of parameterized quantum evolution for pre-fault-tolerant quantum simulation and computation extends beyond
gate-based quantum circuits (e.g.,
\cite{PAT:Babbush-Neven:sublogical-1,Henry-Thabet-Dalyac-Henriet:pasqal-ML:2021,Delyac-Henriet-Jeandel-Lechner-etal:pasqal-Opt:2021}).
In this paper, our interest is in quantum evolutions where the parameters do not enter via simple gates.

\mypar%
As optimization/training algorithms based on estimates of derivatives (e.g., variants of Sto\-chas\-tic Gradient Descent)
outperform derivative-free methods in practice in the training of variational quantum algorithms
(cf.~\cite{Sweke-Wilde-Meyer-Schuld-Fahrmann-MeynardPiganeau-Eisert:SGD:2020} and the references therein), there is the need to
obtain, efficiently, estimates of derivatives with respect to the parameters in a VQA.  Starting with seminal works by Li et
al.~\cite{Li-Yang-Peng-Sun:hyqoptctrl:2017} and Mitarai et al.~\cite{Mitarai-Negoro-Kitagawa-Fujii:qcirclearn:2018}, unbiased
estimators for derivatives have been obtained efficiently using so-called shift rules: If $f$ denotes the expectation-value
function dependent on, w.l.o.g., a single parameter, a shift rule is a relation
\begin{subequations}
  \begin{equation}\label{eq:intro:finite-support-shift-rule}
    f'(x) = \sum_j u_j\, f(x-s_j)
  \end{equation}
  where the $u_j$ (coefficients) and~$s_j$ (shifts) are fixed real numbers, and the equation holds for all $x$.  This is a
  convolution of the expectation-value function with a finite-support measure, and by replacing ``finite-support'' with
  ``finite'', this extends to shift rules with a continuum of shifts:
  \begin{equation}\label{eq:intro:the-shift-rule}
    f'(x) = \int f(x-s) \,d\phi(s) =: f*\phi(x),
  \end{equation}
\end{subequations}
where $\phi$ is a finite measure on~$\RR$, and ``$*$'' denotes convolution.

The notion of a shift rule has been extended, e.g.~\cite{Banchi-Crooks:stoch-param-shift:2021} speak of a Stochastic Shift Rule
for a method that involves other modifications of the quantum evolution than merely changing the parameters.\footnote{%
  There is another sense in which Banchi \& Crooks term shift rule differs from our ``proper'' shift rule in
  Eqn.~\eqref{eq:intro:the-shift-rule}: Banchi \& Crooks include the \emph{estimator} for the resulting value into the term
  stochastic shift rule, while we, following Mitarai et al.\ keep the relation~\eqref{eq:intro:the-shift-rule} (the ``proper''
  shift rule)) separate from an algorithm implementing an estimator of the quantity $f*\phi(x)$.
} %
To avoid ambiguity, we use the term Feasible \emph{Proper} Shift Rule (feasible PSR) for the~$\phi$ in
relation~\eqref{eq:intro:the-shift-rule}.

We add the qualification ``feasible'' as we also study a couple of notions of approximate PSRs
where~\eqref{eq:intro:the-shift-rule} holds up to an approximation error\footnote{Difference quotients in numerical
  differentiation are examples of PSRs that are ``not quite feasible''.}, leading to biased estimators.
% (we'll give two definitions of the term ``nearly feasible'' in \S\ref{ssec:overview:trunc} and
% \S\ref{ssec:overview:Impossible})

The present paper studies PSRs for derivatives of the type of quantum evolutions which is the subject
of~\cite{Banchi-Crooks:stoch-param-shift:2021}: We expect that the parameter, $x$ enters in the following form
\begin{equation}\label{eq:intro:the-unitary}
  e^{i(x A + B)/\hbar}.
\end{equation}
We refer to~\eqref{eq:intro:the-unitary} as ``perturbed-parametric'' unitary, and we speak of a perturbed-parametric
ex\-pec\-ta\-tion-val\-ue function if a measurement depends on~$x$ via a perturbed-parametric unitary.

Banchi \& Crooks~\cite{Banchi-Crooks:stoch-param-shift:2021} propose a stochastic shift rule for estimating derivatives of
per\-turb\-ed-parametric expectation-value functions, building on an equation by Feynman~\cite{Feynman:op-calc:1951,
  Wilcox:exp-op-param-diff:1967}.  As said above, their rule modifies the underlying quantum evolution itself: To obtain an
estimate of the derivative, the unitary~\eqref{eq:intro:the-unitary} must be replaced by a concatenation
\begin{equation}\label{eq:BC-unitaries}
  e^{i t_3\, (x_3 A + B)/\hbar}
  \cdot e^{i \beta\,(x_2/\beta\cdot A + B)/\hbar}
  \cdot e^{i t_1\, (x_1 A + B)/\hbar},
\end{equation}
for certain values of $t_*$, $x_*$, and a small $\beta>0$; if the $B$-term in the evolution can be switched off, $\beta \to 0$
effects the unitary $e^{i x_2 A / \hbar}$.  For $\beta>0$, Banchi \& Crooks refer to their method as Approximate Stochastic
Parameter Shift Rule.

Modifying the underlying quantum evolution has some small technical disadvantages, as it requires a re-calculation of the
schedule of quantum-control pulses.

Moreover, Banchi \& Crooks's shift rule has the disadvantage that it cannot give an unbiased estimate of the derivative in
situations where the $B$-term in the perturbed-parametric unitary~\eqref{eq:intro:the-unitary} is unavoidable: A $O(\beta)$
approximation error results.

In many situations where perturbed-parametric unitaries arise, a certain small bias in the estimate of the derivative may be
allowed.
% In a fault-tolerant quantum computing context (e.g., \cite{Berry-Childs-Cleve-Kothari-Somma:exp-improve:2014,
% Berry-Childs-Cleve-Kothari-Somma:truncTaylor:2015, Low-Chuang:HamSim-QSP:2017, Low-Chuang:HamSim-Qbtztn:2019}), a precision
% parameter will generally be present, and influence the complexity of running the perturbed-parametric unitary.
Indeed, in pre-fault-tolerant settings, expressions such as~\eqref{eq:intro:the-unitary} are approximations up to control
inaccuracies.

But in the case where the $B$-term is unavoidable, Banchi-Crooks's Approximate Stochastic Parameter Shift Rule requires, to
achieve a good approximation, a small $\beta>0$, so the parameter value, $1/\beta$, is large.  In practice, large parameter
settings may not be desirable (e.g., they could result in more cross talk) or even technically possible.\footnote{The methods in
  the present paper cannot substantially alleviate the problem of large magnitudes of parameter-values, but in preliminary
  numerical simulations (\S\ref{ssec:overview:numsim}) they appeared to be more efficient in that regard.}

Moreover, in some situations where perturbed-parametric quantum evolutions arise, the factor $\beta$ in the middle factor
in~\eqref{eq:BC-unitaries} is a duration of some pulse, and effecting unitaries such as
$e^{i \beta\,(x_2/\beta\cdot A + B)/\hbar}$ involves a narrow frequency band for that pulse (e.g., the resonant frequency of a
qubit).  In these situations, the Fourier Uncertainty Principle may make it difficult in principle to choose an arbitrarily
small~$\beta$.

\subsubsection*{Contributions of this paper}
We present a PSR feasible for expectation-value functions where the parameter enters in the form
of~\eqref{eq:intro:the-unitary}; we call it the Nyquist Shift Rule.  Our method is exact, so that estimators without bias
can be constructed; one such estimator, that we present below, has the same variance as Banchi \& Crooks's Stochastic
Parameter Shift Rule.

% We restrict ourselves to first derivatives, and consider (w.l.o.g.)\ a single parameter.

% \mypar%
% If~$\phi$ is a feasible PSR with infinite support, it would be according
% to~\cite{Banchi-Crooks:stoch-param-shift:2021} to refer to it~$\phi$ as a \textit{stochastic} PSR, but, of course,
% there is absolutely nothing stochastic in equation~\eqref{eq:def-feasible--proper-shift-rule}.  Instead, we distinguish between
% the (feasible) PSR, and various estimators that may be designed to obtain samples from it.  To obtain an unbiased
% estimate of $f*\phi(x)$ for a given~$x\in\RR$, one Single-Shot Estimator is the following (stated here informally; see
% Fig.~\ref{fig:simple-estimator} in\S\ref{ssec:overview:opt:cost} for the mathematically rigorous description):
% \begin{quote}
%   Sample a random point~$A\in\RR$ with the probability of~$A=a$ proportional to $\abs{\phi(a)}$, for all~$a\in\RR$; run the VQA
%   with parameter set to~$A$, call the resulting (single shot) measurement result $F(A)$; return $F(A)$, correcting for the sign
%   and magnitude of $\phi(a)$.
% \end{quote}

% %Distinguishing between a PSR and a ``feasible'' PSR allows us to also treat ``approximate'' proper
% %shift rules, where the equation in~\eqref{eq:def-feasible--proper-shift-rule} is satisfied up to some approximation error.

As the tag ``Nyquist'' suggests, our PSR is based on Fourier analysis.
The connection between PSRs and Fourier spectra was probably first observed in~\cite{GilVidal-Theis:CalcPQC:2018}, and exploited
in~\cite{Wierichs-Izaac-Wang-Lin:gen-shift:2021} and elsewhere; the present paper shows ways to exploit Fourier analytic
properties of perturbed-parametric expectation-value functions.  Indeed, our results are based on the observation that the
perturbed-parametric expectation-value function is --- in finite dimension --- band limited: Its Fourier spectrum (i.e., support
of the Fourier transform) is contained in
an interval $[-K,+K]$, where~$K$ is determined by the difference between the largest and smallest eigenvalues of~$A$.
% the interval $[\lambdamin(A)/\hbar - \lambdamax(A)/\hbar, \lambdamin(A)/\hbar - \lambdamax(A)/\hbar]$.
For functions like that, immediately the Nyquist-Shannon Sampling Theorem comes to mind, here stated in the form of the
Shannon-Whittaker Interpolation Formula: E.g., for functions~$f$ with Fourier spectrum contained in $[-\nfrac12,+\nfrac12]$
\begin{subequations}
  \begin{equation}\label{eq:intro:ShannonWhittaker}
    f = \sinc \hexstar f := \sum_{n\in\ZZ} \sinc(\place - n) \, f(n)
  \end{equation}
  where ``$\place$'' is placeholder for the variable, $\sinc := \sin(\pi\place)/(\pi\place)$ and ``$\hexstar$'' refers to
  sum-con\-vo\-lu\-tion\footnote{Note: ``$\hexstar$'' $\ne$ ``$*$''; we won't use ``$\hexstar$'' in the main parts of the paper.} as
  defined with the RHS.  From here, obtaining a derivative\footnote{We use ``$\partial g$'' and ``$g'$'' interchangeably for the
    derivative of a function~$g$ of a single argument.} seems straight forward:
  \begin{equation}\label{eq:intro:∂ShannonWhittaker}
    \partial f = \partial(\sinc \hexstar f) = (\partial\sinc) \hexstar f = f \hexstar (\partial\sinc).
  \end{equation}
  The resulting PSR\footnote{Warning: It isn't one, as the measure is not finite.} $\sinc'(\place -n)$ $(n\in\ZZ)$ decays only
  as $1/\abs{n}$ for $\abs{n}\to\infty$.  This can be fixed by picking a sweet spot point~$x$ for which $\sinc'(x-n))$
  decays as $1/\abs{n}^2$, namely $x=\nfrac12$.  Restarting with
  \begin{equation}\label{eq:intro:derif-at-1/2}
    f'(\nfrac12) = \sum_{n\in\ZZ} \sinc'(\nfrac12-n) \, f(n)
  \end{equation}
  and then performing a reflection of~$f$ (the details are in~\S\ref{ssec:overview:𝛍-𝛟}), results in, for every $x\in\RR$,
  \begin{equation}\label{eq:intro:the-shift-rule--K=1/2}
    f'(x) = \sum_{a\in\nfrac 12 + \ZZ} \frac{ (-1)^{a+\nfrac12} }{ \pi a^2 } f(x-a),
  \end{equation}
  as $\sinc'(\nfrac12-n) = (-1)^{n-1}/(\pi(n-\nfrac12)^2)$.
\end{subequations}

Generalizing to other values of~$K$ for the interval $[-K,+K]$ containing the Fourier spectrum, one obtains the following family
of PSRs, which we call Nyquist Shift Rules:
\begin{equation}\label{eq:intro:general-shift-rule}
  \phi_K := 2K \cdot \sum_{a\in \nfrac12+\ZZ} \frac{ (-1)^{a+\nfrac12} }{ \pi a^2 } \, \delta_{a/2K}, \text{ where $K>0$;}
\end{equation}
with $\delta_x$ denoting the Dirac point measure at the point~$x\in\RR$.

\mypar%
While this hand-waiving argument captures the starting point of the research presented in this paper, there are a number of
problems with it.  First of all, the Shannon-Whittaker formula~\eqref{eq:intro:ShannonWhittaker} as presented doesn't work:
Plugging in $f=\cos(\pi\place)$ at the point~$\nfrac12$ gives infinity
%%%
% sin𝝅(½-𝑛) cos𝝅𝑛  = 1
%%%
on the RHS.\footnote{Indeed, for it to converge for all bounded band-limited functions~$f$, we must require that the Fourier
  spectrum of~$f$ is contained in $[-\nfrac12+\delta,+\nfrac12-\delta]$, and the speed of convergence will depend
  on~$\delta>0$.} %
Secondly, in~\eqref{eq:intro:∂ShannonWhittaker}, we are taking the derivative under a sum which (even if it converged) doesn't
converge absolutely, causing headaches and high blood pressure.  However, if we are able to arrive
at~\eqref{eq:intro:derif-at-1/2}, then~\eqref{eq:intro:the-shift-rule--K=1/2} will follow, and the feasibility of the Nyquist
shift rules~\eqref{eq:intro:general-shift-rule} will be established.

\mypar%
\emph{The contributions one by one.}  %
\textcal{(1)}~Counting the observation about the Fourier spectra of perturbed-parametric expectation-value functions (see
\S\ref{sec:overview:function-spaces}) as the first contribution of this paper, \textcal{(2)} the second contribution is a
rigorous proof that the Nyquist shift rule~\eqref{eq:intro:general-shift-rule} is feasible for expectation-value functions where
the difference between largest and smallest eigenvalues of~$A$ is at most\footnote{The factor $2\pi\hbar$ comes from the choice
  of Fourier transform in this paper.} $K/2\pi\hbar$; see~\S\ref{ssec:overview:feasibility}.

\textcal{(3)} \emph{En passant,} in \S\ref{ssec:overview:𝛍-𝛟}, we characterize the set of all feasible PSRs (for fixed~$K>0$) in
terms of their Fourier transforms --- mirroring the characterization via a system of linear equations
in~\cite{Theis:opt-shiftrules:2021} but with infinitely many equations.  As a first consequence of that characterization, we can
show (\S\ref{sssec:overview:𝛍-𝛟:quantitative}) that the space\footnote{This is an affine space of functions.} of all feasible
PSRs has infinite dimension (for each fixed~$K$) --- so you could definitely say that there are many of them.

\textcal{(4)} As a second consequence of the characterization in \S\ref{ssec:overview:𝛍-𝛟}, we can prove some non-existence
results for particularly nice feasible PSRs (\S\ref{ssec:overview:Impossible}), e.g., it's a one-liner to see that there is no
PSR with compact support that is feasible for the frequency band $[-\nfrac12,+\nfrac12]$.  It takes a little more effort to
prove that there is no feasible PSR which is exponentially concentrated, indeed, there isn't even a family of exponentially
concentrated PSRs that are, in a wide sense,``nearly feasible'' (definition in
Corollary~\ref{cor:overview:Impossible:non-exist-expo-concentr}): You cannot let the approximation error tend to~$0$ without
blowing up either the tail or the ``cost'' (see next item) of the PSR.

\textcal{(5)} The ``cost'' of PSRs: In \S\ref{ssec:overview:opt:cost}, in parallel with the results
in~\cite{Theis:opt-shiftrules:2021}, we show that the (total-variation) norm of a PSR equals the \emph{worst-case standard
  deviation} of a single-shot estimator for the convolution of the expectation-value function with the PSR (see
Fig.~\ref{fig:simple-estimator} below).  Moreover, we prove that, for each $K>0$, there is no feasible PSR that has smaller norm
than our Nyquist shift rule~$\phi_K$, i.e., the Nyquist shift rules are optimal; see \S\ref{ssec:overview:opt}.  The
Stochastic Parameter Shift Rule of Banchi \& Crooks~\cite{Banchi-Crooks:stoch-param-shift:2021} has the same worst-case
standard deviation (but it isn't a ``proper'' shift rule).

\myparsmall%
While from a mathematical point of view, the Nyquist shift rules in~\eqref{eq:intro:general-shift-rule} are the ``right'' ones,
they require to query the expectation-value function at arbitrarily large parameter settings.  Indeed, the expected magnitude of
a parameter setting is~$\infty$ as the harmonic sum diverges.
% Recall that the Banchi-Crooks ``Approximate Stochastic Parameter Shift Rule'' also suffers from high-magnitude ($1/\beta$)
% parameter settings, but also requires a tiny time duration ($\beta$), which in some practical realizations of controllable
% quantum systems requires concentration both in the time and in the frequency domain.
%
The usual remedy for this situation in the context of Nyquist-Shannon-Whittaker theory is to truncate the sum, which introduces
an approximation error.
As this paper's contribution~\textcal{(6)}, in~\S\ref{ssec:overview:trunc}, we discuss truncation of Nyquist shift rules to
shifts of finite magnitude, and upper-bound the error: In terms of big-Os, the decay in the approximation error is comparable to
the Approximate Stochastic Parameter Shift Rule from~\cite{Banchi-Crooks:stoch-param-shift:2021}, namely $O(1/\Delta)$,
where $\Delta$ is the maximum magnitude of a shift.
\textcal{(7)}~We present results from preliminary numerical simulations based on an implementation of the Approximate Stochastic
Parameter Shift Rule and a simplified version of the truncated Nyquist shift rule, see~\S\ref{ssec:overview:numsim}; the
computer code is available
online\footnote{\label{fn:notebook-source-code}\texttt{\href{https://dojt.github.io/storage/nyquist/}{https://dojt.github.io/storage/nyquist/}}}
for experimenting.

\textcal{(8)}~Finally, in~\S\ref{ssec:overview:fold}, we discuss a more sophisticated way of confining parameter values to
finite magnitudes, which we call Folding.  We discuss a Parameter Folding method that exploits a deeper Fourier-analytic
property of perturbed-parametric expectation-value functions, based on eigenvalue perturbation theory.  Parameter Folding
our Nyquist shift rule results in an approximation error $O(1/c^2)$ for points~$x$ that have magnitude~$(1-\Omega(1))c$,
where~$c$ is the maximum magnitude of a parameter value.

We summarize the results about various types of proper, improper, and folded shift rules in Table~\ref{tab:compare-methods}.

\subsubsection*{Organization of the paper}
In the next Section, we give a technical, mathematically rigorous overview of the results of this paper, with an emphasis on
motivation and easy on the proofs.
The more technical proofs are in the Sections~3--8.
Section~\ref{sec:discussion} discusses some questions that arise, and points to future work.

Appendices \ref{apx:math} and~\ref{apx:perturb-th} hold math that is well known or easily derived, added for convenience; in
Appendix~\ref{apx:Half-deriv:linsyseq-lemma} we prove, for the sake of completeness of the presentation, the characterization of
feasible proper shift rules in a more general case than is needed for expectation-value functions.  Appendix~\ref{apx:numsim}
has additional graphs from numerical simulations (cf.~\S\ref{ssec:overview:numsim}).

\mypar%
The author aims for the content of this paper, \emph{from this point on,} to be fully mathematically unambiguous and rigorous,
and welcomes any criticism that points out where this goal has been missed.

\begin{table}[tp]\centering\small
  \begin{tabular}{ll@{\hspace{4em}}c@{\hspace{3em}}cc}%{ll@{}c@{}cc}%
    Method        &Variant                               & Apx err       & Max MoPV                      & Avg MoPV  \\
    \hline
    Banchi-Crooks &approximate                           & $O(\beta)$    & $4/\beta$                     & $4/\beta$       \\
    Nyquist       &exact                                 & $0$           & n/a                           & $\infty$         \\
    {}            &truncated                             & $\eps$        & $\abs{x}+$ $2/\eps\pi + 1/2K$ & $\abs{x}+$ $\ln(1/\eps) + O_K(1)$\\
    \multicolumn{2}{l@{}}{$\to$ parameter-folded, $x$ unconstrained}& $\eps$  & $<$ $4\pi K/\eps+p$      & \\
    \multicolumn{2}{l@{}}{$\to$ parameter-folded, $x\in[-p,+p]$}    & $\eps$  & $O_K(1/\sqrt{\eps})$     & $<$ $p + O_K( \ln(1/\sqrt{\eps}+2p) )$\\
  \end{tabular}
  \caption[]{\textbf{Comparison of methods.}  %
    \small%
    The table shows the approximation error (``apx err''), maximum magnitude of parameter values (``max MoPV'') and average
    magnitude of parameter values when simple estimators are used.  ``Banchi-Crooks'' refers to the Approximate Stochastic
    Parameter Shift Rule in~\cite{Banchi-Crooks:stoch-param-shift:2021}, which depends on a small parameter, $\beta>0$, that
    determines the query points.  The methods presented in this paper are dubbed ``Nyquist''; ``exact'' refers
    to~\eqref{eq:intro:general-shift-rule} with the Single-Shot Estimator, Fig.~\ref{fig:simple-estimator}; ``truncated'' refers to
    the method presented in \S\ref{ssec:overview:trunc} with a variant of the Single-Shot Estimator; ``parameter-folded'' refers to
    the method presented in \S\ref{ssec:overview:fold:quadratic} with the estimator from \S\ref{ssec:overview:fold:def}.  In the
    parameter-folded methods, the maximum and average magnitudes of parameter values depend not on the parameter value~$x$ at
    which the derivative is requested; instead it is influenced by a common divisor (cf.~Fn.~\ref{fn:def:common-divisor}),
    $1/p$, of the eigenvalues of~$A$.  For simplicity, we have set $\hbar=1/2\pi$.}%
  \label{tab:compare-methods}
\end{table}

\section{Technical overview}\label{sec:overview}%----------------------------------------------------------------------------
This section presents the results of the paper in full mathematical rigor, with an emphasis on motivation: we state the results
and discuss their relationships.  In terms of proofs, we give a few that help to motivate, banning the more technical or longer
ones to later sections.

\subsection{Math notations, definitions and preliminary facts}\label{ssec:overview:math}
In this paper, a \textit{finite measure} is a finite (not necessarily positive) regular Borel measure on~$\RR$; we use the
phrase \textit{signed measure} and \textit{complex measure,} resp., to emphasize that only real or also complex values are
allowed.  In this paper, finiteness is implied in the term signed/complex measure.

Complex measures have a canonical decomposition into two signed measures, which in turn have a canonical decompositions into two
positive measures (Jordan decomposition).  The sum of the resulting four (or 2) positive measures in the decomposition of a
finite measure~$\mu$ is called the \textit{total-variation measure} and denoted by $\abs{\mu}$.  The \textit{total-variation
  norm} of a finite measure~$\mu$ is defined as $\Nm{\mu} := \int \One \,d\abs\mu$, and satisfies
\begin{equation}\label{eq:total-variation-sup-norm}
  \Nm{\mu} = \sup_f \int f\,d\mu,
\end{equation}
where the supremum ranges over all measurable functions~$f$ with $\Nm{f}_\infty \le 1$.  This equation shows that convergence
$\mu_n \xrightarrow{n\to\infty} \mu$ in total variation implies convergence for every bounded measurable function~$f$:
$\int f\,d\mu_n \xrightarrow{n\to\infty} \int f\,d\mu$ --- we use that implicitly all the time.

\myparsmall%
Convergence in the total-variation norm of the sum of measures defining the Nyquist shift
rule~\eqref{eq:intro:general-shift-rule} can easily be checked.

\myparsmall%
For a measurable mapping $\tau\colon\RR\to\RR$ and a finite measure~$\mu$ on~$\RR$, we denote by $\tau(\mu)$ the image of the
measure under~$\tau$ (aka push-forward measure).  It is defined by $\tau(\mu)(E) = \mu(\tau^{-1}(E))$ for every Borel
set~$E$, or, equivalently by $\int f \,d\tau(\mu) = \int f\circ \tau\,d\mu$ for all bounded measurable functions~$f$.  For
example,
\begin{equation}\label{eq:overview:tau-of-delta}
  \tau(\delta_a) = \delta_{\tau(a)}.
\end{equation}
We will repeatedly use the following fact, without reference to it: For every measurable $\tau\colon\RR\to\RR$, the mapping
$\mu\mapsto\tau(\mu)$ takes complex measures to complex measures, is linear, and is continuous in the total-variation norm; in
particular, if $\mu := \sum_{j\in\NN} \alpha_j \mu_j$ is a total-variation-norm convergent series, where the $\mu_*$ are complex
measures and the $\alpha_*$ are complex numbers, then $\tau(\mu) = \sum_j \alpha_j \tau(\mu_j)$.

\mypar%
The terms \textit{Fourier transform,} \textit{support,} and \textit{Fourier spectrum} (the support of the Fourier transform)
refer to the concepts for tempered distributions; in the case of finite measures, the Fourier transform coincides (up to
technicalities) with the Fourier-Stieltjes transform:
\begin{equation*}
  \hat \mu(\xi) := \int e^{-2\pii\xi x} \,d\mu(x) \text{ ($\mu$ finite measure on~$\RR$, $\xi\in\RR$);}
\end{equation*}
note the $2\pi$ in the exponent.  The inverse Fourier transform is denoted by~$\check\place$.  We also use the notations
$(\dots)^\Hat$ and $(\dots)^\vee$ for the (inverse) Fourier transform of the function / measure / distribution in the place of
``\dots''.

\mypar%
While the results in this paper can be extended to multi-parameter functions, here, we focus solely on a single parameter.  As a
consequence, all function spaces will be spaces of functions on the real line.  For that reason, when using standard
function-space notation, we omit ``$(\RR)$'': I.e., $L^1$ (space of absolutely integrable functions on~$\RR$), $L^2$ (space of
square integrable functions on~$\RR$), $\mathscr{C}_0$ (space of continuous functions on~$\RR$ which vanish towards
$\pm\infty$), $\mathscr{C}^1$ (space of continuously differentiable functions on~$\RR$), $\mathscr{C}^1_b$ (space of bounded
continuously differentiable functions on~$\RR$), $\Schwartz$ (space of Schwartz-functions on~$\RR$), $\Schwartz'$ (space of
tempered distributions on~$\RR$), etc.

We use the word \textit{smooth} to mean continuously differentiable.

\subsection{Perturbed-parametric evolutions and proper shift rules}
The author apologizes to all physicists for choosing the barbaric normalization\footnote{This normalization simplifies the link
  to Fourier Analysis (with our Fourier transform $\hat f(\xi) = \int e^{-2\pii\xi x}f(x)\,dx$).} $h=1$, i.e., $\hbar:=1/2\pi$,
in the following definition.

\begin{definition}[Perturbed-parametric unitaries and expectation-value functions]\label{def:perturbed-parametric}
  For Hermitian operators $A,B$ on the same finite-dimensional Hilbert space, we refer to the operator-valued function
  \begin{equation}\label{eq:def:perturbed-parametric-unitary}
    x \mapsto e^{2\pii(xA+B)}
  \end{equation}
  as a \textit{perturbed-parametric unitary} or a \textit{perturbed-parametric unitary function.}
  A \textit{perturbed-pa\-ra\-met\-ric expectation-value function} is a function of the following form:
  \begin{equation}\label{eq:def:perturbed-parametric:PQC-EVfun}
    \RR\to\RR \colon
    x \mapsto \tr\bigl( M\, e^{2\pii(xA+B)} \varrho \, e^{-2\pii(xA+B)} \bigr)
  \end{equation}
  where $\varrho$ is a positive operator and~$M$ a Hermitian operator, both on the same Hilbert space as $A,B$.
  To avoid trivial border cases, we require that~$A$ is not a scalar multiple of the identity operator.
\end{definition}

This definition captures the typical setting in which VQAs are used today:
\begin{enumerate}[label=\arabic*.]
\item Prepare an $n$-qubit system in an initial state, say $\ket{0^n}$;
\item Subject the system to a time-dependent evolution (possibly depending on other parameters, but not~$x$);
\item Subject the system to a evolution with Hamiltonian $-(xA+B)$ for one unit of time;
\item Subject the system to further time-dependent evolution (possibly depending on other parameters, but not~$x$);
\item Measure an observable.
\end{enumerate}
Here, the state~$\varrho$ would be reached after step~\#2, and the observable~$M$ would be ~$\mathcal{E}^\dag(M_0)$,
where~$\mathcal{E}$ is the quantum operation resulting from the evolution in step~\#4 and~$M_0$ the observable in step~\#5.

\mypar%
In parallel\footnote{%
  In~\cite[\S\ref{OPT:ssec:overview:def-feasible-shift-rule}]{Theis:opt-shiftrules:2021}, this definition occurs with the
  (additional) restriction that the \emph{support} of~$\phi$ must be finite, targeting, specifically, shift rules as
  in~\eqref{eq:intro:finite-support-shift-rule}.  %
} %
with~\cite{Theis:opt-shiftrules:2021}, we make the following definitions.

\begin{definition}[Proper shift rule]\label{def:proper-SR}
  A \textit{proper shift rule (PSR)} is a finite signed Borel measure on~$\RR$; in a \textit{complex PSR} a complex measure is
  allowed.
\end{definition}
\begin{definition}[Feasible proper shift rule]\label{def:feasible-PSR}
  Let $\GenericFun$ be a vector space of real-valued, bounded, smooth functions defined on~$\RR$.
  A complex PSR~$\phi$ is called \textit{feasible for $\GenericFun$,} if for all $f\in\GenericFun$ and all $x\in\RR$, we have
  \begin{equation}\label{eq:def-feasible--proper-shift-rule}
    f'(x) = f*\phi (x),
  \end{equation}
  where ``$*$'' is convolution: For all $x\in\RR$,
  \begin{equation*}
    f*\phi(x) := \int f(x-s) \,d\phi(s).
  \end{equation*}
\end{definition}

Let's take as an example the symmetric difference quotient, $f'(x) \approx (f(x+\eps)-f(x-\eps))/2\eps$, for fixed $\eps>0$: The
PSR is $(\delta_{-\eps} - \delta_{\eps})/2\eps$.  It is feasible for the space\footnote{Slightly cheating here: Degree 1 and~2
  polynomials are, of course, not bounded functions.} of polynomials of degree at most~2, but it incurs an approximation error
(i.e., is not feasible) on any non-constant expectation-value function.

\subsection{Function spaces}\label{sec:overview:function-spaces}
In this paper, as in~\cite{Theis:opt-shiftrules:2021}, we address questions about parameterized quantum evolutions by proving
theorems about function spaces, and about membership of expectation-value functions in these spaces.  In this section, we define
the spaces and present the facts regarding membership of expectation-value functions.  For convenient reference,
Table~\ref{tab:function-spaces} gives an overview of all spaces we use.

\begin{table}[tp]
  \centering\small
  \begin{tabular}{cl}
    Symbol                      & Semantic \\[.2ex]
    \hline
    $\displaystyle\GenericFun$  & Generic function space (of smooth bounded real-valued functions)       \\[1ex]
    {}                          & \it Space of smooth bounded real-valued functions\dots                 \\[.5ex]
    $\displaystyle\oldFun_\Xi$  & \dots\ with Fourier spectrum contained in the frequency set $\Xi$.     \\[0.5ex]
    $\displaystyle\Fun_K$       & \dots\ with Fourier spectrum contained in the interval $[-K,+K]$.      \\[0.5ex]
    $\displaystyle\decFun_K$    & \dots\ in $\displaystyle\Fun_K$, which are of linear decay.            \\[0.5ex]
    $\displaystyle\sumFun_\Xi$  & \dots\ in the (non-orthogonal) direct sum $\displaystyle\oldFun_\Xi + \decFun_{\max\Xi}$.
  \end{tabular}
  \caption[]{\textbf{Overview of function spaces.}  %
    \small%
    A \textit{frequency set} is a finite, symmetric subset $\Xi \subset \RR$ with $\abs{\Xi}\ge 2$
    (Def.~\ref{def:frequency-set}).  For frequency sets, %%
    \(\displaystyle%
    \max\Xi := \max_{\xi\in\Xi}\xi = \max_{\xi\in\Xi}\abs{\xi} \) %
    holds.}
  \label{tab:function-spaces}
\end{table}

For convenience, and to avoid trivial border cases, we make the following definition.

\begin{definition}[Frequency set]\label{def:frequency-set}
  A \textit{frequency set} is a \emph{finite} set $\Xi\subset\RR$ that is symmetric (i.e., $-\Xi=\Xi$), and that has at least
  two elements, $\abs{\Xi}\ge 2$.
\end{definition}

Let $\Xi$ be a frequency set.  The space of real-valued, bounded, smooth functions with Fourier spectrum contained in~$\Xi$ is
denoted by\footnote{The equation follows from standard tempered distribution theory, based on the boundedness of the functions
  in~$\oldFun_\Xi$; cf.~Appendix~\ref{apx:math:diophant:finite-Fou-spec}.}
\begin{equation}\label{eq:def:oldFun}
  \oldFun_\Xi
  =
  \biggl\{
  f\colon x\mapsto \sum_{\xi\in\Xi} b_\xi e^{2\pii\xi x}
  \Bigm|
  b\in\CC^\Xi \text{ with $\forall\xi\in\Xi$: } b_{-\xi}=b_\xi^*
  \biggr\},
\end{equation}
as in~\cite{Theis:opt-shiftrules:2021}: There, convex optimization techniques and computations are explored, based on the fact
that this space has finite dimension.

The spaces in the present paper, though, are not finite dimensional: For positive real~$K$, we denote by $\Fun_K$ the set of
smooth, bounded real-valued functions on~$\RR$ with Fourier spectrum contained in $[-K,+K]$
\begin{equation*}
  \Fun_K := \lt\{ f\in\mathscr{C}^1_b  \bigm| \text{ $f$ real valued, } \Supp \hat f \subseteq [-K,+K] \rt\}.
\end{equation*}
From Paley-Wiener theory \cite[Theorem~ IX.12]{Reed-Simon:MathPhys-2-Fourier:1975} we know that every tempered distribution with
compact Fourier spectrum is indeed a function which is (extendable to a) holomorphic (function) with polynomial growth (in the
real part).  Hence, the set $\Fun_K$ contains all tempered distributions with two constraints: \textcal{(1)} on the Fourier
spectrum, and \textcal{(2)} boundedness.  The boundedness is necessary, or otherwise the convolution with finite measures would
not be well defined.

We find it more convenient to work with the function space, but the prime example that we are interested in are the
perturbed-parametric expectation-value functions from Def.~\ref{def:perturbed-parametric}.  The following proposition
gives the connection.  (We denote by $\lambdamin(\centerdot),\lambdamax(\centerdot)$ the smallest and largest, resp.,
eigenvalues of an operator.)

\begin{proposition}\label{prop:PQC-Fourier-spec}
  With the notations of Def.~\ref{def:perturbed-parametric}, set $K := \lambdamax(A) - \lambdamin(A)$.  The
  ex\-pec\-ta\-tion-val\-ue function is a bounded analytic function whose Fourier spectrum is contained in $%
  [-K,+K] %
  $, %
  and hence it is a member of~$\Fun_K$.
\end{proposition}

As everybody will surmise, the proof of the proposition, in \S\ref{ssec:fourier:fou-spec}, is based on the Lie Product Formula,
aka ``Trotterization''.

The work in~\cite{Theis:opt-shiftrules:2021} is based on the function space~$\oldFun_\Xi$, which is justified as every function
in that space is an expectation-value function of a variational quantum circuit
\cite[Prop.~\ref{OPT:prop:overview:FourierComputability}]{Theis:opt-shiftrules:2021}.
In the present paper, that is not the case: There are functions in $\Fun_K$ that are not perturbed-parametric expectation-value
functions\footnote{\label{fn:function-in-F_K-setminus-EVFs}%
  For example, the function $x \mapsto \sum_{j=2}^\infty 1/j^2 \cos(2\pi x/j)$ is in $\Fun_\nfrac12$, but does not arise from
  Def.~\ref{def:perturbed-parametric}.
}.  %
Working with the spaces $\Fun_*$ is nothing else but a convenient abstraction, simplifying the reasoning for much of
what we are doing in this paper --- particularly the non-existence proofs.  At some points, a more refined
Fourier-analytic abstraction of perturbed-parametric expectation-value (and unitary) functions is more convenient or necessary.
We prepare that with the following definition.

\begin{definition}[Linear decay]\label{def:overview:function-spaces:linear-decay}
  For non-negative real constants $c,C$, we say that a continuous function~$f$ of a real variable \textit{decays linearly} (or
  is of \textit{linear decay}) \textit{with decay constants~$(c,C)$}, if for all $x\in\RR$ with $\abs{x} \ge c$ we have
  \begin{equation*}
    \abs{ f(x) } \le C/\abs{x}
  \end{equation*}
\end{definition}

Defining, as usual, the \textit{difference set} of a set~$\Lambda$ as
$\Diff\Lambda := \{\lambda-\lambda' \mid \lambda,\lambda'\in \Lambda\}$, textbook eigenvalue perturbation theory (in finite
dimension!) will give us the following.

\begin{proposition}[Fourier-decomposition for perturbed-parametric EVFs]\label{prop:fou-decomp-evf}
  With the notations of Def.~\ref{def:perturbed-parametric}, let~$f$ be a perturbed-parametric expectation value function, and
  denote $\Xi := \Diff\Spec A$.

  There exists a function $f_1\in\oldFun_\Xi$ such that $f_0 := f - f_1$ is of linear decay.
\end{proposition}

In Section~\ref{sec:fourier}, we obtain Prop.~\ref{prop:fou-decomp-evf} as a consequence of a corresponding
Fourier-de\-com\-po\-si\-tion for perturbed-parametric unitary functions (Prop.~\ref{prop:fourier:decomp}).  Details and the
proofs of both propositions are in~\S\S\ref{ssec:fourier:decomp}--\ref{ssec:fourier:proof-decomp}.

With the notations of Prop.~\ref{prop:fou-decomp-evf}, as the tempered distributions with support contained in a given set
form a subspace, and since both $f$ (by Prop.~\ref{prop:PQC-Fourier-spec}) and~$f_1$ have Fourier spectra that are
contained in $[-K,+K]$ for $K:=\max\Xi$, we find that~$f_0$ also has a Fourier spectrum that is contained in $[-K,+K]$.  This
motivates the following definition.

\begin{definition}\label{def:overview:function-spaces:decFun}
  For $K>0$, we denote by $\decFun_K$ the vector space of analytic functions that are of linear decay for some choice of decay
  constants, and whose Fourier spectrum is contained in the interval~$[-K,+K]$.
\end{definition}

Combining Propositions \ref{prop:PQC-Fourier-spec} and~\ref{prop:fou-decomp-evf}, we obtain the following as a direct
consequence.

\begin{corollary}\label{cor:EVF-direct-sum}
  With the notations of Def.~\ref{def:perturbed-parametric}, set $\Xi := \Diff\Spec A$.
  Expectation-value functions are members of the (non-orthogonal) direct sum
  \begin{equation*}
    \sumFun_\Xi := \oldFun_\Xi + \decFun_{\max\Xi}.
  \end{equation*}
  In other words, every perturbed-parametric expectation value function~$f$ can be \emph{uniquely} decomposed as $f = f_1+f_0$
  where
  \begin{itemize}
  \item $f_1 \in \oldFun_\Xi$, i.e., a (finite) linear combination of $\sin(2\pi\xi\place)$ and $\cos(2\pi\xi\place)$ with
    frequencies $\xi\in\Xi$;
  \item $f_0\in\decFun_{\max\Xi}$, i.e., an analytic function of linear decay whose $L^2$-Fourier transform is a
    square-integrable function with support in $[\lambdamin(A) - \lambdamax(A), \lambdamax(A) - \lambdamin(A)]$.
  \end{itemize}
\end{corollary}

The only part of Corollary~\ref{cor:EVF-direct-sum} that we have not discussed yet is the directness of the sum (which is
equivalent to the uniqueness of the decomposition); we refer to Prop.~\ref{prop:apx:math:diophant:direct-sum} in
Appendix~\ref{apx:math:diophant}.

\mypar%
The convenience gain of the Fourier-decomposition theorem lies in the fact that functions of linear decay are square integrable.
So the Fourier transform $\widehat{f_0}$ is a function (not an evil tempered distribution), and, thanks to the compact support,
the inverse Fourier transform is realized simply by the integral:
\begin{equation*}
  f_0(x) = \int_{-K}^K e^{2\pii\xi x} \widehat{f_0}(\xi) \,d\xi \quad\text{for all $x\in\RR$}
\end{equation*}
--- no tempered distribution theory required, not even improper integrals.
The consequence is that, in some occasions, $\sumFun_\Xi$ is technically easier to work with in terms of the Fourier transform
(e.g., see the discussion following Lemma~\ref{lem:overview:linsyseq} in \S\ref{ssec:overview:𝛍-𝛟:charac} below).  In other
occasions, the additional knowledge in $\sumFun_*$ does not seem to translate into reduced technical complexity of the proofs
(e.g., the proof of Prop.~\ref{prop:PQC-Fourier-spec} above).

\subsubsection{Feasibility for $\Fun_{\max\Xi}$ vs feasibility for $\sumFun_\Xi$}\label{sssec:overview:function-spaces:equiv-feasability}
As
\begin{equation*}
  \sumFun_\Xi \subsetneq \Fun_{\max\Xi},
\end{equation*}
every PSR feasible for~$\Fun_{\max\Xi}$ is also feasible for~$\sumFun_\Xi$, and it is conceivable that there could be PSRs that
are feasible for~$\sumFun_\Xi$ but not feasible for~$\Fun_{\max\Xi}$.  This is not the case.  Based on the characterization of
PSRs feasible for $\Fun_{\max\Xi}$ in \S\ref{ssec:overview:𝛍-𝛟:charac} below, we will show the
following.

\begin{proposition}\label{prop:overview:function-spaces:equiv-feasability}
  Let $\Xi$ be a frequency set.
  Every complex PSR feasible for~$\sumFun_\Xi$  is also feasible for $\Fun_{\max\Xi}$.
\end{proposition}

(The proof is in \S\ref{ssec:Half-deriv:FDT}.)
Hence, for every frequency set~$\Xi$, the PSRs feasible for the smaller space~$\sumFun_\Xi$ of
Corollary~\ref{cor:EVF-direct-sum} are exactly the same as those feasible for the larger space~$\Fun_{\max\Xi}$ used in
Prop.~\ref{prop:PQC-Fourier-spec}.

\mypar%
In terms of practical use of the PSRs, the structure of $\sumFun_\Xi$ can be exploited: In \S\ref{ssec:overview:fold} below, we
will discuss the concept of Folding which exploits the linear decay condition to achieve a shift-rule-ish method with an
approximation error.  The approximation error decays quickly (quadratically) with the magnitudes of the parameter values at
which the perturbed-parametric expectation-value function is queried.

\subsection{Reflection, dilation, and the space of feasible proper shift rules}\label{ssec:overview:𝛍-𝛟}
If~$\phi$ is a PSR feasible for a space~$\GenericFun$ (see Def.~\ref{def:feasible-PSR}) then, for $f\in\GenericFun$ and
all~$x\in\RR$ we have $f'(x) = \int f(x-s)\,d\phi(s)$.  Choosing $x=0$ as an anchor point, as it were, we find
$f'(0) = \int f\,d\bar\phi$ where $\bar\phi := (-\place)(\phi)$ is the image of~$\phi$ under the measurable mapping
$(-\place)\colon \RR\to\RR\colon x\mapsto -x$, the reflection on~$0$.  Conversely, any complex measure~$\bar\phi$ satisfying
that integral equation (for all~$f$) gives rise to a feasible PSR, via $\phi := (-\place)(\bar\phi)$.
This approach leads to the characterization of the set of (finite-support) shift rules via the system of equations
in~\cite[Eqn.~\eqref{OPT:primal:GlSys:hatphi-equals-2piixi}]{Theis:opt-shiftrules:2021}.

In the present paper, we choose an anchor point different from~$0$ --- simply because it is convenient for the concrete
Nyquist Shift Rules that we will work with: The anchor point will be~$\nfrac12$.  Moreover, as discussed in the introduction
(and as is evident in~\eqref{eq:intro:general-shift-rule}), the Nyquist shift rule takes its simplest form when the Fourier
spectrum is $[-\nfrac12,\nfrac12]$.

For these reasons, we will consider measures~$\mu$ with the following property:
\begin{equation}\label{eq:overview:deriv-half--measure}
  g'(\nfrac12) = \int g \, d\mu
  \qquad
  \text{for all $g\in\Fun_{\nfrac12}$;}
\end{equation}
the space which~$g$ is allowed to be in will be swapped out for $\sumFun_\Xi$ in some places below.  In any case, we will speak
of a \textit{derivative-computing measure.}

We start by providing the details about the connection between $\mu$'s and PSRs.  First of all, note that
if~$\phi$ is a PSR for a function space containing a certain function~$g$, then
\begin{equation*}
  g'(\nfrac12) = g*\phi(\nfrac12) = \int g(\nfrac12-s) \,d\phi(s) = \int g\circ\tau \,d\phi \int g \,d\tau(\phi),
\end{equation*}
where $\tau\colon\RR\to\RR\colon s\mapsto \nfrac12-s$ is the reflection on~$\nfrac12$.  This means: If~$\phi$ is a PSR feasible
for~$\Fun_\nfrac12$, then $\mu:=\tau(\phi)$ satisfies~\eqref{eq:overview:deriv-half--measure}.  Dilation then gives feasibility
for~$\Fun_K$ in the case $K\ne\nfrac12$.  We summarize in the form of the following proposition (proof details in
\S\ref{ssec:𝛍-𝛟:general}).

\begin{proposition}\label{prop:overview:𝛍-𝛟}
  For every $K>0$, there is a canonical isomorphism between the (real or complex, resp.)\ affine space\footnote{As, strictly
    speaking, we ``don't know yet'' whether these spaces are empty or not, we are abusing terminology here, allowing the empty
    set as an affine space.} of (signed or complex, resp.)\ measures~$\mu$ satisfying~\eqref{eq:overview:deriv-half--measure},
  and the affine space of (real or complex, resp.)\ PSRs feasible for $\Fun_K$.
\end{proposition}

\subsubsection{The space of PSRs}\label{ssec:overview:𝛍-𝛟:charac}
We now come to the characterization of the set of PSRs feasible for $\Fun_K$, for any $K>0$, and for $\sumFun_\Xi$ for any
frequency set~$\Xi$: We mentioned in Prop.~\ref{prop:overview:function-spaces:equiv-feasability} that these sets are the same,
and we will now develop the machinery to prove it.  By what we just summarized, the task is to characterize the
derivative-computing measures, i.e., the finite measures satisfying~\eqref{eq:overview:deriv-half--measure}.

The following lemma is the infinite-support extension of the corresponding statement on finite
support~\cite[Eqn.~\eqref{OPT:primal:GlSys:hatphi-equals-2piixi}]{Theis:opt-shiftrules:2021}.

\begin{lemma}[Fourier-analytic characterization of derivative-computing measures]\label{lem:overview:linsyseq}
  Let~$\mu$ be a finite measure on~$\RR$.
  For~\eqref{eq:overview:deriv-half--measure} to hold, it is necessary and sufficient that
  \begin{equation}\label{eq:overview:linsyseq}
    \text{for all $\xi\in[-\nfrac12,+\nfrac12]$}\colon \qquad \hat\mu(\xi) = -2\pii\xi e^{-i\pi\xi}.
  \end{equation}
\end{lemma}
Recall that the Fourier transform of a tempered distribution which arises from integrating against a complex measure coincides
with integrating against the Fourier-Stieltjes transform of the measure.  As the Fourier-Stieltjes transform of a complex
measure is a continuous function, evaluating $\hat\mu$ at individual frequencies~$\xi$ is well defined.

Let us consider the necessity of the condition.

\begin{proof}[Proof that~\eqref{eq:overview:deriv-half--measure} implies \eqref{eq:overview:linsyseq}]
  For all $\xi\in[-\nfrac12,+\nfrac12]$, the real and imaginary parts of the function $x\mapsto e^{-2\pii\xi x}$ lie in
  $\Fun_{\nfrac12}$.  Assume that~$\mu$ is a complex measure satisfying~\eqref{eq:overview:deriv-half--measure}.  By the complex
  linearity of both the derivative at~$\nfrac12$, $f\mapsto f'(\nfrac12)$, and of the integral $f\mapsto \int f\,d\mu$, by
  applying both to the real- and imaginary parts of $x\mapsto e^{-2\pii\xi x}$, we find that
  \begin{multline*}
    -2\pii\xi e^{-i\pi\xi}
    = (e^{-2\pii\xi\place})'(\nfrac12)
    = (\cos(-2\pii\xi\place))'(\nfrac12) + i\, (\sin(-2\pii\xi\place))'(\nfrac12)
    \\
    = \int \cos(-2\pii\xi\place)\,d\mu + i \int \sin(-2\pii\xi\place) \,d\mu
    = \int e^{-2\pii\xi\place} \,d\mu
    =
    \hat\mu(\xi).
  \end{multline*}
  This proves the condition~\eqref{eq:overview:linsyseq}, demonstrating its necessity in Lemma~\ref{lem:overview:linsyseq}.
\end{proof}

As for the proof of the sufficiency-direction, it is based on Fourier Inversion.  Performing that rigorously for
only~$\sumFun_*$ is less technical than the general case as stated in the lemma, which requires tempered distribution arguments
and Paley-Wiener-Schwartz theory.  As, in view of Corollary~\ref{cor:EVF-direct-sum}, the general case is mostly of academic
interest, in \S\ref{ssec:Half-deriv:linsyseq-lemma}, we prove the lemma only for the special case of~$\sumFun_*$, and demote the
general case, $\Fun_*$, to Appendix~\ref{apx:Half-deriv:linsyseq-lemma}.

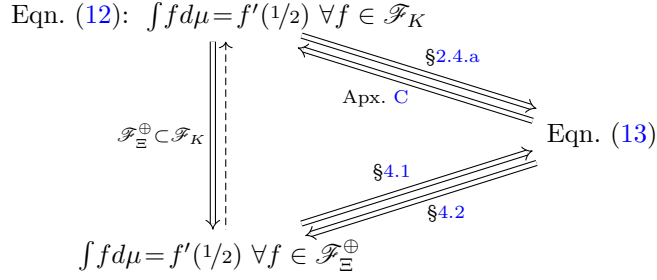
\begin{figure}[tp]
  \begin{equation*}
    \begin{tikzcd}[sep=large]
      \text{Eqn.~\eqref{eq:overview:deriv-half--measure}:\ }\; \text{$\int\! f d\mu\!=\!f'(\nfrac12)$ } \forall f\in\Fun_K
      \arrow[dr, Rightarrow, shift left, "\text{\S\ref{ssec:overview:𝛍-𝛟:charac}}"]
      \arrow[dd, dashleftarrow, shift left]                                         & \\
      {}                                                                            & \text{Eqn.~\eqref{eq:overview:linsyseq}}
      {}                                                                              \arrow[ul, Rightarrow, shift left, "\text{Apx.~\ref{apx:Half-deriv:linsyseq-lemma}}"]
      {}                                                                              \arrow[dl, Rightarrow, shift left, "\text{\S\ref{ssec:Half-deriv:linsyseq-lemma}}"]   \\
      \text{$\int\! f d\mu\!=\!f'(\nfrac12)$ } \forall f\in\sumFun_\Xi
      \arrow[ur, Rightarrow, shift left, "\text{\S\ref{ssec:Half-deriv:FDT}}"]
      \arrow[uu, Leftarrow, shift left, "\sumFun_\Xi\subset\Fun_K"]        &       %\\
    \end{tikzcd}
  \end{equation*}
  \caption[]{\textbf{Fourier-analytic characterization of derivative-computing measures.  }  %
    \small%
    This diagram shows the logical implications between (the versions of) the statements in Lemma~\ref{lem:overview:linsyseq}:
    Eqn.~\eqref{eq:overview:linsyseq}, ``$\hat\mu(\xi)=\dots$'', and variants of Eqn.~\eqref{eq:overview:deriv-half--measure},
    ``$\int f\,d\mu=f'(\nfrac12)$''.  The labels on the arrows give the locations of the proofs.
    \\
    The dashed arrow is implied by the north-east and north-west implications; it yields Prop.~\ref{prop:overview:function-spaces:equiv-feasability} (stated in
    \S\ref{sssec:overview:function-spaces:equiv-feasability}, proof in \S\ref{ssec:Half-deriv:FDT}).
    Lemma~\ref{lem:overview:linsyseq} as stated consists of the south-east and north-west implications: Although the south-east
    implication is implied by the north-east implication proven in Section~4 together with the south implication, its
    presentation serves as motivation in the technical overview, Section~\ref{sec:overview}.  }\label{fig:FouA-charact-logic}
\end{figure}

As the proofs of the versions of Lemma~\ref{lem:overview:linsyseq} are somewhat spread out, we refer to
Fig.~\ref{fig:FouA-charact-logic} for a visual guide.

\mypar%
Revisiting \S\ref{sssec:overview:function-spaces:equiv-feasability}, suppose $\Xi \subset [-\nfrac12,\nfrac12]$ is a frequency
set with $\max\Xi=\nfrac12$.  In Lemma~\ref{lem:Half-deriv:FDT} of \S\ref{ssec:Half-deriv:FDT} we will prove
that~\eqref{eq:overview:linsyseq} is already implied by requiring that the equation $g'(\nfrac12) = \int g \, d\mu$
in~\eqref{eq:overview:deriv-half--measure} holds only for $g\in\sumFun_\Xi$ --- instead of for all $g\in\Fun_\nfrac12$.  This
will prove Prop.~\ref{prop:overview:function-spaces:equiv-feasability} together with the reflection on~$\nfrac12$, translation,
and dilation techniques laid out in \S\ref{ssec:overview:𝛍-𝛟} above.

\subsubsection{``Quantitative'' view}\label{sssec:overview:𝛍-𝛟:quantitative}
Using Lemma~\ref{lem:overview:linsyseq}, we are now ready to make a ``quantitative'' statement about the set of measures~$\mu$
satisfying~\eqref{eq:overview:deriv-half--measure}.  At this point, as it were, we ``don't know yet'' whether a single
such~$\mu$ exists, but if a single one exists, the set is large.

\begin{proposition}[Space of feasible PSRs]\label{prop:overview:feasSR-space}
  The (real) affine space of signed measures~$\mu$ satisfying~\eqref{eq:overview:linsyseq} is either empty or has infinite
  dimension.
\end{proposition}

In \S\ref{ssec:Half-deriv:feasSR-space}, the proof of the proposition proceeds by adding to a single such signed measure
infinitely many linearly independent signed measures whose Fourier transform vanishes on $[-\nfrac12,+\nfrac12]$, and invoking
Lemma~\ref{lem:overview:linsyseq}.

\subsection{Feasibility of the Nyquist shift rules}\label{ssec:overview:feasibility}
It's about time to discuss the feasibility of the Nyquist shift rules $\phi_K$, $K>0$, from~\eqref{eq:intro:general-shift-rule},
for $\Fun_K$; throughout this section, $\phi_*$ will be as in~\eqref{eq:intro:general-shift-rule}.

As indicated in the introduction, we will not attempt to pursue a strategy based on the Shannon-Whittaker Interpolation Theorem.
Instead, we apply Lemma~\ref{lem:overview:linsyseq} of the previous section for a suitable signed measure~$\mu$, which we will
define in a moment.  This~$\mu$ will allow us to define~$\phi_\nfrac12$ via reflection, and dilation will give us all other
$\phi_K$, $K>0$ (Prop.~\ref{prop:overview:𝛍-𝛟}).

\subsubsection{The derivative-computing measure}
\begin{subequations}\label{eq:overview:def:half--measure}
  With
  \begin{equation}\label{eq:overview:def-u}
    u\colon\ZZ\to\RR\colon
    n \mapsto \frac{ (-1)^{n-1} }{ \pi \, (n - \nfrac12)^2 }
  \end{equation}
  we define the following signed measure on~$\RR$:
  \begin{equation}
    \mu := \sum_{n\in\ZZ} u(n) \delta_n,
  \end{equation}
  where convergence in the total-variation norm can easily be checked.
\end{subequations}
One main technical piece of work in this paper is the following theorem.

\begin{theorem}\label{thm:overview:half--measure}
  If~$\mu$ is as in~\eqref{eq:overview:def:half--measure}, then the condition~\eqref{eq:overview:linsyseq} of
  Lemma~\ref{lem:overview:linsyseq} holds.
\end{theorem}

Applying Lemma~\ref{lem:overview:linsyseq} shows that the~$\mu$ in~\eqref{eq:overview:def:half--measure}
satisfies~\eqref{eq:overview:deriv-half--measure}, i.e., integrating against it gives the derivative at~$\nfrac12$, provided
that the integrand is a bounded smooth function with Fourier spectrum contained in $[-\nfrac12,+\nfrac12]$.

The proof of the theorem is in \S\ref{ssec:Half-deriv:pf-half--measure}.

\subsubsection{From $\mu$ to $\phi_K$}
We now state the feasibility of the Nyquist shift rules; the purely technical proofs, based on the reflection and dilation, are
in \S\ref{ssec:𝛍-𝛟:nyquist}.

\begin{corollary}\label{cor:shift-rule}
  The Nyquist shift rule
  \begin{equation*}
    \phi_{\nfrac12}\ =  \sum_{a\in \nfrac12+\ZZ} \frac{ (-1)^{a+\nfrac12} }{ \pi a^2 } \, \delta_a
  \end{equation*}
  (same as in~\eqref{eq:intro:general-shift-rule}), is feasible for~$\Fun_{\nfrac12}$.
\end{corollary}

\begin{corollary}\label{cor:overview:general-shift-rule}
  For each fixed real number $K > 0$, the Nyquist shift rule~\eqref{eq:intro:general-shift-rule} is feasible for~$\Fun_K$.
\end{corollary}

With Prop.~\ref{prop:PQC-Fourier-spec}, we see that the Nyquist shift rule $\phi_K$, $K>0$,
from~\eqref{eq:intro:general-shift-rule}, when convoluted against an expectation-value function~$f$ as in
Def.~\ref{def:perturbed-parametric} with $\lambdamax(A)-\lambdamin(A) \le K$, gives the derivative of~$f$.

\subsection{Cost concept and optimality}\label{ssec:overview:opt}
So we have PSRs feasible\footnote{Let it be repeated that, due to Prop.~\ref{prop:overview:function-spaces:equiv-feasability} in
  \S\ref{sssec:overview:function-spaces:equiv-feasability}, there's no difference between feasibility for $\Fun_*$ and for
  $\sumFun_*$; we stick to $\Fun_*$ as it is the ``simpler'' space.  }  %
for $\Fun_K$, $K>0$.  But are there better ones?
In this section, in answering that question in the negative, we understand the word ``better'' in a narrow technical sense to
mean smaller norm.  The norm of a PSR will turn out to be the worst-case standard deviation of the Single-Shot Estimator
for it.

This section explains why we have somewhat emphasized real-valued PSRs: It can be seen that, as the functions in our spaces
$\Fun_*$, $\sumFun_*$ are real-valued\footnote{So are, importantly, all expectation-value functions.}, the real part of any
feasible complex PSR is a feasible PSR with smaller norm (cf.\
\cite[Lemma~\ref{OPT:lem:ex+cost:wlog-real-coeffs}]{Theis:opt-shiftrules:2021}).  Hence, there doesn't seem to be an advantage
in allowing complex PSRs.

\subsubsection{The cost of a PSR}\label{ssec:overview:opt:cost}
As in~\cite{Theis:opt-shiftrules:2021}, we define the \textit{cost} of a PSR~$\phi$ simply as the (total-variation) norm of the
measure, $\Nm{\phi}$.  It is elementary that the norm coincides with the operator norm of the linear operator
\begin{equation*}
  f \mapsto f*\phi\colon \mathscr{M}_b \to \mathscr{M}_b
\end{equation*}
on the normed space of bounded, measurable real-valued functions of a real variable,
$\mathscr{M}_b$\label{def:space-measurable-fn}, with the supremum norm $\Nm{\place}_\infty$; for convenient reference, we note
two relevant inequalities in a remark.\footnote{Cf.~\eqref{eq:total-variation-sup-norm} for the first one.}

\begin{remark}\label{rem:overview:integral/convolution-ieqs}
  For every real-valued, bounded, measurable function~$f$ and every signed measure~$\phi$, we have
  \begin{subequations}\label{eq:overview:integral/convolution-ieqs}
    \begin{align}
      \int f \,d\phi
      &\le
        \Nm{f}_\infty \cdot \Nm{\phi},                          \label{eq:overview:integral-nm-ieq}
      \\
      \intertext{and in particular,}
      \Nm{f*\phi}_\infty
      &\le
        \Nm{f}_\infty \cdot \Nm{\phi}.                          \label{eq:overview:convolution-nm-ieq}
    \end{align}
  \end{subequations}
\end{remark}

By applications of standard facts about regular measures (and continuous, differentiable, analytic functions), the statements
remain true if we replace the qualification ``bounded measurable'' by ``bound\-ed continuous'', or ``bounded smooth''
$\mathscr{C}^1_b$, or ``bounded infinitely differentiable'', or ``bounded analytic''.

\mypar%
In addition to the application-side interpretations of the operator norm as the worst-case factor by which noise (e.g.,
higher-frequency contributions to the input) is amplified when taking the convolution, \cite{Theis:opt-shiftrules:2021}
discusses two more motivations for referring to $\Nm{\phi}$ as the cost of the PSR~$\phi$, which make sense only for
finite-support PSRs.  Here, we discuss the following: The norm of the PSR is the standard deviation of a single-shot estimation
algorithm for the convolution, in the worst case over all functions~$f$ and points~$x$.

\subparagraph{The Single-Shot Estimator.} %
As we want the convenience of talking about function spaces, we introduce a theoretical computer science concept that replaces a
``shot'', i.e., a single measurement of the observable~$M$ from Def.~\ref{def:perturbed-parametric} in the parameterized quantum
state:

\begin{definition}[Shot oracle]
  A \textit{shot oracle}, $F$, for a function~$f\colon\RR\to\RR$ takes as input $s\in\RR$ and returns a random number
  $F(s)\in\RR$ such that $\Exp F(s)=f(s)$, with the requirement that runs of the shot oracle (for same or different inputs)
  return independent results.
\end{definition}

Fig.~\ref{fig:simple-estimator} defines the Single-Shot Estimator.

\begin{figure}[tp]
  \begin{algorithm*}[H]\TitleOfAlgo{Single-Shot Estimator}
    \Input{$x\in\RR$                                            \\
      Plus: Access to \emph{shot oracle} $F$}                  %
    \KwOut{Random number $E(x)$ with $\Exp E(x) = f*\phi(x)$}
    \BlankLine %
    \nl Sample a random point~$A\in\RR$ according to the probability measure $\abs{\phi}/\Nm{\phi}$ \\
    \nl Invoke the shot oracle for the point $x-A$ to obtain $F(x-A)$                               \\
    \nl \Return $\displaystyle F(x-A) \cdot (d\phi/d\abs{\phi})(A) \cdot \Nm{\phi}$.
  \end{algorithm*}
  \caption[]{Algorithm \textbf{Single-Shot Estimator.}  %
    \small%
    Simple unbiased estimator for the convolution with a real proper shift rule (PSR) $\phi$.  The expression
    $d\phi/d\abs{\phi}$ stands for the Radon-Nikodym derivative of $\phi$ wrt $\abs{\phi}$; note that
    $d\phi/d\abs{\phi}(A) \in\{\pm1\}$.}\label{fig:simple-estimator}
\end{figure}

Apart from abstracting from quantum evolutions and measurements, there is absolutely nothing novel here: The Single-Shot Estimator
is really the direct generalization of the single shot estimator for finite-support shift rules.

\subparagraph{Variance of the Single-Shot Estimator.}  %
For analyzing the (worst-case) standard deviation of the Single-Shot Estimator, consider the following fact.

\begin{proposition}
  Let $\phi$ be a PSR, and assume a shot oracle\footnote{In terms of Def.~\ref{def:perturbed-parametric}, this
    corresponds to the situation where the observable, $M$, has $\pm1$ eigenvalues only, e.g., a multi-qubit Pauli or so.}  that
  returns values in $\{\pm1\}$.
  With input $x\in\RR$, the Single-Shot Estimator in Fig.~\ref{fig:simple-estimator} has variance
  $\displaystyle \Nm{\phi}^2 - \abs{f*\phi(x)}^2$.
\end{proposition}
\begin{proof}
  We use the notations from Fig.~\ref{fig:simple-estimator}.
  Suppose the Single-Shot Estimator receives $x\in\RR$ as input.  By our assumption that $F(\place)$ is $\pm1$-valued and since
  $d\phi/d\abs{\phi}$ is a function taking $\pm1$-values $\abs{\phi}$-almost everywhere, we have $E(x)^2 = \Nm{\phi}^2$, so that
  $\Exp E(x)^2 = \Nm{\phi}^2$.

  As advertised in Fig.~\ref{fig:simple-estimator}, the Single-Shot Estimator is indeed an unbiased estimator for $f*\phi$, i.e,
  $\Exp E(x) = f*\phi(x)$, and we conclude that $\Var E(x) = \Exp E(x)^2 - (\Exp E(x))^2 = \Nm{\phi}^2 - (f*\phi(x))^2$ as
  claimed.
\end{proof}

For a feasible PSR, the variance is $\Nm{\phi}^2 - f'(x)^2$.  Making a worst-case assumption on $f$ and~$x$, we take $f'(x)=0$
--- barren plateaus \cite{McClean-Boixo-Smelyanskiy-Babbush-Neven:barren:2018, Arrasmith-Holmes-Cerezo-Coles:barren:2021}
raising their flat heads again.  In other words, we upper-bound the variance by $\Nm{\phi}^2$.  This worst-case upper bound
leads to the worst-case standard deviation of $\Nm{\phi}$ for the Single-Shot Estimator.

\subsubsection{Review of weak duality}
While much of the convex optimization theory in~\cite{Theis:opt-shiftrules:2021} breaks down when moving from the finite
dimensional vector spaces $\oldFun_\Xi$ there to our infinite dimensional spaces $\Fun_K$, and moving from finite-support PSRs
to infinite-support ones, the \emph{Weak Duality Theorem,}
\cite[Prop.~\ref{OPT:prop:overview:weak-duality}]{Theis:opt-shiftrules:2021} goes through letter for letter.  We give here
a version of it that is tailored to our needs; the proof is in~\S\ref{ssec:opt:weak-duality}.

\begin{proposition}[Weak duality theorem]\label{prop:overview:weak-duality}
  Let~$\GenericFun$ be a vector space of real-valued, bounded smooth functions, which satisfies $f(-\place)\in\GenericFun$ if
  $f\in\GenericFun$.

  If~$\phi$ is a PSR feasible for $\GenericFun$, and $f\in\GenericFun$ with $\Nm{f}_\infty = 1$, then
  \begin{equation}\label{eq:overview:weak-duality-ieq}
    -f'(0)
    \le
    \Nm{\phi}.
  \end{equation}
  Under the antecedent condition, equality holds in~\eqref{eq:overview:weak-duality-ieq} if, and only if,
  inequality~\eqref{eq:overview:integral-nm-ieq} in Remark~\ref{rem:overview:integral/convolution-ieqs} is satisfied with
  equality for $f,\phi$.
\end{proposition}

The spaces $\Fun_*$ and $\sumFun_*$ satisfy the condition on~$\GenericFun$ in Prop.~\ref{prop:overview:weak-duality}: By the
usual tempered-distribution calculations we have for all tempered distributions~$f$, $\Supp (f(-\place))^\Hat = -\Supp \hat f$,
so that $f(-\place)\in\Fun_K$ if $f\in\Fun_K$.

\subsubsection{Optimality of the PSR}
Now we are ready to prove that, in terms of our cost concept,
% i.e., norm, i.e., worst-case standard deviation of the Single-Shot Estimator,
the Nyquist shift rules~\eqref{eq:intro:general-shift-rule} are optimal for $\Fun_K$.

\begin{theorem}\label{thm:overview:opt}
  Let~$K$ be a positive real number.  The Nyquist shift rule~\eqref{eq:intro:general-shift-rule} has smallest norm among all
  PSRs feasible for~$\Fun_K$.  The norm is $2\pi K$.
\end{theorem}

The proof is in \S\ref{ssec:opt:proof}; it comprises guessing a function~$f$ and a calculating equality of the two sides in the
Weak Duality Theorem Prop.~\ref{prop:overview:weak-duality}.

\subsubsection{Other optimal feasible PSRs}\label{sec:overview:otheropt}
In Section~\ref{sec:discussion}, the question will arise whether we can replace~$\phi_K$ as defined
in~\eqref{eq:intro:general-shift-rule} by another optimal feasible PSR --- maybe one whose support is more concentrated
near~$0$.  Here we note the following consequence of (the proof of) Theorem~\ref{thm:overview:opt} together with
Prop.~\ref{prop:overview:weak-duality}.

\begin{corollary}\label{cor:overview:otheropt}
  Fix $K>0$.  If~$\phi$ is a PSR feasible for $\Fun_K$, then~$\phi$ has smallest norm (i.e., is optimal) if, and only if,
  $\phi = -\phi^{0} + \phi^1$ where for $i=0,1$, $\phi^i$ is the restriction of~$\abs{\phi}$ to the set
  \begin{equation}\label{eq:overview:otheropt:supp}
    \lt\{  (\nfrac12 + m)/2K
    \bigm|
    m \equiv i \bmod 2
    \rt\}
  \end{equation}
\end{corollary}

This means that for all $K>0$, a PSR feasible for $\Fun_K$ is optimal for $\Fun_K$ if, and only if, its support is contained in
the set of numbers of the form $(\nfrac12+m)/2K$ where $m\in\ZZ$, and the signs alternate in the right way.

The proof of the corollary is sketched in \S\ref{ssec:opt:otheropt}.

\subsection{Truncation}\label{ssec:overview:trunc}
As discussed in the introduction, the Nyquist shift rules~\eqref{eq:intro:general-shift-rule} require to apply the
perturbed-parametric unitary from Def.~\ref{def:perturbed-parametric} for arbitrarily large values of the parameter~$x$
in~\eqref{eq:def:perturbed-parametric-unitary}, which may not be physically desirable, or even possible.  In this section, we
discuss ways to truncate our PSRs to a bounded set of shifts, at the expense of introducing a (hopefully small) approximation
error in the derivative.  The next section \S\ref{ssec:overview:fold} pursues the same goal using a technique that we call
Folding.  A third way to keep the values of the parameter small is described and used in the section on numerical simulations,
\S\ref{ssec:overview:numsim}.

\subparagraph{Preparations.}  %
We start by discussing the approximation error.
\begin{definition}[Near feasibility]\label{def:near-feasibility}
  For a space of real-valued, bounded, smooth functions~$\GenericFun$, and $\eps\ge 0$, let us say that a PSR~$\phi$ is
  \textit{$\eps$-nearly feasible for~$\GenericFun$,} if we have
  \begin{equation*}
    \Nm{ f*\phi - f'  }_\infty \le \eps \, \Nm{f}_\infty \quad\text{for all $f\in\GenericFun$.}
  \end{equation*}
\end{definition}

Feasibility is the same as being $0$-nearly feasible, of course.

While convenient for our calculations (see the next lemma), the example of the symmetric difference quotients shows the
limitations of this definition.  The approximation guarantee is well-known: For at least 3-times differentiable functions~$f$,
we have (for variable~$\eps$)
\begin{equation*}
  \absB{ f*\bigl(  (\delta_{-\eps} - \delta_{+\eps})/2\eps  \bigr)(x)   - f'(x)  }
  = \eps^2 \abs{f'''(x)} + o_{\eps\to0}(\eps^3);
\end{equation*}
the norm (cost) can readily be computed:
\begin{equation*}
  \Nmb{  (\delta_{-\eps} - \delta_{+\eps})/2\eps  } = 1/\eps.
\end{equation*}
(It has been observed before (e.g., \cite{Banchi-Crooks:stoch-param-shift:2021}) that the standard deviation of estimating based
on this PSR is $\Omega(1/\eps)$.)  We see that improvement in approximation error is bought at the cost of increasing standard
deviation of the estimator: As the sampling complexity (i.e., number of samples to reach a precision) grows quadratic in the
standard deviation, cost and benefit are of equal order.

However, according to our Def.~\ref{def:near-feasibility}, the symmetric difference quotient is only $O(1)$-nearly
feasible --- which erroneously\footnote{%
  Cf.\ the non-existence result in \S\ref{ssec:overview:Impossible:near-feasible}, though: There, a much weaker concept of
  ``$\eps$-nearly feasible'' is used (Corollary~\ref{cor:overview:Impossible:non-exist-expo-concentr}) which includes difference
  quotients.
} %
suggests that it is completely useless.

\mypar%
The following lemma shows how near-feasibility is being used.  Its proof explains why we get $\Nm{f}_\infty$ on the RHS.

% The advantage of the error type in Def.~\ref{def:near-feasibility} is the following: In situations where an absolute error
% is sufficient, it can be bounded effectively, as $\Nm{f}_\infty$ is usually known in the situation of
% Def.~\ref{def:perturbed-parametric} (it is the largest eigenvalue of~$M$).  This means that, if a near-feasibility result
% is available, decisions about truncation or folding can be made \emph{a priori,} without any knowledge of the magnitude of the
% actual function or its derivatives.\footnote{Of course, \emph{upper} bounds on the magnitudes of derivatives can be obtained
%   from the knowledge that the Fourier spectrum is bounded.}

\begin{lemma}\label{lem:overview:trunc:tv}
  Let $\phi,\tilde\phi$ be PSRs, and~$\GenericFun$ a space of smooth, bounded, real-valued functions.

  If~$\phi$ is feasible for $\GenericFun$, then $\tilde\phi$ is $\Nm{\tilde\phi - \phi}$-nearly feasible for $\GenericFun$.
\end{lemma}
\begin{proof}
  As~$\phi$ is feasible for~$\GenericFun$, we can replace $f' = f*\phi$, use the bilinearity of convolution, and then use the
  inequality~\eqref{eq:overview:convolution-nm-ieq} of Remark~\ref{rem:overview:integral/convolution-ieqs}:
  \begin{equation*}
    \Nm{ f*\tilde\phi - f'  }_\infty
    =
    \Nm{ f*(\tilde\phi - \phi) }_\infty
    \le
    \Nm{f}_\infty \cdot \Nm{\tilde\phi-\phi}.
  \end{equation*}
  This concludes the proof.
\end{proof}

\subparagraph{Truncation.}  %
Now we can discuss what happens to our $\phi_K$, $K>0$, when we truncate them.  In the next proposition, no effort has been made
to optimize the constant in front of the~$1/N$; the proof is in \S\ref{ssec:truncfold:trunc}.

\begin{proposition}\label{prop:overview:trunc}
  Let $K>0$, let~$N$ be a positive integer, and denote by $\phi_K^{(N)}$ the measure defined by truncating the sum defining
  $\phi_K$ in~\eqref{eq:intro:general-shift-rule} to the~$2N$ terms centered around~$0$, i.e.,
  \begin{equation}\label{eq:overview:trunc:truncated-general-shift-rule}
    \phi^{(N)}_K
    :=
    2K \cdot
    \sum_{\sstack{a\in \nfrac12+\ZZ\\ -N+\nfrac12\le a \le N-\nfrac12}}
    \frac{ (-1)^{a+\nfrac12} }{ \pi a^2 } \, \delta_{a/2K}.
  \end{equation}
  The PSR $\phi_K^{(N)}$ is $\eps$-nearly feasible for $\Fun_K$, where $\eps := \frac{4K}{\pi (N-\nfrac12)}$.
\end{proposition}

As discussed in the introduction, usually, a small error in the derivative is acceptable.  Setting
$N:=\ceil{ 4K/\eps\pi +\nfrac12 }$, Prop.~\ref{prop:overview:trunc} shows us that we can press the error to
below~$\eps$ by using shifts of magnitudes less than $2/\eps\pi + 1/2K$.

% See \S\ref{ssec:overview:numsim} for the numerical simulations.

\subsection{Folding}\label{ssec:overview:fold}
For truncating a PSR, we have taken an interval and treated specially all shifts that didn't fall into that interval: We simply
discarded them.  We discarded shifts, not parameter values: With a truncation interval $[-\Delta,\Delta]$, for the derivative at
a point $x\in\RR$, the parameter values at which the expectation-value function is queried to approximate the derivative at~$x$
were in $[x-\Delta,x+\Delta]$.

For folding a PSR, we can also take an interval, and treat specially either shifts (in shift folding) or parameter
values (in parameter folding) that don't fall into that interval.  Instead of simply discarding, though, we will replace the
shift/parameter value by one that falls into the interval.

\mypar%
It will make sense, in this subsection, to switch our attention from $\Fun_*$ to
$\sumFun_\Xi := \oldFun_\Xi + \decFun_{\max\Xi}$, i.e., we always think of the expectation-value function~$f$ as being
decomposed into $f=f_1+f_0$, as in Corollary~\ref{cor:EVF-direct-sum}.

\subsubsection{Folding with zero perturbation, i.e., $[A,B]=0$}\label{ssec:overview:fold:B=0}
We start by clarifying the relationship between this paper's Nyquist shift rule~\eqref{eq:intro:general-shift-rule} which has
infinite support and is feasible for~$\sumFun_\Xi$, and the known PSR which has finite support and is
feasible~\cite{Wierichs-Izaac-Wang-Lin:gen-shift:2021} for~$\oldFun_\Xi$.

For every positive integer~$L$, there exists a finite-support PSR feasible for $\oldFun_{\{-L,\dots,L\}}$
\cite{Wierichs-Izaac-Wang-Lin:gen-shift:2021}:
% which has norm $2\pi L$: Namely the one described in....  we present it in the same form as in
% \cite[Theorem~\ref{OPT:thm:optimality:d=1:one}]{Theis:opt-shiftrules:2021}:
\begin{equation}\label{eq:overview:recall--finitespec-opt--integer}
  \begin{gathered}
    \frac{1}{2L} \sum_{j=-L}^{L-1} u_j \, \delta_{(j+\nfrac12)/2L}, \qquad\text{where}
    \\
    u_j    := \frac{ \pi \, (-1)^{j+1} }{ \sin^2(\pi\,(j+\nfrac12)/2L) } \quad\text{for $j=-L,\dots,L-1$;}
  \end{gathered}
\end{equation}
the expression in the 2nd line is the first derivative of the modified Dirichlet kernel at $(j+\nfrac12)/2L$.   The
following remark makes this compatible with the notation in this paper, by dilation.

\begin{remark}[\cite{Wierichs-Izaac-Wang-Lin:gen-shift:2021}]\label{rem:overview:trunc:small-fun}
  Let $K,\xi_1$ be positive real numbers with the property that $L := K/\xi_1 \in \NN$, and define
  \begin{equation*}
    \Xi := \Bigl\{ \ell\cdot \xi_1 \Bigm| \ell = -L,-L+1, \dots, L-1,L \Bigr\}.
  \end{equation*}

  The following is a PSR feasible for $\oldFun_\Xi$ with support contained in\footnote{Note that the functions in $\oldFun_\Xi$
    are $\nfrac{1}{\xi_1}$-periodic.} $\lt[-1/2\xi_1, +1/2\xi_1\rt[$ and with norm\footnote{The norm is discussed
    in~\cite{Theis:opt-shiftrules:2021}, but it also follows from Theorem~\ref{thm:overview:fold:B=0} below.} $2\pi K$:
  \begin{equation*}
    \psi
    := 2K \cdot
    \sum_{\sstack{a\in \nfrac12+\ZZ\\ -L+\nfrac12\le a \le L-\nfrac12}}
    \frac{\pi}{(2L)^2} \cdot \frac{ (-1)^{a+\nfrac12} }{ \sin^2(\pi a/2L) } \, \delta_{a/2K}.
  \end{equation*}
\end{remark}

\newcommand{\modh}{\mathbin{\%}}
\newcommand{\modl}{\mathbin{\%_{\scriptscriptstyle+}}}
\newcommand{\modr}{\mathbin{\%_{\scriptscriptstyle-}}}
For a positive real number~$p$ and every $x\in\RR$ we will use the following notations:
\begin{equation}\label{eq:overview:fold:def-mod}
  \lt.
  \begin{array}[c]{l}
  x\modh p \\
  x\modl p \\
  x\modr p
  \end{array}
  \rt\} := \text{ the unique } y\in \lt\{
  \begin{array}[c]{l}
    \lt[-\nfrac{p}{2},+\nfrac{p}{2}\rt[ \\
    \lt[0            ,p            \rt[ \\
    \lt]-p           ,0            \rt]
  \end{array}
  \quad\rt\} \text{ with } x=y \bmod p\ZZ.
\end{equation}
(Recall the definition of $x=y\bmod p\ZZ$, which is: $y-x\in p\ZZ$.)  With the notations of the previous remark, the mapping
$x\to x\modh (1/\xi_1)$ folds its argument into a fundamental region of periodicity of the functions in~$\oldFun_\Xi$.

For a fixed real number $p>0$, by \textit{shift-folding} a PSR \textit{with mod-$p$}, we mean taking the image of the measure
under the mapping $(\place\modh p)\colon \RR \to \lt[-p/2, +p/2\rt[$.  We will give a general definition of shift folding in
Def.~\ref{def:overview:folding} below; here we are only interested in taking the remainder mod-$p$.  With the notations of
Remark~\ref{rem:overview:trunc:small-fun}, by shift-folding with mod~$p$ where $p=\nfrac{1}{\xi_1}$, we compile the measure of
every real number onto the corresponding point in the fundamental region $\lt[-1/2\xi_1, +1/2\xi_1\rt[$ of periodicity of the
function space~$\oldFun_\Xi$.

Now we are ready to establish the relationship between the infinite-support Nyquist shift
rules~\eqref{eq:intro:general-shift-rule} feasible for~$\sumFun_\Xi$ and known finite-support PSRs feasible for $\oldFun_\Xi$:
With the notation of Remark~\ref{rem:overview:trunc:small-fun}, if we take a Nyquist shift rule and shift-fold with
mod-$1/\xi_1$, then we obtain the known shift rules feasible~\cite{Wierichs-Izaac-Wang-Lin:gen-shift:2021} and optimal
\cite[Theorem~\ref{OPT:thm:optimality:d=1:one}]{Theis:opt-shiftrules:2021} for $\oldFun_\Xi$.

\begin{theorem}\label{thm:overview:fold:B=0}
  For all~$K,\xi_1,\psi$ as in Remark~\ref{rem:overview:trunc:small-fun}, and with~$\phi_K$ the Nyquist shift
  rule~\eqref{eq:intro:general-shift-rule}, we have $(\place\modh (\nfrac{1}{\xi_1}))(\phi_K) = \psi$.
\end{theorem}

The proof in~\S\ref{ssec:truncfold:fold-B=0} is based on partial fraction decomposition of $1/\sin^2$.

This theorem is relevant not only to make the point that, just as the PSRs~$\psi$ from Remark~\ref{rem:overview:trunc:small-fun}
are the right\footnote{Case of $\Xi$'s without gaps, otherwise see~\cite{Theis:opt-shiftrules:2021}.} PSRs for $\oldFun_\Xi$,
our Nyquist shift rules~\eqref{eq:intro:general-shift-rule} are, mathematically speaking, the right PSRs for $\sumFun_\Xi$.  The
theorem also motivates our general concept of folding, as defined in the next subsection, \S\ref{ssec:overview:fold:def}.

\mypar%
We have taken the approach here to present the special case where~$\Xi$ consists of equi-spaced frequencies with no gaps.  It
should be clear to the attentive reader that Theorem~\ref{thm:overview:fold:B=0} applies more generally in the case of frequency
sets~$\Xi$ that have a common divisor\footnote{%
  \label{fn:def:common-divisor}
  A \textit{common divisor} of a set~$S\subset\RR$ is a positive real number $\gamma_1$ such that $S\subset\gamma_1\ZZ$.}.  %
The frequency set~$\Xi$ having a common divisor is equivalent to the functions in $\oldFun_\Xi$ being periodic (the periods are
the reciprocals of the common divisors).  In~\cite{Theis:opt-shiftrules:2021} we prove that in the case of frequency sets with
gaps, the finite-support shift rule~$\psi$ from Remark~\ref{rem:overview:trunc:small-fun} is still feasible and optimal, but
there is a wealth of optimal finite-support shift rules, notably some with smaller support.

\subsubsection{General definitions for shift and parameter folding}\label{ssec:overview:fold:def}
We treat folding abstractly.  Here's the definition.

\begin{definition}[Folding function]\label{def:overview:folding}
  For a positive real number~$p$, we call a measurable function $\tau\colon\RR\to\RR$ a \textit{folding function} or a
  \textit{$p$-folding,} if~$\tau$ is equal to the identity modulo~$p\ZZ$, i.e.,
  \begin{equation*}
    \text{for all $s\in\RR$:}\quad \tau(s) = s \bmod p\ZZ;
  \end{equation*}
  in remainder calculus: $\tau(s)\modh p = s\modh p$ for all~$s$.
\end{definition}

The intuition behind a folding function is that in the decomposition $f = f_1 + f_0$ from Corollary~\ref{cor:EVF-direct-sum},
the folding function with $p:=1/\xi_1$ (notations from Remark~\ref{rem:overview:trunc:small-fun}) banks on the $p$-periodicity
of the $f_1$-term to recover the known shift rules~\cite{Wierichs-Izaac-Wang-Lin:gen-shift:2021} for~$f_1$.  It ignores the
$f_0$-term, introducing an approximation error.  The hope is that as $f_0$ decays towards $\pm\infty$, the approximation error
introduced through the folding can be made small.  That hope can be brought to fruition, as the next subsection shows.

The formal definitions are in the following Lemma~\ref{lem:overview:fold:folding-types}: we refer to
Item~\ref{lem:overview:fold:shift-folding} as \textit{shift folding,} and to Item~\ref{lem:overview:fold:param-folding} as
\textit{parameter folding.}
Fig.~\ref{fig:simple-folding-estimator} shows the estimator for parameter-folded PSRs that corresponds to the Single-Shot
Estimator for unfolded ones, Fig.~\ref{fig:simple-estimator}.

\begin{lemma}[Fundamental folding-lemma]\label{lem:overview:fold:folding-types}
  Let~$\Xi$ be a frequency set, and $p>0$ a real number such that $1/p$ is a common divisor of~$\Xi$.  Let $\phi$ be
  a PSR and~$\tau$ a $p$-folding.

  If $\phi$ is feasible for~$\oldFun_\Xi$, then the following hold.
  \begin{enumerate}[label=(\alph*)]
  \item\label{lem:overview:fold:shift-folding}%
    The PSR $\tau(\phi)$ is feasible for~$\oldFun_\Xi$, i.e., for all $f\in\oldFun_\Xi$ and all $x\in\RR$ we have
    \begin{equation*}
      f'(x) = \int f(x-\tau(s))\,d\phi(s);
  \end{equation*}
  and
  \item\label{lem:overview:fold:param-folding}%
    For all $f\in\oldFun_\Xi$ we have $f' = (f\circ\tau)*\phi$, i.e., for all $x\in\RR$,
    \begin{equation*}
      f'(x) = \int f(\tau(x-s))\,d\phi(s).
    \end{equation*}
  \end{enumerate}
\end{lemma}

The proof of Lemma~\ref{lem:overview:fold:folding-types} is given in \S\ref{ssec:truncfold:fold--fun-lemma}, as it illuminates
the relationships between the technical aspects of Def.~\ref{def:overview:folding} and shift/parameter folding.

\begin{figure}[tp]
  \begin{algorithm}[H]\TitleOfAlgo{Simple Folding Estimator}
    \Input{$x\in\RR$                                              \\
      Plus:         Access to \emph{shot oracle} $F$.}             %
    \KwOut{Random number $E(x)$ with $\Exp E(x) = (f\circ\tau)*\phi$}
    \BlankLine %
    \nl Sample a random point~$A\in\RR$ according to the probability measure $\abs{\phi}/\Nm{\phi}$      \\
    \nl Apply the folding function to obtain $B := \tau(x-A)$                                            \\
    \nl Invoke the shot oracle for the point $B$ to obtain $F(B)$                                        \\
    \nl \Return $\displaystyle F(B) \cdot (d\phi/d\abs{\phi})(A) \cdot \Nm{\phi}$.
  \end{algorithm}
  \caption[]{Algorithm \textbf{Simple Folding Estimator.}  %
    \small%
    Simple unbiased estimator for the parameter-folded PSR~$\phi$ with folding function~$\tau$.  Cf.\
    Fig.~\ref{fig:simple-estimator}.%
  }\label{fig:simple-folding-estimator}
\end{figure}

% Motivated by Fourier-Decomposition, Prop.~\ref{prop:fou-decomp-evf}, and the previous section, the idea of folding is to
% ``move'' evaluation points of~$f$ around in such a way that the resulting method gives an exact derivative of the
% $\oldFun_{\Diff\Spec A}$-component of the expectation-value function, has a small error on the linearly decaying component, and
% has a small support.

\subsubsection{An example of parameter folding with quadratic decay in approximation error}\label{ssec:overview:fold:quadratic}
There are many ways to chose the folding function.  Here we present one of them, for which we can prove its effectiveness in
parameter folding using our Fourier-decomposition toolkit.  The definition of the folding function is inside the following
lemma.  (The proof is mere arithmetic and therefore omitted.)  We use the $\%$-notations defined
in~\eqref{eq:overview:fold:def-mod} above.

\begin{lemma}\label{lem:overview:fold:quadratic:def-tau}
  For positive real numbers $p,c$ satisfying $p|c$ (i.e., $c/p \in\NN$), define the following function:
  \begin{equation}
    \tau_{p,c}
    \colon \RR \to     \RR
    \colon s   \mapsto
    \begin{cases}
      -c + s\modr p, &\text{if }  s \le -c-p; \\
      s,             &\text{if }  s \in \lt] -c-p, +c+p \rt[; \\
      +c + s\modl p, &\text{if }  s \ge +c+p.
    \end{cases}
  \end{equation}
  The function $\tau_{p,c}$ is a $p$-folding function with image $\lt] -c-p, +c+p \rt[$.
  \qed
\end{lemma}

For the remainder of this section, and in \S\ref{ssec:truncfold:fold-quadratic} (which contains the proofs), we use the
following big-O notation: We write $g(c) = O(h(c))$ for functions $g=g(c)$, $h=h(c)\ge0$, to indicate the existence of an
absolute constant~$C$ such that $\abs{g(c)}\le C\cdot h(c)$ for all allowed~$c$; if the constant is not absolute but depends on,
say, ``$K$'', we write $O_K(\centerdot)$.  We also use the corresponding big-$\Omega$ notation.

Corollary~\ref{cor:EVF-direct-sum} motivates the conditions in the following proposition, which quantifies the approximation
error, operator norm (i.e., cost), and maximum parameter values.

\begin{proposition}\label{prop:overview:fold:quadratic}
  Let~$\Xi$ be a frequency set, $p$ a positive real number such that~$\nfrac1p$ is a common divisor of~$\Xi$, and~$c$ a
  positive real number with $p|c$; let $\tau_{p,c}$ as in Lemma~\ref{lem:overview:fold:quadratic:def-tau}.

  The Nyquist shift rule~$\phi_K$ parameter-folded by $\tau_{p,c}$ as in
  Lemma~\ref{lem:overview:fold:folding-types}\ref{lem:overview:fold:param-folding},
  \begin{equation}\label{prop:overview:fold:quadratic:param-folding:def}
    f \mapsto (f\circ\tau_{p,c})*\phi_K
  \end{equation}
  has the following properties.

  \begin{enumerate}[label=(\alph*)]
  \item\label{prop:overview:fold:quadratic:param-folding:apx}%
    If $f = f_1+f_0 \in \sumFun_\Xi$ where $f_1 \in \oldFun_\Xi$ and~$f_0$ decays linearly with constants $(c,C)$ for some~$C$,
    then
    \begin{equation*}
        \absB{  (f\circ\tau_{p,c})*\phi_K (x)   - f'(x)  }
        \le
        \lt\{
        \begin{tabular}{lll}
          \multicolumn{2}{l}{$\displaystyle \frac{8KC}{\pi c(2Kc-1)},$}  &for $x\in[-p,+p]$; \\
            \\
            $\displaystyle \frac{8\pi K C}{c}$,                         &&for all $x\in\RR$.
          \end{tabular}
          \rt.
    \end{equation*}
  \item\label{prop:overview:fold:quadratic:param-folding:norm}%
    The operator norm of the linear operator\footnote{On the space of measurable bounded real-valued functions with
      $\Nm{\centerdot}_\infty$, cf.\ page~\pageref{def:space-measurable-fn}.} $\mathscr{M}_b \to \mathscr{M}_b$ defined
    in~\eqref{prop:overview:fold:quadratic:param-folding:def} is smaller than the norm of~$\phi_K$, i.e., $2\pi K$
    (cf.~Theorem~\ref{thm:overview:opt}).
  \item\label{prop:overview:fold:quadratic:param-folding:max-value}%
    The maximum magnitude of a parameter value at which~$f$ is evaluated in~\eqref{prop:overview:fold:quadratic:param-folding:def}
    is less than $c+p$.
  \item\label{prop:overview:fold:quadratic:param-folding:avg-value}%
    In the Simple Folding Estimator, Fig.~\ref{fig:simple-folding-estimator}, with input $x\in[-p,+p]$, the expected magnitude
    of parameter values at which the shot oracle is queried, is at most
    \begin{equation*}
      \abs{x} + \frac{ \ln(K(c+2p)) + O(1) }{ \pi^2 K }.
    \end{equation*}
  \end{enumerate}
\end{proposition}

To parse the expressions involving products of $p$ or $c$ with~$K$ (in denominators or under the logarithm), note that
\begin{equation}\label{eq:overview:fold:quadratic:ck-ge-1}
  cK \ge pK \ge 1.
\end{equation}
The second inequality follows from the fact that $1/p$ divides the elements of~$\Xi$ one of which is~$K$, and $K>0$ holds
as~$\Xi$ is a frequency set (Def.~\ref{def:frequency-set}).

\mypar%
To summarize, first note that the expression on the RHS in Item~\ref{prop:overview:fold:quadratic:param-folding:apx} in the case
``$x\in[-p,+p]$'' is $O_{K,C}(1/c^2)$, i.e., we have quadratic decay of the approximation error in terms of the largest
parameter value queried.

As for the value of~$p$, in the situation of Def.~\ref{def:perturbed-parametric}, a greatest common divisor of the difference
set of the spectrum of~$A$ would be expected to be known, and a smaller common divisor can be chosen to cover a larger interval
in the order-$1/c^2$ case of Item~\ref{prop:overview:fold:quadratic:param-folding:apx}.  The number~$c$ is the main quantity to
play with: In the interval $[-p,+p]$, the approximation error goes down quadratically in~$c$, while the maximum magnitude of a
parameter value increases linearly, and the expected magnitude of a parameter value increases only logarithmically.  Note that
increasing~$c$ might allow to decrease~$C$, as for each fixed~$f$, $C$ decays linearly as $c\to\infty$.

The strict condition ``$x\in[-p,+p]$'' in Item~\ref{prop:overview:fold:quadratic:param-folding:apx} is merely to make the proof
less onerous: The interested reader will, upon inspection of the proof in \S\ref{ssec:truncfold:fold-quadratic}, realize that
the quadratic decay of the approximation error holds if $\abs{x} \le p + (1-\Omega(1))c$, and hence in particular for
$\abs{x}=o(c)$.  Generally speaking, it should be understood that the approximation error decreases as $\abs{x}\ll c+p$ and
increases as~$\abs{x}$ approaches $c+p$.  A look at the proof reveals that a more fine grained analysis would have in the
denominator the term $c+p-\abs{x}$ instead of~$2Kc$, provided that $\abs{x} \le c+p-\Omega_K(1)$; this would lead to decays
between $(1/c)^1$ and~$(1/c)^2$ in that region of~$x$'s.

\subsection{Non-existence results}\label{ssec:overview:Impossible}
In this section, we briefly discuss a few results of non-existence of the optimist's feasible PSRs, e.g., those
with compact support.  As non-existence is so frustrating, we won't spend too much time on it; the proofs are mostly only
sketched.

Before we start, two things must be emphasized.

Firstly, the non-existence results are only for \emph{feasible} (as opposed to approximate) PSRs: As we already know from the
discussions about truncation in~\S\ref{ssec:overview:trunc}, for every $\eps>0$ there exists an $\eps$-almost-feasible PSR with
compact support (and decent norm/cost).  Having said that, note that the size of the support in these examples grows beyond all
bounds as~$\eps$ tends towards~$0$; we will prove below that that is unavoidable.

Secondly, recalling the discussion in \S\ref{sssec:overview:function-spaces:equiv-feasability}, the reader should be aware:
While the results of this section pertain to feasibility of PSRs for $\Fun_\nfrac12$, with proofs making use of all frequencies
$\xi\in[-\nfrac12,+\nfrac12]$, by Prop.~\ref{prop:overview:function-spaces:equiv-feasability}, non-existence of the
described types of PSRs feasible for $\sumFun_\Xi$ is implied for all frequency sets~$\Xi$ with $\max\Xi=\nfrac12$.  (And from
there, of course, by dilation, for all frequency sets.)

\mypar%
We start with some observations involving slight modifications of our Nyquist shift rules.

\subsubsection{Other optimal PSRs}
An inspection of the proof of Theorem~\ref{thm:overview:half--measure} in \S\ref{ssec:Half-deriv:pf-half--measure} shows that,
for all~$K>0$, the Nyquist shift rule~$\phi_K$ is the unique complex PSR that is feasible for~$\Fun_K$ and whose support is
contained in the support of~$\phi_K$.  Indeed, the function~$u$ in~\eqref{eq:overview:def:half--measure} is the inverse periodic
Fourier transform of the RHS of equation~\eqref{eq:overview:linsyseq} from Lemma~\ref{lem:overview:linsyseq}.  (The Fourier
Inversion Theorem applies as both~$u$ and $\hat u$ are absolutely integrable.)

In particular, there is no other complex PSR that is \emph{optimal} for $\Fun_K$, as, by Corollary~\ref{cor:overview:otheropt},
the support of such a PSR would have to be contained in the support of~$\phi_K$.

\subsubsection{PSRs with compact support}\label{ssec:overview:Impossible:cpct}
By standard Paley-Wiener theory, a complex measure~$\mu$ with compact support will have a Fourier transform that is a
holomorphic function on a connected domain containing the real line.  From standard complex variable function theory we know
that two holomorphic functions on a connected domain must be equal if they coincide on a set that contains an accumulation
point.  This means that Lemma~\ref{lem:overview:linsyseq} restricts the choices for~$\hat\mu$ to only a single one:
\begin{equation*}
  \hat\mu = \Bigl(  \RR\to\CC\colon  \xi\mapsto \hat\mu(\xi) = -2\pii\xi \, e^{-i\pi\xi}  \Bigr),
\end{equation*}
so that~$\mu$ is the tempered distribution ``for every Schwartz function, take the derivative at~$\nfrac12$''.  But this is a
contradiction, as that particular tempered distribution is not a finite measure.\footnote{The equality
  $[\varphi\mapsto \partial_{\nfrac12}\varphi] = \int\place d\mu$ for a finite measure~$\mu$ would imply, for all positive
  integers~$k$, $2\pi k = \int \sin(2\pi k(x-\nfrac12))\,d\mu(x)$, which would contradict the finiteness of the measure.}

Hence, there is no compactly supported complex PSR feasible for $\Fun_*$.

\subsubsection{Exponential concentration}
Now, using nothing but standard tools, we show that complex PSRs with exponential concentration cannot exist.

\begin{definition}[Exponential concentration]\label{def:overview:Impossible:expo-concentr}
  We say that a complex measure~$\mu$ on the real line is \textit{exponentially concentrated} if there exist $C,r>0$ such that
  \begin{equation}\label{def-eq:overview:Impossible:expo-concentr}
    \abs{\mu}(\RR\setminus[-T,+T]) \le C \cdot e^{-rT} \quad\text{for all $T\ge0$.}
  \end{equation}
\end{definition}

With a similar argument as in the case of compact support, we will prove the following in~\S\ref{ssec:Impossible:expo-concentr}.

\begin{theorem}\label{thm:overview:Impossible:non-exist-expo-concentr}
  Let $\Xi\subseteq[-\nfrac12,+\nfrac12]$ be an infinite set.
  There exists no complex PSR~$\phi$ with exponential concentration that is feasible for the function space~$\GenericFun$
  spanned by the functions $\cos(2\pi\xi\place)$, $\sin(2\pi\xi\place)$, $\xi\in\Xi$.
\end{theorem}

We note that for $\Xi := [-\nfrac12,+\nfrac12]$, the space~$\Fun_\nfrac12$ contains the space~$\GenericFun$ in
Theorem~\ref{thm:overview:Impossible:non-exist-expo-concentr}, and hence no exponentially concentrated PSR is feasible
for~$\Fun_\nfrac12$.

\subsubsection{Exponentially concentrated nearly feasible PSRs}\label{ssec:overview:Impossible:near-feasible}
We can also give impossibility results for complex PSRs with non-zero approximation error.

For this context, the concept of Nearly Feasible from Def.~\ref{def:near-feasibility} is too limiting.  Instead we
simply take convergence of the approximations of the derivative in the point~$0$ to the correct value.  The proof is sketched
in~\ref{ssec:Impossible:apx-error}.

\begin{corollary}\label{cor:overview:Impossible:non-exist-expo-concentr}
  Let $\Xi$ and~$\GenericFun$ be as in Theorem~\ref{thm:overview:Impossible:non-exist-expo-concentr}.

  The following does not exist: Numbers $C, r>0$ and a norm-bounded family $\phi_j$, $j\in\NN$, of complex PSRs all
  satisfying~\eqref{def-eq:overview:Impossible:expo-concentr} from the definition of exponential concentration (all with the
  given~$r$), such that $f*\phi_j(0) \xrightarrow{j\to\infty} f'(0)$ for every~$f\in\GenericFun$.
\end{corollary}

Note that the boundedness condition, $\Nm{\phi_j} = O(1)$ ($j\to\infty$) is needed, otherwise the symmetric difference quotient
would be a counterexample.

\subsection{Numerical simulation}\label{ssec:overview:numsim}
To make an attempt at understanding the practical utility of the Nyquist shift rule vs Banchi-Crooks's method, the author has
designed a small \emph{Pluto}\footnote{\texttt{\href{https://plutojl.org}{plutojl.org}}}-notebook containing code in the mathematical
computation programming language \emph{Julia}\footnote{\texttt{\href{https://julialang.org/}{julialang.org}}}, which simulates
the two methods numerically and allows us to produce colorful pictures.

The \emph{Approximate Stochastic Parameter Shift Rule} (ASPSR) has been implemented as
in~\cite{Banchi-Crooks:stoch-param-shift:2021}, accommodating our idiosyncratic choice of $\hbar=1/2\pi$.  It is compared to a
version of the truncated Nyquist shift rule (STNySR), tailored especially for the comparison to Banchi-Crooks: A positive real
number~$T$ can be provided by the user which has the effect of limiting to the interval $[-T,+T]$ all parameter values in
queries to the expectation-value function.

As the stochastic properties of the two estimators are identical, the numerical simulation ignores that aspect.  It allows to
query
\begin{itemize}
\item In the case of the STNySR: Directly the expectation-value function as in~\eqref{def:perturbed-parametric} at a
  given parameter value $x\in\RR$;
\item In the case of the ASPSR: For given $s,x,\eps$, directly the expectation value
  \begin{gather*}
    \tr\bigl( M\, U_\pm(s,x,\eps) \varrho \,U_\pm(s,x,\eps)^\dag \bigr), \qquad\text{where}\\
    U_\pm(s,x,\eps) := e^{2\pii s(x A+B)} e^{2\pii\eps (\pm A/8\eps + B)} e^{2\pii (1 - s)(x A+B)}.
  \end{gather*}
\end{itemize}
The truncation parameter, $T$, mentioned above is set to $1/8\eps$, to ensure that both methods use parameter values in the same
interval.

The \emph{Julia} code produces random perturbed-parametric unitary instances as follows: A random matrix~$M$ with $\pm1$
eigenvalues; a random positive-semidefinite trace-1 matrix~$\varrho$; a random matrix~$A$ with $\pm1$-eigenvalues; a
random standard Gaussian Hermitian matrix~$B$.

The Pluto notebook makes it is easy to make numerical simulations to compare absolute and relative errors between ASPSR and
STNySR graphically in plots.  While in the following we discuss some typical features and noteworthy behavior based on plots
created with the code, the reader is invited to play with the notebook\footref{fn:notebook-source-code} and judge for her or
himself.

\mypar%

\begin{figure}[tp]
  \centering
  \subfloat[]{%
    \begin{minipage}[b]{0.45\textwidth}\centering%
      \includegraphics[width=1.1\linewidth]{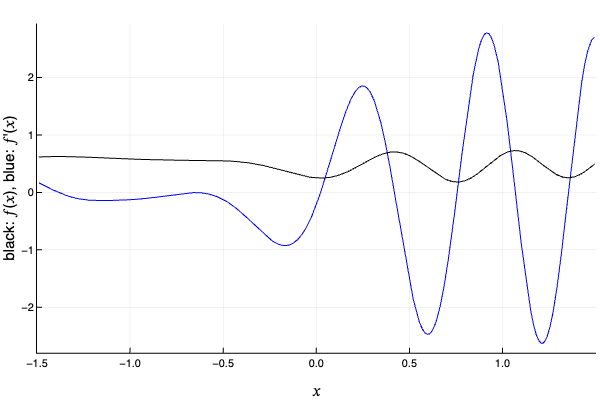}
    \end{minipage}%
  }
  \hfill
  \subfloat[]{%
    \begin{minipage}[b]{0.45\textwidth}\centering%
      \includegraphics[width=1.1\linewidth]{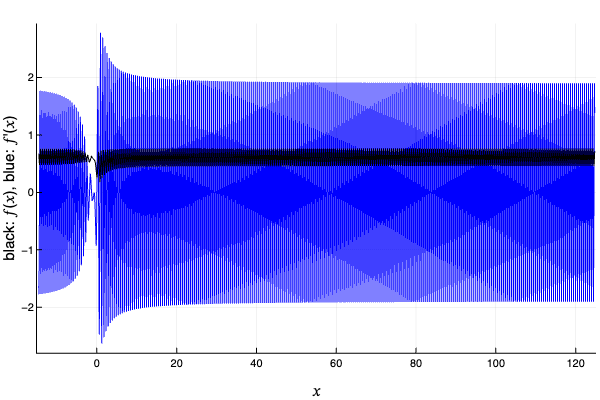}
    \end{minipage}%
  }
  \\
  \subfloat[]{%
    \begin{minipage}[b]{0.45\textwidth}\centering%
      \includegraphics[width=1.1\linewidth]{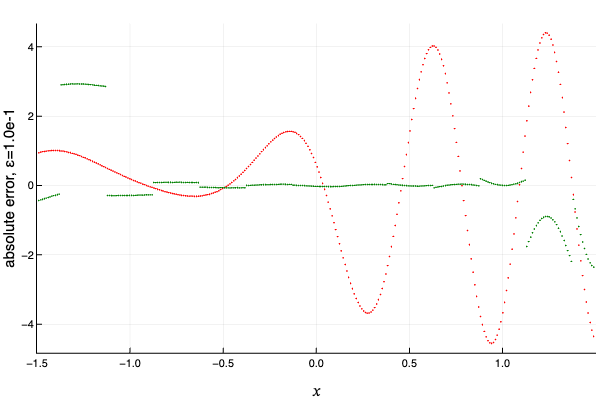}
    \end{minipage}%
  }
  \hfill
  \subfloat[]{%
    \begin{minipage}[b]{0.45\textwidth}\centering%
      \includegraphics[width=1.1\linewidth]{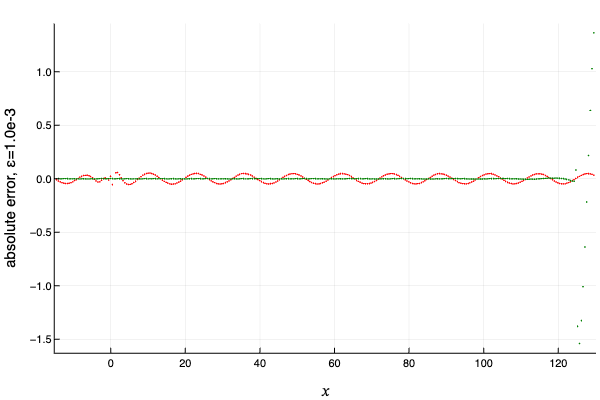}
    \end{minipage}%
  }
  \\
  \subfloat[]{%
    \begin{minipage}[b]{0.45\textwidth}\centering%
      \includegraphics[width=1.1\linewidth]{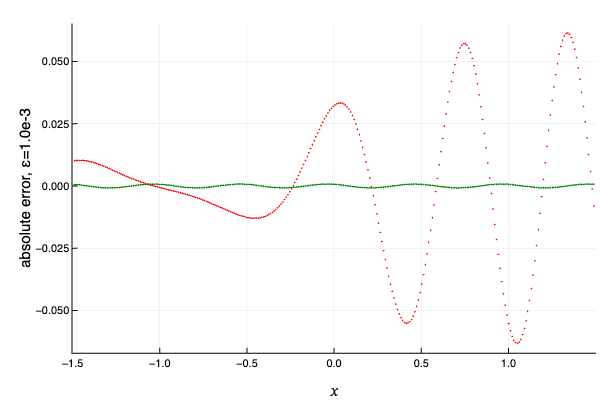}
    \end{minipage}%
  }
  \hfill
  \subfloat[]{%
    \begin{minipage}[b]{0.45\textwidth}\centering%
      \includegraphics[width=1.1\linewidth]{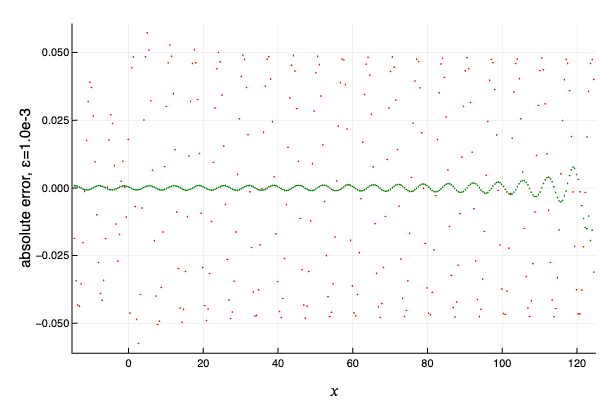}
    \end{minipage}%
  }
  \caption[]{\textbf{Example of approximations provided by this paper's STNySR and BC's ASPSR.  }  %
    \small%
    Data from a randomly created (and arbitrarily selected) expectation-value function~$f$ is presented (with matrices of size
    $4\times 4$).  Graphs (a),(b) show \textcolor{darkgray}{$f$ (black)} and \textcolor{blue}{$f'$ (blue)}, on different
    parameter intervals.  In (c,d,e,f) absolute errors of the \textcolor{red}{ASPSR (red)} and the \textcolor{green}{STNySR
      (green)} are plotted.  In (c) we took $\eps=10^{-1}$ (leading to the cut-off $T=1.25$ for the STNySR) and in (d), (e),
    (f)) we took $\eps=10^{-3}$ (leading to the cut-off $T=125$).}\label{fig:bc-ny-comparison:abserrB}
\end{figure}

Fig.~\ref{fig:bc-ny-comparison:abserrB}, shows a single expectation-value function randomly constructed in that way, and the
absolute errors of ASPSR and the STNySR.  Typical features are that,
\textcal{(1)}~due to the cutoff at $T:=1/8\eps$ of parameter values in the STNySR, near the ends of the interval $[-T,+T]$ the
approximation errors explode;
and
\textcal{(2)}~STNySR is better than ASPSR when there is sufficient gap between the parameter value~$x$ and the cut-off
$\pm 1/8\eps$.

As~$A$ has $\pm1$ eigenvalues we have $K=2$, and hence the query points of the Nyquist shift rules are $1/4$ apart.
Sub-figure~(c) of Fig.~\ref{fig:bc-ny-comparison:abserrB} shows break points in the green STNySR-line, which are caused by
changes in the set of shifts that are queried.

\begin{figure}[tp]
  \centering
  \subfloat[]{%
    \begin{minipage}[b]{0.45\textwidth}\centering%
      \includegraphics[width=1.1\linewidth]{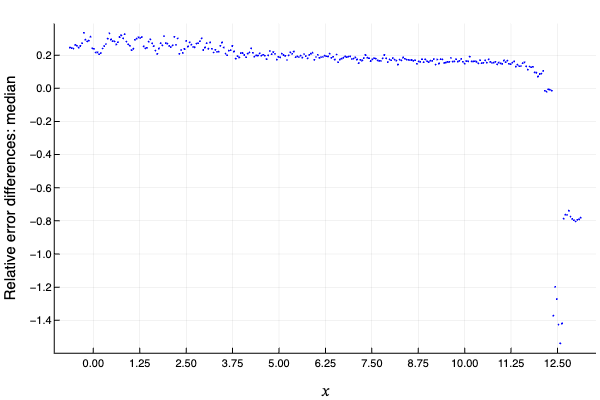}
    \end{minipage}%
  }
  \hfill
  \subfloat[]{%
    \begin{minipage}[b]{0.45\textwidth}\centering%
      \includegraphics[width=1.1\linewidth]{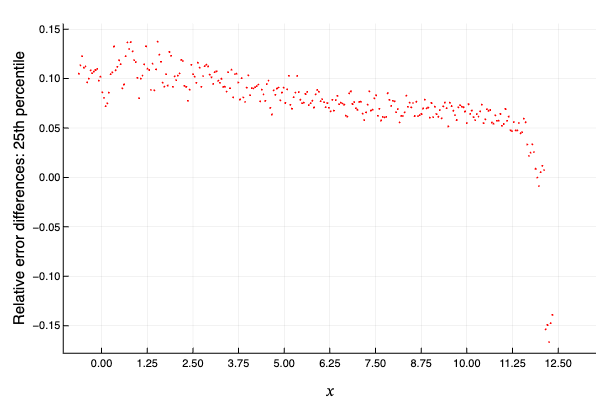}
    \end{minipage}%
  }
  \\
  \subfloat[]{%
    \begin{minipage}[b]{0.45\textwidth}\centering%
      \includegraphics[width=1.1\linewidth]{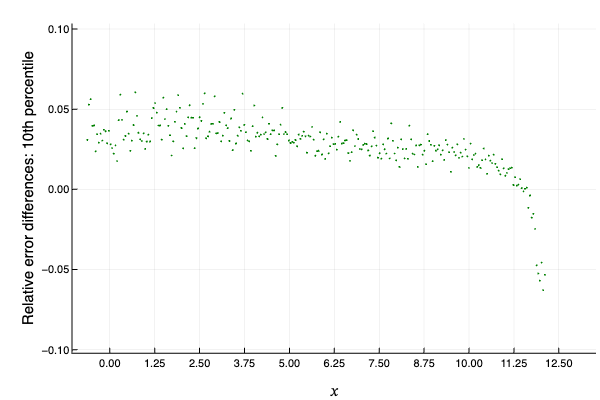}
    \end{minipage}%
  }
  \hfill
  \subfloat[]{%
    \begin{minipage}[b]{0.45\textwidth}\centering%
      \includegraphics[width=1.1\linewidth]{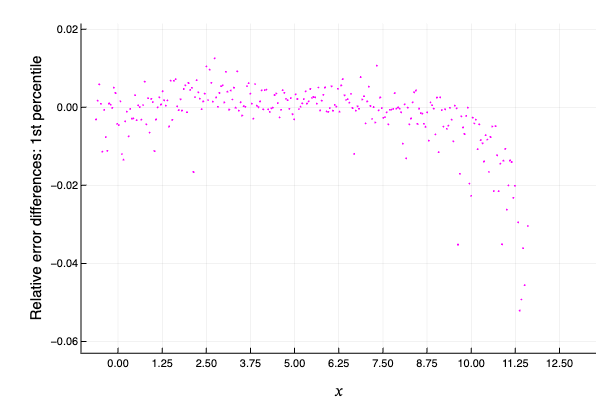}
    \end{minipage}%
  }
  \caption[]{\textbf{Comparison of relative errors between STNySR and BC's ASPSR.  }  %
    \small%
    Plots of 200 random instances of expectation-value functions~$f$ chosen randomly as described in the text (with $8\times 8$
    matrices), considered at 300 parameter values, $x$.  For each combination of parameter value~$x$ and expectation-value
    function~$f$, the difference between the relative error of the ASPSR-approximation (with $\eps=10^{-2}$) of $f'(x)$ minus
    the relative error of the STNySR-approximation of $f'(x)$ (with parameter cut-off $1/8\eps=12.5$) is computed.  For each
    parameter value~$x$, the \textcolor{blue}{median (blue)}, \textcolor{red}{25th percentile (red)}, \textcolor{green}{10th
      percentile (green)}, \textcolor{magenta}{1st percentile (magenta)} of the 200 random instances are plotted, in different
    scales.  }\label{fig:bc-ny-comparison:relerr2}
\end{figure}

Fig.~\ref{fig:bc-ny-comparison:relerr2} shows the differences of relative errors between the two methods.  Positive values mean
STNySR is better.  It can be seen (positive green 10th-percentile points) that in about 90\% of the random instances, STNySR was
better than the ASPSR, at least where the point~$x$ at which the derivative is requested is sufficiently far away from the
cut-off $1/8\eps = 12.5$.

As the query points of the Nyquist shift rules are $1/4$ apart, if $x>12.25$, no query point of STNySR is to the right of~$x$
--- leading to a noise-only ``approximation'' of the derivative.

It can be seen (positive blue median points) that for $x\le 12$, i.e., when there are at least~3 query points to the right
of~$x$, STNySR gives at least as good an approximation of the derivative as ASPSR, in at least half of the cases.  It should be
noted that the mean (not plotted) lies above the median, indicating that the advantage is, on average, substantial.

For a considerable region of the parameter, STNySR is roughly at least as good ASPSR in 99\% of the instances (magenta 1st
percentile data points hugging zero).

\mypar%
Appendix~\ref{apx:numsim} has plots of the results of more numerical simulations.  The reader should understand that the
presented numerical results are preliminary, and that more refined and extensive numerical simulations and statistical analysis
are necessary for a comprehensive comparison of the methods.  In particular, which (proper or not) shift rules are preferable in
which parameter regions when the parameter values are constrained to an interval might be a topic of future research.

\section{Fourier analysis of perturbed-parametric unitaries}\label{sec:fourier}%----------------------------------------------
In this section, we will deal with the Fourier-analytic properties of the functions in Def.~\ref{def:perturbed-parametric}:
Perturbed-parametric unitary functions and expectation-value functions.  We start by reviewing the (standard) notation we use.

\subsection{Notations}\label{ssec:fourier:notation}
Section~\ref{sec:fourier} is somewhat demanding in terms of notations, as we have to deal with operator-valued tempered
distributions.

\subsubsection{Tempered distributions}
We denote the space of (complex-valued) Schwartz functions on~$\RR$ by $\Schwartz$, and the tempered distributions on~$\RR$ by
$\Schwartz'$.
We denote the duality between Schwartz functions (on the left) and tempered distributions (on the right) by
``$\dual{\centerdot}{\centerdot}$''; e.g., for a Borel measure~$\mu$ of at most polynomial growth we have
$\dual{\varphi}{\mu} := \int \varphi(x) \,d\mu(x)$, and for a measurable function~$f$ of at most polynomial growth we have
$\dual{\varphi}{f} := \int \varphi(x) f(x)\,dx$.

As we are using angle-brackets ``$\dual{\centerdot}{\centerdot}$'' for tempered distributions, we revert to parentheses
``$\cip{\centerdot}{\centerdot}$'' for the Hilbert-space inner product; it is \emph{linear in the right argument,} anti-linear
in the left argument.

\subsubsection{Operator-valued functions}
The space of linear operators on a finite-dimensional Hilbert space~$\cH$ is $\LL(\cH)$.

In all of Section~\ref{sec:fourier}, we denote by $\opNm{\place}$ the operator / spectral norm (of operators on a Hilbert space
or of square matrices, resp.).  If~$T$ is a set and~$F$ is an operator-valued function defined on~$T$, we let
$\opNm{F}_{\infty,T} := \sup_{x\in T} \opNm{F(x)}$; for $T=\RR$ we omit the subscript~$T$ on the norm.

For the concepts of linear decay and square integrability of operator-valued functions we use $\opNm{\place}$ (although they are
norm independent in finite dimension).  Integration of operator-valued functions against complex-valued Schwartz functions and
the Fourier transform are defined as usual: element-wise.  This means, for example, the Fourier transform~$\hat F$ of an
operator-valued function~$F$ is defined through the condition: For all vectors $\phi,\psi$ in the underlying Hilbert space,
\begin{gather*}
  \cip{ \phi }{ \hat F \psi }
  =
  \Bigl( x\mapsto \cip{ \phi }{ F(x)\psi } \Bigr)^\Hat
\end{gather*}

We will also need a small addition to the ``$\oldFun_\Xi$'' notation: Denoting the underlying Hilbert space by~$\cH$, for an
arbitrary finite non-empty set $\Lambda\subset\RR$, the complex vector space of bounded functions with values in $\LL(\cH)$
whose Fourier spectrum is contained in~$\Lambda$ is denoted by
\begin{equation}\label{eq:def:oldFunU}
  \oldFunOp_\Lambda
  =
  \bigl\{   f\colon x\mapsto \sum_{\lambda\in\Lambda} e^{2\pii\lambda x} B_\lambda
  \bigm|
  B\in\LL(\cH)^\Lambda \bigr\};
\end{equation}
We emphasize that $\Lambda$ is not required to be symmetric.

\subsection{The Fourier-decomposition theorems}\label{ssec:fourier:decomp}
Now we are ready to formulate the Fourier-decomposition theorem.

\begin{proposition}[Fourier-decomposition for perturbed-parametric unitaries]\label{prop:fourier:decomp}
  With the notations of Def.~\ref{def:perturbed-parametric}, let $U\colon x \mapsto e^{2\pii(xA+B)}$, and
  $\Lambda := \Spec A$.

  There exists a unitary-operator-valued function $\tilde U$ with the following properties.
  \begin{enumerate}[label=(\alph*)]
  \item\label{prop:fourier:decomp:decomp}%
    $\tilde U \in \oldFunOp_\Lambda$ and $U - \tilde U$ is of linear decay.
  \item\label{prop:fourier:decomp:commu}%
    If $A$, $B$ commute, then $U = \tilde U$.
  \item\label{prop:fourier:decomp:unique}%
    For any finite-Fourier-spectrum, bounded, operator-valued function~$\tilde V$:
    If $lim_{x\to\infty} ( U(x) - \tilde V(x) ) = 0$, then $\tilde V = \tilde U$.
  \end{enumerate}
\end{proposition}

Item~\ref{prop:fourier:decomp:unique} is a corollary to Item~\ref{prop:fourier:decomp:decomp}, and serves as a sanity check: It
establishes that $\tilde U$ is the \emph{unique} bounded, finite-Fourier-spectrum operator-valued function that approximates~$U$
for all large parameter values.

The proof of Prop~\ref{prop:fourier:decomp} is eigenvalue perturbation theory in finite dimension; the details
fill~\S\ref{ssec:fourier:proof-decomp}; now we derive Prop.~\ref{prop:fou-decomp-evf} from Prop.~\ref{prop:fourier:decomp}.

\subsubsection{Proof of Prop.~\ref{prop:fou-decomp-evf} from Prop.~\ref{prop:fourier:decomp}}\label{sssec:fourier:decomp:evf-pf}
We now derive the Fourier-decomposition theorem for perturbed-parametric expectation-value functions
(\S\ref{sec:overview:function-spaces}) from Fourier-decomposition for perturbed-parametric unitaries.

\begin{proof}[Proof of Prop.~\ref{prop:fou-decomp-evf}]
  We simply expand the definition of the perturbed-parametric expectation-value function in
  Def.~\ref{def:perturbed-parametric}.  With $D\colon x \mapsto U(x)-\tilde U(x)$, for all $x\in\RR$,
  \begin{multline*}
    \tr\bigl( M\, U(x) \varrho U(x)^\dag \bigr)
    =
    \tr\Bigl( M\, (\tilde U(x) + D(x))\varrho (\tilde U(x)^\dag + D(x)^\dag) \Bigr)
    =
    \\
      \tr\Bigl( M\, \tilde U(x) \varrho \tilde U(x)^\dag \Bigr)
    + \tr\Bigl( M\, \tilde U(x) \varrho D(x)^\dag \Bigr)
    + \tr\Bigl( M\, D(x) \varrho \tilde U(x)^\dag \Bigr)
    + \tr\Bigl( M\, D(x) \varrho D(x)^\dag \Bigr).
  \end{multline*}
  We set $f_1 := \tr\Bigl( M\, \tilde U(\place) \varrho \tilde U(\place)^\dag \Bigr)$.  This is a bounded function with Fourier
  spectrum contained in $\Diff\Spec A$ (e.g., \cite{GilVidal-Theis:CalcPQC:2018, Wierichs-Izaac-Wang-Lin:gen-shift:2021}, or see
  the proof of Lemma~\ref{lem:fourier:fou-spec:trotter} below).  By standard estimates, we find that each of the remaining terms
  has linear decay.  As sums of functions of linear decay have linear decay, this completes the proof of the proposition.
\end{proof}

\subsection{Proof of Fourier-decomposition for unitaries}\label{ssec:fourier:proof-decomp}
\textsl{Notational conventions for \S\ref{ssec:fourier:proof-decomp}.$*$.}  %%
This section is somewhat hard on the alphabet.  For that reason, we adopt the following notation convention:
We denote matrices by typewriter-font letters $\tA,\tB,\tC,\dots$.  Moreover, for operators denoted by $A,B,U,\dots$, when an
orthonormal basis has been fixed, we denote the matrix corresponding to the operator by the typewriter version of the letter,
e.g., $A\leftrightarrow\tA$, $B\leftrightarrow\tB$, \dots.  We denote the identity matrix by $\tI$.

\mypar%
The proof of Prop.~\ref{prop:fourier:decomp} is by degenerate (i.e., eigenvalues are not necessarily simple) eigenvalue
perturbation theory, as everybody has learned it in their quantum mechanics class.  To enable a mathematically rigorous proof,
Appendix~\ref{apx:perturb-th} reviews the mathematical foundation, Rellich's theorem, based on which it also summarizes textbook
perturbation theory in matrix notation (Corollary~\ref{cor:perturbation-theory}), for convenient reference.

\newcommand{\tLambdacoeff}[1]{\tLambda^{[#1]}}            %
\newcommand{\tWcoeff}[1]{\tW^{[#1]}}                      %
\newcommand{\ErrW}{\tE_{_W}}                              %
\newcommand{\ErrLambda}{\tE_{_\Lambda}}                   %
\newcommand{\tildetU}{\widetilde{\tU}}                    %

\subsubsection{Proof of Prop.~\ref{prop:fourier:decomp} \ref{prop:fourier:decomp:decomp},\ref{prop:fourier:decomp:commu}}\label{sssec:fourier:proof-decomp:ab}
This subsection holds the proof of Fourier-decomposition, Prop.~\ref{prop:fourier:decomp}.  We make some definitions and we will
work with them throughout the subsection.

We use the notations of Def.~\ref{def:perturbed-parametric}, and denote by~$d$ the dimension of the underlying Hilbert
space.  Define $U(x) := \exp(2\pii( xA + B ))$ for $x \in \RR$.

We use the following matrix big-O %and little-o
notation: By $\bigOfrom{R}(1/x^2)$ we refer to an unnamed square-matrix valued function $x\mapsto \tM(x)$ defined on
$\RR\setminus[-R,+R]$, with the property that
\begin{equation*}
  \sup_{x\in \RR\setminus[-R,+R]}  \opNm{ x^2 \tM(x) } \quad < \infty.
\end{equation*}
(Recall that $\opNm{\centerdot}$ is the spectral norm, but the property is independent of the norm, as we are in finite
dimension.)
% By $\littleo(1/\abs{x})$ we refer to an unnamed square-matrix valued function $x\mapsto \tM(x)$ defined on
% $\RR\setminus[-R,+R]$ for some $R>0$, with the property that
% \begin{equation*}
%   \lim_{\abs{x}\to\infty} \opNm{ x \tM(x) } \quad = 0.
% \end{equation*}

\begin{lemma}\label{lem:fourier:proof-decomp:exp}
  We use the notations above.
  %%
%  \begin{enumerate}[label=(\alph*)]
%  \item\label{lem:fourier:proof-decomp:exp:decomp}%
    There exists an orthonormal basis of eigenvectors of~$A$, and a bounded $d$-by-$d$ matrix-valued function $\tC$ on~$\RR$
    with the following property.

    \begin{quotation}
      Denoting $\tA,\tB,\tU$ the $(d\times d)$-matrices of the operators $A,B,U$, resp., wrt the basis, and setting
      \begin{equation*}
        \tildetU(x) := \exp\bigl( 2\pii(x\tA + \Diag(\tB)) \bigr)
      \end{equation*}
      we have for all~$x$ (with $\abs{x}\ge 1$),
      \begin{equation*}
        \tU(x)
        =
        \tildetU(x) + \frac{\tC(x)}{x} + \bigOfrom{1}(1/x^2).
      \end{equation*}
    \end{quotation}
    If $A,B$ commute, then the basis can be chosen to consist of eigenvectors of~$B$, so that $U=\tilde U$.
%  \end{enumerate}
\end{lemma}
\begin{proof}
  The remark about the case when $A,B$ commute is trivial.

  In the general case, we apply Corollary~\ref{cor:perturbation-theory} from Appendix~\ref{apx:perturb-th}, and use the notation
  therein.  In view of Items \ref{cor:perturbation-theory:W0}, \ref{cor:perturbation-theory:Lambda0}
  and~\ref{cor:perturbation-theory:Lambda1} of Corollary~\ref{cor:perturbation-theory} we define
  \begin{equation*}
    \begin{split}
      \ErrW(z)      &:= \tW(z)      - \tI - z\tWcoeff{1}                              \\
      \ErrLambda(z) &:= \tLambda(z) - \tA - z\Diag(\tB)   - z^2\tLambdacoeff{2};
    \end{split}
  \end{equation*}
  and note that $z\mapsto \ErrW(z)/z^2$ and $z\mapsto \ErrLambda(z)/z^3$ are bounded on\footnote{The factor $1/2$ ensures that
    the power series are bounded as $\abs{z}\to r/2$.} $[-r/2,+r/2]$, by Items \ref{cor:perturbation-theory:W-conv}
  and~\ref{cor:perturbation-theory:Lambda-conv}, resp., of Corollary~\ref{cor:perturbation-theory}.

  We set $R := \max(1, 2/r)$ and calculate, for $\abs{x}>R$,
  \begin{align}
    \tU(x)
    &=
      \exp(2\pii( x\tA + \tB ))                                              \notag\\
    &=
      \exp(x 2\pii (\tA + \tfrac1x \tB))                                     \notag\\
    &=
      \tW(\nfrac1x) \, \exp(x 2\pii \tLambda(\nfrac1x) )\, \tW(\nfrac1x)^\dag.  \tag{$*$}\label{eq:RNDqfcobg8y}
  \end{align}
  From here, we treat the terms separately.  We find:
  \begin{equation*}
    \begin{split}
      \tW(\nfrac1x)      &= \tI + \tfrac1x \tWcoeff{1} + \ErrW(\nfrac1x)        \\
      \tW(\nfrac1x)^\dag &= \tI - \tfrac1x \tWcoeff{1} + \ErrW(\nfrac1x)^\dag
    \end{split}
  \end{equation*}
  where the first equation is just the definition of~$\ErrW$, and the second follows from the first using the fact the
  $\tWcoeff{1} = \tW'(0)$ is anti-hermitian, by Corollary~\ref{cor:perturbation-theory}\ref{cor:perturbation-theory:W-unitary}, as
  $\tW(0)=\tI$.

  For the middle factor in~\eqref{eq:RNDqfcobg8y}, we find
  \begin{align*}
    \exp(x 2\pii \tLambda(\nfrac1x) )
    &=
      \exp\Bigl(x 2\pii \bigl( \tA + \tfrac1x\Diag(\tB)   + \tfrac1{x^2}\tLambdacoeff{2} + \ErrLambda(\nfrac1x) \bigr) \Bigr)
      \\
    &=
      \exp\Bigl(2\pii \bigl( x\tA + \Diag(\tB)   + \tfrac1x\tLambdacoeff{2} + x\ErrLambda(\nfrac1x) \bigr) \Bigr)
      \\
    &=
      \tildetU(x) \cdot \exp(2\pii\tLambdacoeff{2}/x) \cdot \exp(2\pii(x\ErrLambda(\nfrac1x))
  \end{align*}
  Items \ref{cor:perturbation-theory:diagonalization} and~\ref{cor:perturbation-theory:W-unitary} of
  Corollary~\ref{cor:perturbation-theory} imply that all entries of all~$\tLambdacoeff{*}$ are real numbers, and hence, for the
  spectral norm of the matrix exponentials, we find
  \begin{equation*}
    \begin{split}
      \exp(2\pii\tLambdacoeff{2}/x) - \tI                 & =  2\pii \tLambdacoeff{2}/x + \bigOfrom{1}(1/x^2)     \\
      \opNm{ \exp(2\pii x\ErrLambda(\nfrac1x)) - \tI    } & \le 2\pi \opNm{\ErrLambda(\nfrac1x)}\cdot x = O^{\ge R}(1/x^2),
    \end{split}
  \end{equation*}
  where in the first line we just use the Taylor remainder bound of the exponential function, whereas in the second line we use
  the bound on $\ErrLambda$ mentioned above.

  Now, continuing~\eqref{eq:RNDqfcobg8y}, we calculate
  \begin{equation*}
    \tU(x)
    =
    \begin{aligned}[t]
      &
        \Bigl( \tI + \tfrac1x \tWcoeff{1} + \ErrW(\nfrac1x) \Bigr)
      \\
      &\qquad
        \cdot
        \tildetU(x)
        \cdot \bigl(\tI + (\exp(2\pii \tLambdacoeff{2}/x   )-\tI) \bigr)
        \cdot \bigl(\tI + (\exp(2\pii x\ErrLambda(\nfrac1x))-\tI) \bigr)
      \\
      &\qquad\qquad\qquad\qquad
        \cdot
        \Bigl( \tI - \tfrac1x \tWcoeff{1} + \ErrW(\nfrac1x)^\dag \Bigr)
    \end{aligned}
  \end{equation*}
  Collecting terms, we conclude
  \begin{equation*}
    \tU(x)
    = \tildetU(x)
    + \frac{\tC(x)}{x}
    + \bigOfrom{R}(1/x^2),
  \end{equation*}
  with
  \begin{equation}\label{eq:fourier:proof-decomp:exp:decomp:def-of-C}
    \tC(x) := [\tWcoeff{1},\tildetU(x)] + 2\pii \tildetU(x)\, \tLambdacoeff{2}.
  \end{equation}
  This completes the proof of Lemma~\ref{lem:fourier:proof-decomp:exp}, noting that $\bigOfrom{1}$ arises from $\bigOfrom{R}$ by
  the continuity of the three summands on $\RR\setminus\{0\}$.
\end{proof}

We can now derive Prop.~\ref{prop:fourier:decomp} directly from Lemma~\ref{lem:fourier:proof-decomp:exp}.

\subsubsection{Proof of Prop.~\ref{prop:fourier:decomp}\ref{prop:fourier:decomp:unique}}\label{sssec:fourier:proof-decomp:c}
We have to show that
\begin{equation*}
  U(x)  - \tilde V(x)  \xrightarrow{\abs{x}\to\infty} 0
\end{equation*}
implies $\tilde V = \tilde U$.  But by Item~\ref{prop:fourier:decomp:decomp} of the proposition, we see that
$D := \tilde U - \tilde V$ is a bounded, finite-Fourier-spectrum, operator-valued function which satisfies
\begin{equation*}
  D(x) =  \xrightarrow{\abs{x}\to\infty} 0.
\end{equation*}
This implies that $D=0$, by invoking, e.g., Prop.~\ref{prop:apx:math:diophant:direct-sum} in Appendix~\ref{apx:math:diophant}.

\mypar%
This completes the proof of Prop.~\ref{prop:fourier:decomp}, Fourier-decomposition for perturbed-parametric unitaries.

\subsection{The Fourier spectrum}\label{ssec:fourier:fou-spec}
\subsubsection{Operator-valued tempered distributions}
We set out to prove Prop.~\ref{prop:fourier:fou-spec}.  The annoyingly technical proof is based on the Lie Product Formula
\begin{equation*}
  e^{2\pii(xA+B)} = \lim_{n\to\infty} \bigl( e^{2\pii xA/n} e^{2\pii B/n} \bigr)^n, \quad\text{ for all~$x\in\RR$;}
\end{equation*}
we emphasize that the limit is taken pointwise.

We will prove convergence of the Fourier transforms of the finite products to the Fourier transform of the RHS.  The technical
nuisance is that the Fourier transforms of the finite products are linear combinations of Dirac measures, while the infinite
product has a continuous, square-integrable component. (We know that from the Fourier-decomposition theorem,
Prop.~\ref{prop:fourier:decomp}.)  Unfortunately, that means that the convergence can only happen in the sense of tempered
distributions.  Moreover, as we reason about the unitary-valued function (i.e., not the expectation-value function), we need
operator-valued tempered distributions.  The author wishes to emphasize that they don't add difficulty, just abstraction.

Operator-valued tempered distributions are, basically, matrices with entries in~$\Schwartz'$; but while speaking of matrices
requires fixing an orthonormal basis of the Hilbert space, we give the equivalent coordinate-free definition: An operator-valued
tempered distribution~$\tau$ is a sesqui-linear\footnote{Linear on the right side, anti-linear on the left.}  mapping
\begin{equation*}
  \cH\times\cH \to \Schwartz'
  \colon
  (\phi,\psi) \mapsto \tau_{\phi,\psi}.
\end{equation*}

For every given $\varphi\in\Schwartz$, as $(\phi,\psi) \mapsto \dual{ \varphi }{ \tau_{\phi,\psi} }$ is a sesqui-linear form
on~$\cH$, there exists a unique linear operator, denoted by $\dual{ \varphi }{ \tau }$, that satisfies, for all $\phi,\psi$
\begin{equation*}
  \cip{ \phi }{ \dual{\varphi}{\tau} \psi }
  =
  \dual{ \varphi }{ \tau_{\phi,\psi} };
\end{equation*}
the mapping $\varphi \to \dual{\varphi}{\tau}$ is linear.  Moreover, the set of operator-valued tempered distributions is a
complex vector space with the standard arithmetic operations.

With this machinery, we can now get to work.

\subsubsection{Fourier spectra of perturbed-parametric unitary functions}
The Fourier spectrum of the expectation-value function~\eqref{eq:def:perturbed-parametric:PQC-EVfun} relates to that of the
unitary function~\eqref{eq:def:perturbed-parametric-unitary} as in the following lemma.  We sketch the proof for the sake of
completeness.

\begin{lemma}\label{lem:fourier:op-val'd-fou-helper}
  Let $n\in\NN$ and $F$, $F_1,\dots,F_n$ bounded smooth operator-valued functions, and $M_0,\dots,M_n$ arbitrary operators.
  \begin{enumerate}[label=(\alph*)]
  \item\label{lem:fourier:op-val'd-fou-helper:trace}%
    The Fourier spectrum of the complex-number valued function $x\mapsto \tr(F(x))$ is contained in $\Supp\hat F$;
  \item\label{lem:fourier:op-val'd-fou-helper:dagger}%
    The Fourier spectrum of the operator-valued function $x \mapsto F(x)^\dag$ is
    \begin{equation*}
      -\Supp\hat F := \{ -\xi \mid \xi\in \Supp\hat F \};
    \end{equation*}
  \item\label{lem:fourier:op-val'd-fou-helper:multi}%
    If each of $F_1,\dots,F_n$ has compact Fourier spectrum, then the Fourier spectrum of the operator-valued function
    $x \mapsto M_0 F_1(x) M_1 \dots F_n(x) M_n$ is contained in
    \begin{equation*}
      \Supp\widehat{F_1} + \dots + \Supp\widehat{F_n},
    \end{equation*}
    (where the sum of sets $S_1,\dots,S_n$ is defined as $S_1+\dots+S_n := \{s_1+\dots+s_n \mid s_i\in S_i, i=1,\dots,n\}$).
  \end{enumerate}
\end{lemma}
\begin{proof}[Sketch of proof.]\mbox{}
  \begin{txtitem}[label=\textbullet]
  \item Item~\ref{lem:fourier:op-val'd-fou-helper:trace} follows from the fact that $\tr F(\place)$ is a sum of matrix
    elements, each with Fourier spectrum contained in $\Supp\hat F$.

  \item For Item~\ref{lem:fourier:op-val'd-fou-helper:dagger}, we consider the matrix elements: For $\phi,\psi\in\cH$ we have,
    for all~$x$,
    \begin{equation*}
      \cip{ \phi }{ F(x)^\dag\psi } = \cip{ \psi }{ F(x)\phi }^*,
    \end{equation*}
    and the statement follows from standard Fourier analysis: The equation $\widehat{f^*}=\hat f^*(-\place)$ holds also for the
    Fourier transform of tempered distributions~$f$; hence $\dual{ \varphi }{ \widehat{f^*} } = 0$ for all Schwartz
    functions~$\varphi$ with $\Supp\varphi\subseteq \RR\setminus(-\Supp\hat f)$.

  \item For Item~\ref{lem:fourier:op-val'd-fou-helper:multi}, consider the function of~$n$ variables
    \begin{equation*}
      Y\colon (x_1,\dots,x_n) \mapsto M_0 F_1(x_1) M_1 \dots F_n(x_n) M_n;
    \end{equation*}
    we are interested in the Fourier spectrum of the operator-valued function $x\mapsto Y(x,\dots,x)$.  The matrix elements
    of~$Y$ are linear homogeneous polynomials of degree~$n$ in the matrix elements of the $F_j$, $j=1,\dots,n$, i.e., they are
    of the form $g_1 \cdot \dots \cdot g_n$, where $\Supp\widehat{g_j} \subseteq \Supp\widehat{F_j}$, $j=1,\dots,n$, and all
    these supports are compact.

    The statement now follows from standard Fourier analysis.  Indeed, from \cite{Rudin:FA:1991}, Def.~6.36, following
    remarks, and Theorems 6.37 and~7.19, if $g_1,\dots,g_n$ are $\CC$-valued bounded measurable functions with compact Fourier
    spectrum, then
    \begin{equation*}
      (g_1 \cdot \dots \cdot g_n)^\Hat
      =
      \widehat{g_1} * \dots * \widehat{g_n};
    \end{equation*}
    the support of the last distribution is $\Supp\widehat{g_1} + \dots + \Supp\widehat{g_n}$
    (\cite[Theorem~6.37(b)]{Rudin:FA:1991}, note that all distributions that occur are tempered).
  \end{txtitem}
\end{proof}

The central mini-result of \S\ref{ssec:fourier:fou-spec} is the following fact.

\begin{proposition}[Fourier spectrum of perturbed-parametric unitary]\label{prop:fourier:fou-spec}
  With the notations of Def.~\ref{def:perturbed-parametric}, the Fourier spectrum of the perturbed-parametric unitary
  function is contained in the interval $[\lambdamin(A),\lambdamax(A)]$.
\end{proposition}

Before we prove the proposition, let us derive Prop.~\ref{prop:PQC-Fourier-spec} (\S\ref{sec:overview:function-spaces})
from it.

\begin{proof}[Proof of Prop.~\ref{prop:PQC-Fourier-spec}]
  Firstly, boundedness and analyticity of the expectation-value function are obvious.

  Secondly, we consider the Fourier spectrum.
  With the notations of Definition~\ref{def:perturbed-parametric}, denote the unitary function by
  $U\colon x \mapsto e^{2\pii(xA+B)}$, and the expectation-value function by $f\colon x \mapsto \tr( M U(x) \varrho U(x)^\dag)$.
  By Prop.~\ref{prop:fourier:fou-spec}, the Fourier spectrum of~$U$ is contained in
  $[\lambdamin(A),\lambdamax(A)]$.  Lemma~\ref{lem:fourier:op-val'd-fou-helper} then gives that the Fourier spectrum of~$f$ is
  contained in
  \begin{multline*}
    [\lambdamin(A),\lambdamax(A)] - [\lambdamin(A),\lambdamax(A)]
    \\
    =
    [\lambdamin(A) - \lambdamax(A) , \lambdamax(A) - \lambdamin(A)].
   \end{multline*}
   This concludes the proof.
\end{proof}

\subsubsection{Proof of Prop.~\ref{prop:fourier:fou-spec}}
We pull the following lemma out for easier readability.

\begin{lemma}\label{lem:fourier:fou-spec:trotter}
  With the notations of Def.~\ref{def:perturbed-parametric}, letting, for $n\in\NN$,
  \begin{equation*}
    U_n \colon x \mapsto \bigl( e^{2\pii xA/n} e^{2\pii B/n} \bigr)^n,
  \end{equation*}
  we have $\Supp\widehat{U_n} \subseteq [\lambdamin{A},\lambdamax{A}]$.
\end{lemma}
\begin{proof}
  Same calculation as always: Let $P_\lambda$ be the spectral projection onto the eigenspace for eigenvalue $\lambda$ of~$A$.
  The equation
  \begin{equation*}
    e^{2\pii \place A/n}
    =
    \sum_{\lambda\in\Spec A} P_\lambda \cdot e^{2\pii \lambda/n \cdot \place}
  \end{equation*}
  implies, by the linearity of the Fourier transform,
  \begin{equation*}
    \bigl( e^{2\pii \place A/n} \bigr)^\Hat = \sum_{\lambda\in\Spec A} P_\lambda \cdot \delta_{\lambda/n},
  \end{equation*}
  and hence $\Supp\lt( \bigl( e^{2\pii \place A/n} \bigr)^\Hat \rt) = \Spec(A/n)$.

  Now, by Lemma~\ref{lem:fourier:op-val'd-fou-helper}\ref{lem:fourier:op-val'd-fou-helper:multi}, we find that
  \begin{equation*}
    \Supp\widehat{U_n}
    \subseteq
    \underbrace{   \Spec(A/n) + \dots + \Spec(A/n)   }_{n\text{ terms}}.
  \end{equation*}
  Noting that $\Spec(A/n) = (\Spec A)/n$, the proof of the lemma is concluded by observing that the sum of sets on the RHS is
  contained in the interval $[\lambdamin{A},\lambdamax{A}]$.
\end{proof}

\begin{proof}[Proof of Prop.~\ref{prop:fourier:fou-spec}]
  With the notations of Def.~\ref{def:perturbed-parametric}, denote the unitary function by $U\colon x \mapsto e^{2\pii(xA+B)}$,
  and let~$U_n$, $n\in\NN$, as in Lemma~\ref{lem:fourier:fou-spec:trotter}.
  By the Lie Product Formula, for every fixed~$x\in\RR$, the sequence of operators $(U_n(x))_{n\in\NN}$ converges to~$U(x)$,
  i.e., the sequence of operator-valued functions $(U_n)_n$ converges pointwise to~$U$.

  \begin{Claim}
    The sequence converges to~$U$ in the tempered-distribution topology.
  \end{Claim}
  This is a known fact, modulo technicalities arising from the functions being operator-valued; the standard arguments are
  below, for the sake of completeness.

  From the claim, as the Fourier transform is continuous, we find that $\hat U_n \xrightarrow{n\to\infty} \hat U$, and hence
  \begin{equation}\tag{$*$}\label{eq:RNDefwicfwu}
    \Supp\hat U \subseteq \bigcup_{n\in\NN} \Supp\widehat{U_n};
  \end{equation}
  indeed: if~$\varphi$ is a Schwartz function with support disjoint from $\bigcup_{n\in\NN} \Supp\widehat{U_n}$, then
  $\dual{\varphi}{\widehat{U_n}} = 0$ for all~$n$, and hence $\dual{\varphi}{\widehat{U}} = 0$.

  From~\eqref{eq:RNDefwicfwu} and Lemma~\ref{lem:fourier:fou-spec:trotter}, we arrive at
  $\Supp\hat U \subseteq [\lambdamin(A),\lambdamax(A)]$.

  \medskip\noindent%
  \textit{Proof of the claim.} %
  Let $\varphi\colon\RR\to\CC$ be a Schwartz function.  We have to show that the sequence of operators $(\dual{\varphi}{U_n})_n$
  converges to $\dual{\varphi}{U}$.  It is sufficient (finite-dimensionality of~$\cH$) to show convergence for every matrix
  element, i.e., for every $\phi,\psi\in\cH$, we have to show
  $\cip{\phi}{\dual{\varphi}{U_n}\psi} \xrightarrow{n\to\infty} \cip{ \phi }{ \dual{\varphi}{U}\psi }$.  By definition,
  \begin{equation*}
    \cip{ \phi }{ \dual{\varphi}{U_n}\psi }
    =
    \dual{\varphi}{ x\mapsto \cip{\phi}{U_n(x)\psi} }
    =
    \int \cip{\phi}{U_n(x)\psi}\varphi(x) \,dx;
  \end{equation*}
  similarly for~$U$ in place of~$U_n$.  As we have
  $\absb{ \cip{\phi}{U_n(x)\psi}\varphi(x) } \le \Nm{\phi}\Nm{\psi}\cdot\abs{\varphi(x)}$ for all~$x$ and $\abs{\varphi}$ is
  integrable, the condition in the dominated convergence theorem applies, and we have
  $\cip{ \phi }{ \dual{\varphi}{U_n}\psi } \xrightarrow{n\to\infty} \cip{ \phi }{ \dual{\varphi}{U}\psi }$.
  This completes the proof of the claim, and hence of Prop.~\ref{prop:fourier:fou-spec}.
\end{proof}

\section{The derivative at $\nfrac12$}\label{sec:Half-deriv}%----------------------------------------------------------------
We continue using the tempered-distributions notations defined in \S\ref{ssec:fourier:notation}.

In this section, we discuss the Fourier-analytic characterization of measures computing the derivative at~$\nfrac12$ of
functions in~$\Fun_*$ and of functions in~$\decFun_*$; cf.~\eqref{eq:overview:deriv-half--measure}.  We derive consequences for
the space of feasible shift rules, and, last not least, we'll prove that the Nyquist shift rules are feasible.

We start by discussing the potential consequences of the Fourier-Decomposition theorem in the form of
Corollary~\ref{cor:EVF-direct-sum}, i.e., the ``$\Fun_*$ vs.~$\sumFun_*$''.

\subsection{Impact of the Fourier-decomposition theorem}\label{ssec:Half-deriv:FDT}
In the next subsection \S\ref{ssec:Half-deriv:linsyseq-lemma}, we will prove the characterization, in terms of~$\hat\mu$, of the
finite complex measures~$\mu$ which satisfy~\eqref{eq:overview:deriv-half--measure}: Integrating a against~$\mu$ gives the
derivative at~$\nfrac12$ for all functions in~$\Fun_{\nfrac12}$ (Lemma~\ref{lem:overview:linsyseq}).  The current section proves
that restricting to the smaller space $\sumFun_*$ of Corollary~\ref{cor:EVF-direct-sum} does not yield any additional feasible
PSRs, i.e., we prove Prop.~\ref{prop:overview:function-spaces:equiv-feasability}.

Fix a frequency set $\Xi\subset[-\nfrac12,+\nfrac12]$ with $\max\Xi=\nfrac12$.  As both taking the derivative at~$\nfrac12$ and
integration against~$\mu$ are linear, a characterization of derivative-computing measures for $\sumFun_\Xi$
(cf.~\eqref{eq:overview:deriv-half--measure}) must consist of the following two parts:
\begin{enumerate}[label=\textcal{(\arabic*.)}]
\item For all $\xi\in\Xi$, we need $\int e^{2\pii\xi x} ,d\mu(x) = 2\pii\xi e^{i\pi\xi}$;
\item\label{enum:Half-deriv:FDT:decFun}%
  For all $f_0\in \decFun_\nfrac12$, we need $\int f_0(x)\,d\mu(x) = f_0'(\nfrac12)$.
\end{enumerate}
The first part takes care of~$f_1$ in the notation of the Fourier-decomposition theorem, Prop.~\ref{prop:fou-decomp-evf}.  It
corresponds to condition~\eqref{eq:overview:linsyseq} in Lemma~\ref{lem:overview:linsyseq}, but it is now a condition on only a
finite number of frequencies, exactly like for the unperturbed theory in Eqn.~\eqref{OPT:primal:GlSys:hatphi-equals-2piixi}
of~\cite{Theis:opt-shiftrules:2021} (only that we have chosen the anchor point~$\nfrac12$ here, instead of~$0$ in
\cite{Theis:opt-shiftrules:2021}).

As for part~\ref{enum:Half-deriv:FDT:decFun}, the hope would be that, even though in the functions in~$\decFun_\nfrac12$ all
Fourier frequencies in $[-\nfrac12,+\nfrac12]$ are allowed to occur, the linear decay condition
%%
%\FinishThis{%
%  (and potentially additional conditions based on Prop.~\ref{prop:fourier:decomp}\ref{prop:fourier:decomp:powerseries}) } %
%%
would result in a larger set of feasible PSRs.

Unfortunately, that does not seem to be the case: Not even restricting the space in part~\ref{enum:Half-deriv:FDT:decFun} to
only translated $\sinc$-functions,
\begin{equation*}
  f_0 = \sinc(\place - x_0)\text{ for some $x_0\in\RR$,}
\end{equation*}
increases the space of feasible PSRs, as the following lemma shows.

\begin{lemma}\label{lem:Half-deriv:FDT}
  Let~$\mu$ be a complex measure.
  If
  \begin{equation}\tag{weak-\ref{eq:overview:deriv-half--measure}}\label{eq:Half-deriv:FDT:deriv-half--measure}
    \sinc'(\nfrac12 - x_0) = \int \sinc(\place - x_0) \,d\mu
    \qquad
    \text{for all $x_0\in\ZZ$},
  \end{equation}
  then $\mu$ satisfies the condition~\eqref{eq:overview:linsyseq} of Lemma~\ref{lem:overview:linsyseq}.
\end{lemma}

Clearly, \eqref{eq:overview:deriv-half--measure} implies~\eqref{eq:Half-deriv:FDT:deriv-half--measure};
Lemma~\ref{lem:Half-deriv:FDT} together with Lemma~\ref{lem:overview:linsyseq} show that the two conditions are in fact
equivalent.

\begin{proof}[Proof of Lemma~\ref{lem:Half-deriv:FDT}]
  We use a basic fact \cite[Theorem~2.1]{Stein-Shakarchi:Fourier:2003} about Fourier transforms of periodic functions in the
  following form: If $h_1,h_2\colon\RR\to\CC$ are two continuous functions such that, for all~$x_0\in\ZZ$,
  \begin{equation}\tag{$*$}\label{eq:RNDf2kfqb8y}
    \int_{-\nfrac12}^\nfrac12 e^{2\pii\xi x_0} h_1(\xi) \,d\xi
    =
    \int_{-\nfrac12}^\nfrac12 e^{2\pii\xi x_0} h_2(\xi) \,d\xi,
  \end{equation}
  then $h_1,h_2$ agree on~$[-\nfrac12,+\nfrac12]$.

  For $h_2$ we take the RHS in condition~\eqref{eq:overview:linsyseq}: $h_2(\xi) = -2\pii\xi e^{-i\pi\xi}$.  With~$\mu$
  satisfying Condition~\eqref{eq:Half-deriv:FDT:deriv-half--measure}, for $h_1$ we take $\hat\mu$, recalling that the
  Fourier-Stieltjes transform of a finite measure is continuous.

  That's it, now we verify~\eqref{eq:RNDf2kfqb8y} by direct calculation.  For every~$x_0\in\RR$, we have ($\check\place$ is
  inverse Fourier transform):
  \begin{align*}
    \int_{-\nfrac12}^\nfrac12 e^{2\pii\xi x_0} \hat\mu(\xi) \,d\xi
    &=
      \int_{-\nfrac12}^\nfrac12 e^{2\pii\xi x_0} \int_\RR e^{-2\pii\xi x} \,d\mu(x) \,d\xi
    \\
    &=
      \int_\RR \int_{-\nfrac12}^\nfrac12 e^{2\pii\xi x_0} \, e^{-2\pii\xi x} \,d\xi \,d\mu(x)
      &&\cmmt{finite measures, bounded functions}
    \\
    &=
      \int_\RR \int_\RR  e^{-2\pii\xi x} \cdot e^{2\pii\xi x_0} \One_{[-\nfrac12,+\nfrac12]}(\xi) \,d\xi \,d\mu(x)
    \\
    &=
      \int_\RR \sinc(\place-x_0) \,d\mu
    &&\cmmt{$(\One_{[-\nfrac12,+\nfrac12]})^\Hat = \sinc$}
    \\
    &=
      \sinc'(\nfrac12 - x_0)
    &&\cmmt{condition~\eqref{eq:Half-deriv:FDT:deriv-half--measure}}
    \\
    &=
      \int_\RR e^{-2\pii\xi(\nfrac12-x_0)} \cdot (-2\pii\xi) \One_{[-\nfrac12,+\nfrac12]} \,d\xi
    &&\cmmt{$\sinc' = \bigl(-2\pii\place\,\One_{[-\nfrac12,+\nfrac12]}\bigr)^\Hat$}
    \\
    &=
      \int_{-\nfrac12}^\nfrac12 e^{2\pii\xi x_0} \, (-2\pii\xi)e^{-i\pi\xi} \,d\xi.
  \end{align*}
  This completes the proof of Lemma~\ref{lem:Half-deriv:FDT}.
\end{proof}

We can now finish off the proposition in \S\ref{sssec:overview:function-spaces:equiv-feasability}.

\begin{proof}[Proof of Prop.~\ref{prop:overview:function-spaces:equiv-feasability}]
  The claimed feasibility for $\Fun_{\max\Xi}$ follows from the feasibility for $\sumFun_\Xi$ by Lemma~\ref{lem:Half-deriv:FDT}
  and Prop.~\ref{prop:overview:𝛍-𝛟}.
\end{proof}

\subsection{Fourier-analytic characterization of derivative-computing measures, case~$\sumFun_*$}\label{ssec:Half-deriv:linsyseq-lemma}
In \S\ref{ssec:overview:𝛍-𝛟:charac}, we have proven one direction of Lemma~\ref{lem:overview:linsyseq}.  Here, we prove the
other direction in the special case described in Corollary~\ref{cor:EVF-direct-sum}: We replace, in
condition~\eqref{eq:overview:deriv-half--measure}, the qualification ``$g\in\Fun_\nfrac12$'' by ``$g\in\sumFun_\Xi$'', for
arbitrary frequency set $\Xi \subset[-\nfrac12,+\nfrac12]$.

As a matter of fact, we will take a condition that is both more general and allows for a lazier proof --- here is the statement
(cf.~Lemma~\ref{lem:overview:linsyseq}):

\begin{proposition}\label{prop:Half-deriv:linsyseq-lemma}
  Let~$\mu$ be a complex measure on~$\RR$.
  If~\eqref{eq:overview:linsyseq} holds,
  then \(\displaystyle %%
  g'(\nfrac12) = \int g \, d\mu \) %%
  holds for all functions $g\in\Fun_\nfrac12$ which can be decomposed as $g = g_0 + g_1$ where
  \begin{itemize}
  \item $g_0$ is smooth and square integrable; and
  \item $g_1$ has finite Fourier spectrum, i.e.\footnote{Note that boundedness of~$g_1$ is implied by the smoothness of~$g_0$.},
    there exists a finite $\Xi\subseteq[-\nfrac12,+\nfrac12]$ with $g_1 \in\oldFun_\Xi$.
  \end{itemize}
\end{proposition}
\begin{proof}
  Let $\mu,g,g_0,g_1,\Xi$ as described.  There exist complex numbers $b_\xi$, $\xi\in\Xi$, such that
  \begin{equation}\label{eq:RND87qgtbo8:decompose}\tag{$*$}
    \begin{split}
      g      &=    g_0        + \sum_{\xi\in\Xi} b_\xi e^{2\pii\xi\place},
      \\
      \hat g &= \widehat{g_0} + \sum_{\xi\in\Xi} b_\xi \delta_\xi.
    \end{split}
  \end{equation}
  Note that $\widehat{g_0}$ is the usual $L^2$-Fourier transform of~$g_0$, in particular, it is an $L^2$-function (not an evil
  tempered distribution).

  By linearity, it suffices to prove the derivative-computing measure equation for each term of~\eqref{eq:RND87qgtbo8:decompose}
  separately:
  \begin{subequations}
    \begin{align}
      \label{eq:RND87qgtbo8:L2}\tag{A}
      &\text{For the $L^2$-part:   } & g_0'(\nfrac12)          &= \int g_0(x) \,d\mu(x);
      \\
      \label{eq:RND87qgtbo8:xi}\tag{B}
      &\text{For each $\xi\in\Xi$: } & 2\pii\xi \, e^{i\pi\xi} &= \int e^{2\pii\xi x} \,d\mu(x).
    \end{align}
  \end{subequations}
  The equations~\eqref{eq:RND87qgtbo8:xi} correspond to the condition~\eqref{eq:overview:linsyseq}, evaluated in $-\xi$, so
  there is nothing to be done.

  For the $L^2$-part~\eqref{eq:RND87qgtbo8:L2}, we argue based on basic Fourier analysis.
  As~$g_0$ is a smooth $L^2$-function whose Fourier spectrum is contained in $[-\nfrac12,+\nfrac12]$, we have
  $\hat g_0\in L^1([-\nfrac12,+\nfrac12])$, and the following equations:
  \begin{equation*}
    \begin{split}
      g_0(x)  &= \int_{-\nfrac12}^{\nfrac12}            e^{2\pii\xi x} \, \hat g_0(\xi) \,d\xi \qquad\text{for all $x\in\RR$, and}\\
      g_0'(x) &= \int_{-\nfrac12}^{\nfrac12} 2\pii\xi \,e^{2\pii\xi x} \, \hat g_0(\xi) \,d\xi \qquad\text{for all $x\in\RR$.}
    \end{split}
  \end{equation*}
  Using these, we calculate:
  \begin{align*}
    \int g_0(x) \,d\mu(x)
    &=
      \int \int_{-\nfrac12}^{\nfrac12} e^{2\pii\xi x}\,\widehat{g_0}(\xi) \,d\xi \,d\mu(x)
    \\
    &=
      \int_{-\nfrac12}^{\nfrac12} \int e^{2\pii\xi x}\,\widehat{g_0}(\xi) \,d\mu(x) \,d\xi %&&\cmmt{Tonelli+Fubini as $\widehat{g_0}\in L^1$}
    \\
    &=
      \int_{-\nfrac12}^{\nfrac12} \int e^{2\pii\xi x} \,d\mu(x) \widehat{g_0}(\xi) \,d\xi
    \\
    &=
      \int_{-\nfrac12}^{\nfrac12} (2\pii\xi) e^{i\pi\xi} \widehat{g_0}(\xi) \,d\xi
    &&\cmmt{condition~\eqref{eq:overview:linsyseq}}
    \\
    &=
      g_0'(\nfrac12).
  \end{align*}
  This completes the proof.
\end{proof}

As noted in \S\ref{ssec:overview:𝛍-𝛟:charac}, the proof of the general case of Lemma~\ref{lem:overview:linsyseq}, i.e., taking
derivatives of all $\Fun_\nfrac12$-functions, is in Appendix~\ref{apx:Half-deriv:linsyseq-lemma}.

\subsection{Proof of the Space-of-Feasible-PSRs proposition}\label{ssec:Half-deriv:feasSR-space}
We now prove Prop.~\ref{prop:overview:feasSR-space}: The set of all signed measures~$\mu$
satisfying~\eqref{eq:overview:linsyseq} is either empty or an infinite-dimensional (real) affine space.  If it is
empty, there is nothing to prove.  Otherwise let~$\mu$ be such a measure.

As indicated in \S\ref{ssec:overview:𝛍-𝛟:charac}, we will exhibit a set of measures $\nu_a$, $a\in\NN$, that are linearly
independent\footnote{I.e., for every finite(!) linear combination of $\nu_*$'s, if the result equals the $0$-measure, then all
  coefficients in the linear combination must be~$0$.}  and that satisfy $(\Supp\nu_a) \cap [-\nfrac12, +\nfrac12] = 0$.  Each
of the measures $\mu + \nu_a$, $a\in\NN$, satisfies the condition~\eqref{eq:overview:linsyseq} in
Lemma~\ref{lem:overview:linsyseq}, as~$\mu$ satisfies it.

For the $\nu_a$ we take the finite signed measures $\cos(2\pi a\place) \sinc^2$ (density on Lebesgue measure).  We prove the
conditions in the following two lemmas, but first we need a remark.

\begin{remark}\label{rem:𝚲}
  Denote by $\Lambda\colon\RR\to\RR\colon \xi\mapsto \max(0,1-\abs{\xi})$ the \textit{triangle function.}  It is a standard fact
  that $\hat\Lambda = \sinc^2$, and this equation easily implies
  \begin{equation*}
    \check\Lambda
    =
    \sinc^2.
  \end{equation*}
\end{remark}

\begin{lemma}\label{ssec:Half-deriv:feasSR-space:FtSupp}
  For all $a\in\RR$, we have
  $\displaystyle \bigl( \cos(2\pi a\place)\sinc^2 \bigr)^\Hat = \bigl( \Lambda(\place-a) + \Lambda(\place+a) \bigr)/2$.
\end{lemma}
\begin{proof}
  By Remark~\ref{rem:𝚲}, we have $\bigl( \Lambda(\place-a) \bigr)^\vee = e^{2\pii a\place}\sinc^2$, so that
  $\Bigl( \bigl(\Lambda(\place-a) + \Lambda(\place+a) \bigr)/2 \Bigr)^\vee = \cos(2\pi a\place)\sinc^2$.

  The statement about the Fourier transform of $\cos(2\pi a\place)\sinc^2$ follows from the Fourier Inversion Theorem (e.g.,
  \cite[Theorem~4.11]{Folland:abstract-HA:2016}).
\end{proof}

\begin{lemma}\label{ssec:Half-deriv:feasSR-space:linindep}
  The set of functions
  \begin{equation*}
    \Bigl\{ \cos(2\pi (3a+\nfrac12)\place)\,\sinc^2  \Bigm| a \in \NN \Bigr\}
  \end{equation*}
  is linearly independent.
\end{lemma}

Before we prove the lemma, we finish the proof of Prop.~\ref{prop:overview:feasSR-space}.  Let
\begin{equation*}
  \nu_a := \cos(2\pi(3a+\nfrac12)\place)\sinc^2\cdot\lambda,
\end{equation*}
where $\lambda$ is the Lebesgue measure.  By Lemma~\ref{ssec:Half-deriv:feasSR-space:FtSupp}, we have
\begin{equation}\label{rnd:oeufbwoe:supp-of-nu}
  \Supp \widehat{\nu_a} = [-3a-\nfrac32, -3a+\nfrac12] \cup [3a-\nfrac12, 3a+\nfrac32]
\end{equation}
and these sets are disjoint from $[-\nfrac12,+\nfrac12]$.  The proof of Prop.~\ref{prop:overview:feasSR-space} is completed by
proving Lemma~\ref{ssec:Half-deriv:feasSR-space:linindep}: the linear independence of the $\nu_a$, $a\in\NN$.

\begin{proof}[Proof of Lemma~\ref{ssec:Half-deriv:feasSR-space:linindep}]
  The functions $\widehat{\nu_a}$, $a\in\NN$ are linearly independent as their supports are disjoint;
  see~\eqref{rnd:oeufbwoe:supp-of-nu}.  As the inverse Fourier transform is injective and Fourier Inversion holds, the measures
  $\nu_a$, $a\in\NN$, are linearly independent, too.
\end{proof}

\subsection{Proof of feasibility of the Nyquist shift rule}\label{ssec:Half-deriv:pf-half--measure}
In this section, we prove Theorem~\ref{thm:overview:half--measure}: The measure~$\mu$ defined
in~\eqref{eq:overview:def:half--measure} satisfies the condition~\eqref{eq:overview:deriv-half--measure} of
Lemma~\ref{lem:overview:linsyseq}.

The function $u\colon\ZZ\to\RR$ in~\eqref{eq:overview:def:half--measure} plays a special role.  In the next subsection,
\S\ref{ssec:Half-deriv:pf-half--measure:Foutraf-u-pf}, we will prove the following.

\begin{proposition}\label{prop:Half-deriv:pf-half--measure:Foutraf-u}
  With~$u$ as in~\eqref{eq:overview:def-u}, we have,
  \begin{equation}\label{eq:Half-deriv:pf-half--measure:Foutraf-u}
    \text{for all $\xi\in [-\nfrac12,+\nfrac12]$: }\quad  \hat u(\xi) = -2\pii\xi \, e^{-i\pi\xi}.
  \end{equation}
\end{proposition}

Here, we are using the Fourier transform on~$\ZZ$: For all absolutely summable sequences $v\colon\ZZ\to\CC$,
\begin{equation}\label{eq:ZZ-Fourier-transform}
  \hat v(\xi) := \sum_{n\in\ZZ} e^{-2\pii\xi n} v(n);
\end{equation}
the result will be a 1-periodic\footnote{More accurately, a continuous function on $\RR/\ZZ$.} continuous function.\footnote{And, wait for it --- the
  RHS of~\eqref{eq:Half-deriv:pf-half--measure:Foutraf-u} indeed evaluates to $-i\pi$ for both $\xi=\nfrac12$ and
  $\xi=-\nfrac12$.}

Completing the proof of the theorem is now a \emph{piece of cake.}

\begin{proof}[Proof of Theorem~\ref{thm:overview:half--measure}]
  Let~$\mu$ be as in~\eqref{eq:overview:def:half--measure}, and $\xi \in [-\nfrac12,+\nfrac12]$.  A swift calculation:
  \begin{align*}
    \hat\mu(\xi)
    &=
      \int e^{-2\pii\xi s} \mu(s)
    \\
    &=
      \int e^{-2\pii\xi s} \,d\bigl({\textstyle\sum_n u(n)\delta_n}\bigr)(s) &&\cmmt{def.~of $\mu$ in~\eqref{eq:overview:def:half--measure}}
    \\
    &=
      \sum_{n\in\ZZ} \int e^{-2\pii\xi s} u(n)\,d\delta_n(s)
    \\
    &=
      \sum_{n\in\ZZ} u(n) e^{-2\pii\xi n}     &&
    \\
    &=
      \hat u(\xi)
    \\
    &=
      -2\pii\xi \, e^{-i\pi\xi}               &&\cmmt{by Prop.~\ref{prop:Half-deriv:pf-half--measure:Foutraf-u}.}
  \end{align*}
  And the proof is completed.
\end{proof}

\subsubsection{The Fourier transform of~$u\in\ell^1(\ZZ)$}\label{ssec:Half-deriv:pf-half--measure:Foutraf-u-pf}
To compute the $\ZZ$-Fourier transform of~$u$, basically, you have to take the $\ZZ$-Fourier transform of $n\mapsto 1/n^2$ and
work from there.

\begin{subequations}
  We denote by $\Li$ the \emph{dilogarithm} or \emph{Spence's function}~\cite{Zagier:dilogarithm:2007}.  It is a holomorphic
  function on $\CC\setminus \lt[1,\infty\rt[$ with continuous extensions to $\CC\setminus\lt]1,\infty\rt[$.  The eponymous
  equation
  \begin{equation}\label{eq:dilogarithm:sum-def}
    \Li(z) = \sum_{n=1}^\infty \frac{z^n}{n^2} \text{ for all $z\in\CC$ with $\abs{z} \le 1$}
  \end{equation}
  holds (Eqn.~(3.1) in~\cite{Maximon:dilogarithm:2003}, or beginning of Chapter~I in~\cite{Zagier:dilogarithm:2007}, plus Abel's
  theorem).
  Moreover we have the functional equation (\S{I.2} in~\cite{Zagier:dilogarithm:2007}, Eqn.~(3.2)
  in~\cite{Maximon:dilogarithm:2003}):
  \begin{equation}\label{eq:dilogarithm:funeq}
    \Li(z) + \Li(1/z) = \tfrac{-1}{2}\Log^2(-z) - \pi^2/6 \quad \text{ for all $z\notin \lt[0,\infty\rt[$,}
  \end{equation}
  where~$\Log$ is the principle branch of the complex logarithm, i.e., defined on $\CC\setminus\lt]-\infty,0\rt]$.
\end{subequations}

We recycle the notation ``$\modh$''; see~\eqref{eq:overview:fold:def-mod}.

\begin{lemma}\label{lem:Half-deriv:pf-half--measure:Foutraf-u-pf:DiLog-g}
  For all $\xi\in\RR$ we have
  \begin{equation*}
    \Li(e^{2\pii\xi}) + \Li(e^{-2\pii\xi})
    = \frac{- \bigl( 2\pii \cdot ( (\xi+\nfrac12)\modh 1 ) \bigr)^2 }{2} - \pi^2/6.
  \end{equation*}
\end{lemma}
\begin{proof}
  The functional equation~\eqref{eq:dilogarithm:funeq} of the dilogarithm gives the equation for all $\xi\notin\ZZ$.

  We verify it by hand for $\xi\in\ZZ$.  In that case, on the LHS we have $2\cdot\Li(1) = \pi^2/3$
  by~\eqref{eq:dilogarithm:sum-def} and the Basel Problem (or \S{I.1} in~\cite{Zagier:dilogarithm:2007}).

  On the RHS, for integral~$\xi$, we find that $(\xi+\nfrac12)\modh 1 = -\nfrac12$, and hence
  \begin{equation*}
    \frac{- \bigl( 2\pii \cdot ( (\xi+\nfrac12)\modh 1 ) \bigr)^2 }{2}
    =
    \frac{- (-i\pi)^2 }{2}
    =
    \pi^2/2,
  \end{equation*}
  from which we have to subtract $\pi^2/6$ to arrive at the RHS; so LHS and RHS coincide also for integral~$\xi$.
\end{proof}

We define for all $\xi \in\RR$,
\begin{subequations}
  \begin{align}\label{eq:Half-deriv:pf-half--measure:Foutraf-u-pf:def-g}
    g(\xi) &:= \Li(e^{2\pii\xi}) + \Li(e^{-2\pii\xi})
    \\\notag
           &=  \sum_{n\in\ZZ\setminus\{0\}} \frac{e^{-2\pii\xi n}}{n^2} &&\cmmt{by~\eqref{eq:dilogarithm:sum-def}}
    \\\notag
           &=  \frac{- \bigl( 2\pii \cdot ( (\xi+\nfrac12)\modh 1 ) \bigr)^2 }{2} - \pi^2/6 &&\cmmt{by Lemma~\ref{lem:Half-deriv:pf-half--measure:Foutraf-u-pf:DiLog-g}.}
    \\
    g_0(\xi) &:= g(2\xi)/4
    \\\notag
           &= \sum_\sstack{n\in\ZZ\setminus\{0\},\\n\text{ even}}     \frac{  e^{-2\pii\xi n}  }{  n^2  }
    \\
    g_1(\xi) &:= g(\xi) - g_0(\xi)
    \\\notag
           &= \sum_\sstack{n\in\ZZ,\\n\text{ odd}}                    \frac{  e^{-2\pii\xi n}  }{  n^2  }.
  \end{align}
\end{subequations}
(The equation for~$g_0$ follows from the one for~$g$ by applying it for $2\xi$.)

\begin{lemma}\label{lem:Half-deriv:pf-half--measure:Foutraf-u-pf:DiLog-g1}
  For all $\xi\in\RR$,
  \begin{equation*}
    \hat u(\xi)
    =
    \tfrac{-4i}{\pi} \cdot e^{-i\pi\xi} \cdot g_1( (\xi-\nfrac12)/2 ).
  \end{equation*}
\end{lemma}
\begin{proof}
  This is a direct calculation:
  \begin{align*}
    \hat u(\xi)
    &= \sum_{n\in\ZZ} \frac{ - (-1)^n e^{-2\pii\xi n} }{\pi \, (n-\nfrac12)^2 }
    \\
    &= \sum_{n\in\ZZ} \frac{ - (-1)^n e^{-2\pii\xi n} }{\pi \, (2n-1)^2/4 }
    \\
    &= \sum_{n\in\ZZ} \frac{ - e^{-2\pii(\xi -\nfrac12)n} }{\pi \, (2n-1)^2/4 }
    \\
    &= \tfrac{-4}{\pi} \sum_{n\in\ZZ} \frac{ e^{-2\pii(\xi -\nfrac12)n} }{ (2n-1)^2 }
    \\
    &= \tfrac{-4}{\pi} \sum_{n\in\ZZ} \frac{ e^{-2\pii(\xi/2 -\nfrac14)(2n)} }{ (2n-1)^2 }
    \\
    &= \tfrac{-4}{\pi} \cdot e^{-2\pii(\xi/2-\nfrac14)} \cdot \sum_{n\in\ZZ} \frac{ e^{-2\pii(\xi/2 -\nfrac14)(2n-1)} }{ (2n-1)^2 }
    \\
    &= \tfrac{-4i}{\pi} \cdot e^{-i\pi\xi} \cdot g_1( (\xi-\nfrac12)/2 ),
  \end{align*}
  where in the last step we have used the equation for~$g_1$ above.
\end{proof}

We can now complete the proof of the proposition.

\begin{proof}[Proof of Prop.~\ref{prop:Half-deriv:pf-half--measure:Foutraf-u}]
  As $\hat v$ is 1-periodic for every absolutely summable $v\colon\ZZ\to\ZZ$, we fix $\xi\in\lt[ -\nfrac12 , +\nfrac12\rt[$.
  We have
  \begin{align*}
    g_1( (\xi-\nfrac12)/2 )
    &=
      g( (\xi-\nfrac12)/2 ) - g_0( (\xi-\nfrac12)/2 )
    \\
    &=
      g( (\xi-\nfrac12)/2 ) - g( \xi-\nfrac12 )/4
    \\
    &=
      \frac{- \bigl( 2\pii \cdot ( (\xi/2+\nfrac14)\modh 1 ) \bigr)^2 }{2}         -
      \frac{- \bigl( 2\pii \cdot (  \xi            \modh 1 ) \bigr)^2 }{2}\cdot\frac{1}{4}
      \\
    &\qquad
      - \pi^2/6 + \pi^2/24
    \\
    &\eqcmt{(*)}
      2\pi^2 \Bigl(
      (\xi/2+\nfrac14)^2     -   \xi^2/4
      \Bigr)
      - \pi^2/8
    \\
    &=
      \tfrac{\pi^2}{2} (
      \xi + \nfrac14
      )
      - \pi^2/8.
  \end{align*}
  In equation marked with an asterisk, omitting the ``$\%1$'' is allowed as $\xi \in \lt[-\nfrac12,+\nfrac12\rt[$ and hence
  $\xi/2+\nfrac14 \in \lt[-\nfrac12,+\nfrac12\rt[$.

  Now starting from Lemma~\ref{lem:Half-deriv:pf-half--measure:Foutraf-u-pf:DiLog-g1}, we compute, for every $\xi\in\RR$,
  \begin{align*}
    \hat u(\xi)
    &=
      \tfrac{-4i}{\pi} \cdot e^{-i\pi\xi} \cdot g_1( (\xi-\nfrac12)/2 )
    \\
    &=
      \tfrac{-4i}{\pi} \cdot e^{-i\pi\xi} \cdot \bigl( \tfrac{\pi^2}{2} (\xi + \nfrac14) -\pi^2/8 \bigr)
    \\
    &=
      -2\pii     \cdot e^{-i\pi\xi} \cdot (\xi + \nfrac14 - \nfrac14)
    \\
    &=
      -2\pii\xi \cdot e^{-i\pi\xi}.
  \end{align*}
  This completes the proof of the proposition.
\end{proof}

\section{Details about reflection and dilation}\label{sec:𝛍-𝛟}%--------------------------------------------------------
\subsection{Proof of the general facts}\label{ssec:𝛍-𝛟:general}
We prove Prop.~\ref{prop:overview:𝛍-𝛟} stated in \S\ref{ssec:overview:𝛍-𝛟}.  The main work is in the following
lemma.

\begin{lemma}\label{lem:𝛍-𝛟:general}
  Let~$\mu$ be a finite measure.
  \begin{enumerate}[label=(\alph*)]
  \item\label{lem:𝛍-𝛟:general:1/2}%
    With $\tau\colon\RR\to\RR\colon s\mapsto \nfrac12-s$ the reflection on~$\nfrac12$:
    The finite measure~$\mu$ satisfies~\eqref{eq:overview:deriv-half--measure}, if and only if $\phi:=\tau(\mu)$ is feasible for
    $\Fun_{\nfrac12}$.
  \item\label{lem:𝛍-𝛟:general:K}%
    For fixed $K_1,K_2$, with $\tau\colon\RR\to\RR\colon x\mapsto K_1 x / K_2$ the dilation by $K_1/K_2$:
    If the PSR~$\phi$ is feasible for~$\Fun_{K_1}$, then the PSR $\psi := K_2/K_2 \cdot \tau(\phi)$ is feasible for
    $\Fun_{K_2}$.
  \end{enumerate}
\end{lemma}
\begin{proof}
  \emph{Proof of Item~\ref{lem:𝛍-𝛟:general:1/2}:} %%
  Let~$x\in\RR$ and $f,g\in\Fun_{\nfrac12}$ such that $g = f(\place+x-\nfrac12)$, i.e., $f = g(\place-x+\nfrac12)$; note that
  $\Fun_\nfrac12$ is invariant under translations.  We write:
  \begin{alignat*}{2}
    f'(x)
    = g'(\nfrac12)
%    \\
    &&\qquad\eqcmt{(?)}\qquad
       \int g(s)\,d\mu(s)
    &
       = \int f(x-(\nfrac12-s)) \,d\mu(s)
    \\
    &&&
        = \int f(x-\place)\circ\tau \,d\mu
    \\
    &&&
        = \int f(x-\place) \,d\phi
    \\
    &&&
        = f*\phi(x),
  \end{alignat*}
  where all equations except the one tagged with the question mark hold, and ($?$) is equivalent to $f'(x) = f*\phi(x)$.
  As ($?$) is~\eqref{eq:overview:deriv-half--measure}, this proves Item~\ref{lem:𝛍-𝛟:general:1/2}.

  \myparsmall
  \emph{Proof of Item~\ref{lem:𝛍-𝛟:general:K}:} %%
  For $f\in\Fun_{K_2}$ and $x\in\RR$, noting that $f(K_1\place/K_2) \in \Fun_{K_1}$, we calculate, using in~($*$)
  that~$\phi$ is feasible for $\Fun_{K_1}$:
  \begin{multline*}
    f'(x)
    =
    \tfrac{K_2}{K_1} \, \lt( f(K_1\place/K_2) \rt)'        (K_2\, x/K_1)
    \\
    \eqcmt{(*)}
    \tfrac{K_2}{K_1} \,    ( f(K_1\place/K_2)    )*\phi \; (K_2\, x/K_1)
    =
    \tfrac{K_2}{K_1} \, \int f( K_1(K_2\, x/K_1 - s)/K_2 ) d\phi(s)
    \\
    \shoveright{%
    =
    \tfrac{K_2}{K_1} \, \int f( x - K_1 \,s /K_2 ) d\phi(s)
    =
    \int f( x - \place )\circ\tau \, \tfrac{K_2}{K_1} \, d\phi
    =
    \int f(x - \place ) \; d\Bigl(\tfrac{K_2}{K_1}\tau(\phi)\Bigr)
    }%
    \\
    =
    \int f(x - \place ) \, d\psi
    =
    f*\psi (x),
  \end{multline*}
  which proves Item~\ref{lem:𝛍-𝛟:general:K}.
\end{proof}

As a direct corollary of the lemma we obtain the isomorphisms required in Prop.~\ref{prop:overview:𝛍-𝛟}, noting that:
\textcal{(1)} taking images of measures is a $\CC$-linear operation; \textcal{(2)} reflection and dilation map signed measures
to signed measures; \textcal{(3)} reflection is self-inverse, and the inverse of dilation by~$\alpha$ is dilation by $1/\alpha$.

This completes the proof of Prop.~\ref{prop:overview:𝛍-𝛟}.

\subsection{Derivation of the Nyquist shift rules}\label{ssec:𝛍-𝛟:nyquist}

\begin{proof}[Proof of Corollary~\ref{cor:shift-rule}]
  Applying Theorem~\ref{thm:overview:half--measure}, Lemma~\ref{lem:overview:linsyseq}, and
  Lemma~\ref{lem:𝛍-𝛟:general}\ref{lem:𝛍-𝛟:general:1/2} to the~$\mu$ defined
  in~\eqref{eq:overview:def:half--measure}, we find that the measure $\phi := \tau(\mu)$, for $\tau\colon x\mapsto\nfrac12-x$
  the reflection on~$\nfrac12$, is a PSR that is feasible for~$\Fun_\nfrac12$.  We calculate, by linearity and norm-continuity
  of $\nu\mapsto\tau(\nu)$ and Eqn.~\eqref{eq:overview:tau-of-delta},
  \begin{multline*}
    \phi
    = \tau(\mu)
    = \sum_{n\in\ZZ} u(n) \tau(\delta_n)
    = \sum_{n\in\ZZ} u(n) \delta_{\tau(n)}
    = \sum_{n\in\ZZ} u(n) \delta_{\nfrac12-n}
    \\
    =           \sum_{n\in\ZZ}          \frac{ (-1)^{n-1}        }{ \pi \, (n - \nfrac12)^2 } \, \delta_{\nfrac12-n}
    \eqcmt{(*)} \sum_{a\in\nfrac12+\ZZ} \frac{ (-1)^{a+\nfrac12} }{ \pi \, a^2              } \, \delta_a
    = \phi_\nfrac12
  \end{multline*}
  where in the equation marked with the asterisk we perform the change of variables $a:=\nfrac12-n$.  This proves the corollary.
\end{proof}

\begin{proof}[Proof of Corollary~\ref{cor:overview:general-shift-rule}]
  From the previous corollary, we know that~$\phi_\nfrac12$ is a PSR that is feasible for~$\Fun_\nfrac12$.  Fix
  $K>0$.  Applying Lemma~\ref{lem:𝛍-𝛟:general}\ref{lem:𝛍-𝛟:general:K}, we find that the measure
  $2K\cdot (\place/2K)(\phi_\nfrac12)$ is a PSR feasible for~$\Fun_K$.  Again invoking linearity, continuity, and
  Eqn.~\eqref{eq:overview:tau-of-delta}, we calculate
  \begin{equation*}
    2K   \cdot (\place/2K)(\phi_\nfrac12)
    = 2K \sum_{a\in\nfrac12+\ZZ} \frac{ (-1)^{a+\nfrac12} }{ \pi \, a^2 } \, (\place/2K)(\delta_a)
%%    \\
    = 2K \sum_{a\in\nfrac12+\ZZ} \frac{ (-1)^{a+\nfrac12} }{ \pi \, a^2 } \, \delta_{a/2K}
    = \phi_K,
  \end{equation*}
  and the proof of the corollary is complete.
\end{proof}

\section{Optimality}\label{sec:opt}%-----------------------------------------------------------------------------------------
This section has parallels with~\cite{Theis:opt-shiftrules:2021}.
\subsection{The Weak Duality Theorem}\label{ssec:opt:weak-duality}
Compared to \cite[Prop.~\ref{OPT:prop:overview:weak-duality}]{Theis:opt-shiftrules:2021}, the version of the Weak Duality
Theorem in the current paper has the additional statement about the inequality being an equation.

\begin{proof}[Proof of Prop.~\ref{prop:overview:weak-duality}]
  Let $\phi$ be a PSR feasible for~$\GenericFun$ and $f\in\GenericFun$.  We calculate:
  \begin{align*}
    -f'(0)
    &=
      ( f(-\place) )'(0)
    \\
    &=
      (f(-\place))*\phi  (0)              &&\cmmt{by feasibility of~$\phi$ and $f(-\place)\in\GenericFun$}
    \\
    &=
      \int f(s) \, d\phi(s)
    \\
    &\lecmt{(*)}
      \Nm{f}_\infty \cdot \Nm{\phi}      &&\cmmt{\eqref{eq:overview:integral-nm-ieq} in Remark~\ref{rem:overview:integral/convolution-ieqs}}
    \\
    &=
      \Nm{\phi}.                         &&\cmmt{as $\Nm{f}_\infty \le 1$}
  \end{align*}
  In the second equation, we have used the condition that $f(-\place)\in\GenericFun$ if $f\in\GenericFun$.

  The inequality $-f'(0) \le \Nm{\phi}$ --- that we have just proven --- is satisfied with equality if and only if the
  inequality ($*$) holds with equality.
\end{proof}

\subsection{Proof of optimality of the Nyquist shift rules}\label{ssec:opt:proof}
Here we give the proof of Theorem~\ref{thm:overview:opt}.

\begin{proof}[Proof of Theorem~\ref{thm:overview:opt}]
  In view of the Weak Duality Theorem, Prop.~\ref{prop:overview:weak-duality}, we need to present a function
  $f^\star\in\Fun_K$ such that
  \begin{equation}\label{eq:RNDs8odi3fj}\tag{$*$}
    -\partial f^\star(0) = \Nm{\phi_K}.
  \end{equation}
  Clearly, the function $f^\star\colon x\mapsto -\sin(2\pi K x)$ is in $\Fun_K$, and we have $-\partial f^\star(0) = 2\pi K$.

  To compute $\Nm{\phi_K}$, first recall the well-known fact (Lemma~\ref{lem:math:sum-square-odd} in
  Appendix~\ref{apx:math:sum-square-odd}) that the sum of all reciprocals of the squares of all positive odd integers
  is~$\pi^2/8$.

  Regarding~$\phi_K$, we have
  \begin{multline*}
    \Nm{\phi_K}
    =
    2K\cdot \sum_{a\in\nfrac12+\ZZ} \frac{1}{\pi\, a^2}
    =
    2K\cdot \sum_{n\in\ZZ} \frac{1}{\pi\, (n+\nfrac12)^2}
    =
    2K\cdot \sum_{n\in\ZZ} \frac{1}{\pi\, (2n+1)^2/4}
    \\
    =
    2K\cdot \frac{4}{\pi} \cdot \sum_\sstack{m\in\ZZ\\m\text{ odd}} \frac{1}{m^2}
    =
    2K\cdot \frac{4}{\pi} \cdot 2\pi^2/8
    =
    2\pi K.
  \end{multline*}
  This establishes the second part of~\eqref{eq:RNDs8odi3fj}, and completes the proof of the theorem.
\end{proof}

\subsection{The support of optimal proper shift rules}\label{ssec:opt:otheropt}
We now sketch the proof of Corollary~\ref{cor:overview:otheropt}.

Let $K>0$ and assume~$\phi$ is a PSR feasible for~$\Fun_K$.  Moreover, let $f^\star := -\sin(2\pi K \place)$ be the dual
optimal solution established in the proof of Theorem~\ref{thm:overview:opt} above.

As we already know from the proof of Theorem~\ref{thm:overview:opt} that $-\partial f^\star(0)$ is equal to the norm of the
optimal shift rule, by the Weak Duality Theorem applied to $\phi,f^\star$, equality in the
inequality~\eqref{eq:overview:weak-duality-ieq} there is sufficient and necessary for~$\phi$ to be optimal.  The last sentence
of Prop.~\ref{prop:overview:weak-duality} states that that is equivalent to inequality~\eqref{eq:overview:integral-nm-ieq}
of Remark~\ref{rem:overview:integral/convolution-ieqs} holding with equality.

Now, it is elementary (and left to the reader) to check that the condition in the statement of
Corollary~\ref{cor:overview:otheropt} is equivalent to the inequality~\eqref{eq:overview:integral-nm-ieq} of
Remark~\ref{rem:overview:integral/convolution-ieqs} holding with equality for~$\phi$.

\section{Truncation and folding}\label{sec:truncfold}%-----------------------------------------------------------------------
\subsection{Truncation}\label{ssec:truncfold:trunc}
We extract the technical computations as a lemma, for easier readability.

\begin{lemma}\label{ssec:truncfold:trunc:tvdist}
  With $K,N,\phi_K, \phi_K^{(N)}$ as in Prop.~\ref{prop:overview:trunc}, we have
  \begin{equation*}
    \Nmb{ \phi_K^{(N)} - \phi_K }
    =
    \frac{4K}{\pi} \cdot \sum_{n=N}^\infty  \frac{1}{(n+\nfrac12)^2}
    \le
    \frac{4K}{\pi (N-\nfrac12)}.
  \end{equation*}
\end{lemma}
\begin{proof}
  The difference between the two measures is
  \begin{equation*}
    2K\cdot \sum_{n=N}^\infty \frac{(-1)^{n+1}}{\pi\,(n+\nfrac12)^2} \, \delta_{(-\nfrac12-n)/2K}
    +
    2K\cdot \sum_{n=N}^\infty \frac{(-1)^{n+1}}{\pi\,(n+\nfrac12)^2} \, \delta_{(\nfrac12+n)/2K}.
  \end{equation*}
  Taking the norm, we obtain
  \begin{equation*}
    \frac{4K}{\pi} \cdot \sum_{n=N}^\infty  \frac{1}{(n+\nfrac12)^2}.
  \end{equation*}
  Now some high-school math: We lazily upper-bound the sum by the integral
  \begin{equation*}
    \int_{N-\nfrac12}^\infty  \frac{dx}{x^2} = \frac{1}{N-\nfrac12},
  \end{equation*}
  which completes the proof of the lemma.
\end{proof}

\begin{proof}[Proof of Prop.~\ref{prop:overview:trunc}]
  By Lemma~\ref{lem:overview:trunc:tv}, $\phi_K^{(N)}$ is $\Nm{ \phi_K^{(N)} - \phi_K }$-nearly feasible, and
  Lemma~\ref{ssec:truncfold:trunc:tvdist}, the norm is upper-bounded by~$\eps$.
\end{proof}

\subsection{Folding}\label{ssec:truncfold:fold}
\subsubsection{Folding with zero perturbation ($[A,B]=0$)}\label{ssec:truncfold:fold-B=0}
Here we prove Theorem~\ref{thm:overview:fold:B=0}, which states that, for all positive real numbers $K,\xi_1$ with
$L := K/\xi_1 \in \NN$, $\phi_K$ the Nyquist shift rule~\eqref{eq:intro:general-shift-rule} and~$\psi$ as in
Remark~\ref{rem:overview:trunc:small-fun}, shift-folding $\phi_K$ with mod-$\nfrac{1}{\xi_1}$ gives exactly~$\psi$.

The proof relies on the \textit{partial fraction decomposition of $1/\sin^2$:}
\begin{equation}\label{eq:pfd:1/sin^2}
  \text{For all $z\in\CC\setminus\ZZ$:}\quad
  \frac{\pi^2}{\sin^2(\pi z)} = \sum_{\ell\in\ZZ} \frac{1}{(z + \ell)^2};
\end{equation}
this is a textbook identity, see, e.g., \cite[Exercise~12 in Chapter~3]{Stein-Shakarchi:book-complex:2010}.

We note the following elementary fact about modulo-arithmetic for reference:
\begin{equation}\label{eq:modulo-arithmetic}\tag{$\%$}
  \text{For all $\gamma>0$, $x\in\RR$:}\quad x\modh \gamma = \gamma\cdot \bigl( (x/\gamma)\modh  1 \bigr).
\end{equation}

\begin{proof}[Proof of Theorem~\ref{thm:overview:fold:B=0}]
  Let~$K,\xi_1$, $L=K/\xi_1$, and~$\psi$ as in Remark~\ref{rem:overview:trunc:small-fun}, and let~$\phi_K$ as
  in~\eqref{eq:intro:general-shift-rule}.

  Applying Eqn.~\eqref{eq:modulo-arithmetic} twice, first with $\gamma:=1/\xi_1$ and second with $\gamma:=2L$, we find that for all
  $j\in\ZZ$,
  \begin{multline}\label{eq:RNDg8fehoio}\tag{$*$}
    (  (j+\nfrac12)/2K  )\modh (  \nfrac1{\xi_1}  )
    =
    (  ( (j+\nfrac12)/2L )\modh 1  )/\xi_1
    \\
    =
    (  (  j+\nfrac12 )\modh (2L)  )/2L\xi_1
    =
    (  (  j+\nfrac12 )\modh (2L)  )/2K.
  \end{multline}
  Now we calculate
  \begin{align*}
    (\place\modh (\nfrac1{\xi_1}))(\phi_K)
    &=
      2K \cdot \sum_{j\in\ZZ} \frac{ (-1)^{j+1} }{ \pi\, (j+\nfrac12)^2 } \; \delta_{((j+\nfrac12)/2K)\modh (\nfrac1{\xi_1})}
    \\
    &=
      2K \cdot \sum_{j\in\ZZ} \frac{ (-1)^{j+1} }{ \pi\, (j+\nfrac12)^2 } \; \delta_{(  (  j+\nfrac12 )\modh (2L)  )/2K}
    &&\cmmt{by~\eqref{eq:RNDg8fehoio}}
    \\
    &=
      2K \cdot
      \sum_{\sstack{a\in \nfrac12+\ZZ\\ -L+\nfrac12\le a \le L-\nfrac12}}
      \delta_{a/2K} \cdot
      \sum_\sstack{j\in\ZZ\\(j+\nfrac12)\modh (2L)=a} \frac{ (-1)^{j+1} }{ \pi\, (j+\nfrac12)^2 }.
  \end{align*}
  Comparing the last expression with~$\psi$, we find that we have to show, for all $a\in \nfrac12+\ZZ$ with
  $-L+\nfrac12\le a \le L-\nfrac12$, that the following holds:
  \begin{equation}\label{eq:RND97r2wh9d}\tag{$*\!*$}
    \sum_\sstack{j\in\ZZ\\(j+\nfrac12)\modh (2L)=a} \frac{ (-1)^{j+1} }{ \pi\, (j+\nfrac12)^2 }
    \musteq
    \frac{\pi}{(2L)^2} \cdot \frac{ (-1)^{a+\nfrac12} }{ \sin^2(\pi a/2L) }.
  \end{equation}
  Replacing the implicit condition under the sum by an explicit one (i.e., we define $\ell$ through
  $j+\nfrac12 = a+\ell\cdot2L$), we obtain, on the LHS of~\eqref{eq:RND97r2wh9d},
  \begin{equation*}
    \sum_\sstack{\ell\in\ZZ} \frac{ (-1)^{a+\nfrac12 + \ell\cdot2L} }{ \pi\, (a + \ell\cdot2L)^2 }
    =
    \sum_\sstack{\ell\in\ZZ} \frac{ (-1)^{a+\nfrac12} }{ \pi\, (a + \ell\cdot2L)^2 }.
  \end{equation*}
  Replacing this on the LHS of~\eqref{eq:RND97r2wh9d} and, isolating $\pi^2/\sin^2(\pi a/2L)$ on the RHS
  of~\eqref{eq:RND97r2wh9d}, on the LHS we are left with
  \begin{equation*}
    \sum_{\ell\in\ZZ} \frac{ (-1)^{2(a+\nfrac12)} }{ (a + \ell\cdot2L)^2 / (2L)^2 }
    =
    \sum_{\ell\in\ZZ} \frac{ 1 }{ (a/2L + \ell)^2 }.
  \end{equation*}
  We recognize the partial fraction decomposition of $\pi^2/\sin^2(\pi z)$, for $z := a/2L$, as in~\eqref{eq:pfd:1/sin^2} above,
  which completes the proof.
\end{proof}

\subsubsection{Proof of the fundamental folding lemma}\label{ssec:truncfold:fold--fun-lemma}

\begin{proof}[Proof of Lemma~\ref{lem:overview:fold:folding-types}]
  Writing out as integrals, we calculate for $f\in\oldFun_\Xi$, $x\in\RR$, for
  Item~\ref{lem:overview:fold:shift-folding},
  \begin{align*}
    f(x-\tau(s))
    &= f(x - \tau(s)\modh p)               &&\cmmt{as $f(x-\place)$ is $p$-periodic}
    \\
    &=
    f(x - s\modh p)                        &&\cmmt{as $\tau$ is a $p$-folding}
    \\
    &=
    f(x - s)                           &&\cmmt{as $f(x-\place)$ is $p$-periodic;}
  \end{align*}
  and for Item~\ref{lem:overview:fold:param-folding},
  \begin{align*}
    f(\tau(x-s))
    &= f(\tau(x - s)\modh p)               &&\cmmt{as $f$ is $p$-periodic}
    \\
    &=
    f((x - s)\modh p)                      &&\cmmt{as $\tau$ is a $p$-folding}
    \\
    &=
    f(x - s)                               &&\cmmt{as $f$ is $p$-periodic.}
  \end{align*}
  The statements about the derivative then follows if~$\phi$ is feasible for $\oldFun_\Xi$.
\end{proof}

\subsubsection{Proof of Prop.~\ref{prop:overview:fold:quadratic}}\label{ssec:truncfold:fold-quadratic}
Throughout this section we use the notation of Prop.~\ref{prop:overview:fold:quadratic}.
For easier readability, we abbreviate $\phi:=\phi_K$ for the Nyquist shift rule~\eqref{eq:intro:general-shift-rule}, and
$\tau:=\tau_{p,c}$.

\begin{proof}[Proof of Item~\ref{prop:overview:fold:quadratic:param-folding:apx} of Prop.~\ref{prop:overview:fold:quadratic}]
  Let $f,f_0,f_1$ and $(c,C)$ as stated.
  By linearity of~\eqref{prop:overview:fold:quadratic:param-folding:def} and the derivative, followed by the triangle inequality
  we find, for all $x\in\RR$,
  \begin{multline}\label{eq:RND2xinfrhc}\tag{$*$}
    \absb{  (f\circ\tau)*\phi(x)   - f'(x)  }
    \\
    \le
    \absb{  (f_0\circ\tau)*\phi(x)   - f_0'(x)  }
      +
    \absb{  (f_1\circ\tau)*\phi(x)   - f_1'(x)  }
    \\
    =
      \absb{  (f_0\circ\tau)*\phi(x)   - f_0'(x)  },
  \end{multline}
  where the last equation arises from the Fundamental
  Folding-Lemma~\ref{lem:overview:fold:folding-types}\ref{lem:overview:fold:param-folding}.

  Hence, we only have to bound the error of the linearly decaying part, $f_0$, and we use a refined version of the inequalities
  in Remark~\ref{rem:overview:integral/convolution-ieqs}, in a setting similar to the proof of
  Lemma~\ref{lem:overview:trunc:tv}: As $f_0\in\Fun_K$, by Corollary~\ref{cor:overview:general-shift-rule}, we can replace
  $f_0'$ by $f_0*\phi$, replacing and upper-bounding the last expression in~\eqref{eq:RND2xinfrhc} as follows
  \begin{equation*}\label{eq:RNDw489b7wc}\tag{$*\!*$}
    \absb{  (f_0\circ\tau - f_0 )*\phi(x)  }
    \le
    \int \abs{ f_0(\tau(x-s)) - f_0(x-s) } \,d\abs{\phi}(s).
  \end{equation*}
  We split the last integral into a sum $I_0 + I_- + I_+$ by distinguishing three non-overlapping cases on $x-s$
  as follows (using Lemma~\ref{lem:overview:fold:quadratic:def-tau}):
  \begin{center}
    \begin{tabular}{llll}
      Symbol  & $x-s$ interval         & $\tau(x-s)$ interval   & $s$ interval                \\
      \hline
      $I_+$   & $\lt]-\infty,-c-p\rt]$ & $\lt]-c-p, -c\rt]$     & $\lt[x+(c+p),+\infty\rt[$   \\
      $I_0$   & $\lt]-c-p, +c+p\rt[$   & $\lt]-c-p, +c+p\rt[$   & $\lt]x-c-p,x+c+p\rt[$       \\
      $I_-$   & $\lt[c+p,\infty\rt[$   & $\lt[c,c+p\rt[$        & $\lt]-\infty,x-(c+p)\rt]$.
    \end{tabular}
  \end{center}
  We go through them one by one, starting with the easiest: On the interval $\lt]-c-p,+c+p\rt[$, we have $f_0\circ\tau = f_0$,
  so that $I_0=0$.

  \myparsmall%
  \begin{description}
  \item[$I_-$,] i.e., assume $x-s \ge c+p$.  We start by bounding
    \begin{equation*}
      \int_{-\infty}^{x-(c+p)} \abs{f_0(\tau(x-s)) - f_0(x-s)} \,d\abs\phi(s)
      \le
      \Nm{f_0\circ\tau - f_0}_{\infty,\lt[c+p,\infty\rt[} \cdot \abs\phi(\lt]-\infty,x-(c+p)\rt])
    \end{equation*}
    As~$f_0$ is a $(c,C)$-linearly decaying function, we can bound
    \begin{equation*}
      \Nm{f_0\circ\tau - f_0}_{\infty,\lt[c+p,\infty\rt[}
      \le
      2C/c.
    \end{equation*}
    For the $\abs\phi$-part, we distinguish between $x\in[-p,+p]$ and arbitrary~$x$.
    \begin{itemize}
    \item For arbitrary~$x$, we upper-bound by $\abs\phi(\lt]-\infty,x-(c+p)\rt]) \le \abs\phi(\RR) = \Nm{\phi}=2\pi K$ by
      Theorem~\ref{thm:overview:opt}.
    \item For $x\in[-p,+p]$, we can do more.  Noting that $x-(c+p) \le -c$ as $x\le p$, we find
      by the definition of~$\phi$ in~\eqref{eq:intro:general-shift-rule},
      \begin{multline*}
        \abs\phi(\lt]-\infty,x-(c+p)\rt])
        \\
        \le
        \abs\phi(\lt]-\infty,-c\rt])
        =
        2K\, \sum_\sstack{ a \in \nfrac12+\ZZ \\ a/2K \le -c } \frac{1}{ \pi\, a^2 }
        \le
        \frac{2K}{\pi} \, \int_{2Kc-1}^\infty  \frac{dx}{x^2}
        =
        \frac{2K}{\pi \, (2Kc-1)},
      \end{multline*}
      where we have relied on $cK \ge 1$, Eqn.~\eqref{eq:overview:fold:quadratic:ck-ge-1}.
    \end{itemize}
  \item[$I_+$:] The case is symmetric, yielding the same bound.
  \end{description}

  Adding the three terms we find that the RHS in~\eqref{eq:RNDw489b7wc} is
  \begin{equation*}
    I_0 + I_- + I_+
    \le
    2\cdot \frac{2C}{c}\cdot
    \begin{cases}
      \frac{2K}{\pi\,(2Kc-1)}    , &\text{if $x\in[-p,+p]$;} \\
      2\pi K                     , &\text{otherwise.}
  \end{cases}
  \end{equation*}
  This concludes the proof of Prop.~\ref{prop:overview:fold:quadratic}\ref{prop:overview:fold:quadratic:param-folding:apx}.
\end{proof}

\begin{proof}[Proof of Item~\ref{prop:overview:fold:quadratic:param-folding:norm} of Prop.~\ref{prop:overview:fold:quadratic}]
  We have to find the supremum of the number $\Nm{ (f\circ\tau)*\phi }_\infty$, where~$f$ ranges over all real-valued measurable
  functions with $\Nm{f}_\infty\le 1$.

  We will apply Remark~\ref{rem:overview:integral/convolution-ieqs}, to the restriction~$\phi^{(c)}$ of $\phi$ to the image
  of~$\tau$, which, by Lemma~\ref{lem:overview:fold:quadratic:def-tau}, is $\lt] -c-p, +c+p \rt[$.

  \begin{multline*}
    \Nm{ (f\circ\tau)*\phi }_\infty
    =
    \Nm{ (f\circ\tau)*\phi^{(c)} }_\infty
    \\
    \le
    \Nm{ f\circ\tau }_\infty \cdot \Nm{\phi^{(c)}}
    \le
    \Nm{ f }_\infty \cdot \Nm{\phi^{(c)}}
    \le
    \Nm{\phi^{(c)}}
    =
    \Nm{\phi} - \Theta_K(1/c).
  \end{multline*}
  where we have used the notation $\Theta(\centerdot) = O(\centerdot) \cap \Omega(\centerdot)$.

  This concludes the proof of
  Prop.~\ref{prop:overview:fold:quadratic}\ref{prop:overview:fold:quadratic:param-folding:norm}.
\end{proof}

\begin{proof}[Proof of Item~\ref{prop:overview:fold:quadratic:param-folding:max-value} of Prop.~\ref{prop:overview:fold:quadratic}]
  This follows directly from the statement about the support of $\tau$ in Lemma~\ref{lem:overview:fold:quadratic:def-tau}.
\end{proof}

\begin{proof}[Proof of Item~\ref{prop:overview:fold:quadratic:param-folding:avg-value} of Prop.~\ref{prop:overview:fold:quadratic}]
  We have to give an upper bound to the quantity
  \begin{equation*}
    E(x) := \frac{1}{\Nm{\phi}} \, \int  \absb{ \tau(x-s) } \,d\abs\phi(s),
  \end{equation*}
  over $x\in[-p,+p]$.

  We split the value into three summands by distinguishing the non-overlapping cases
  \begin{align*}
    x-s&\le -c-p,        &         -c-p &< x-s < c+p,          & c+p &\le x-s,
  \end{align*}
  and upper-bound each separately.

  For the case $x-s \le -c-p$ we upper bound as follows (recall from Theorem~\ref{thm:overview:opt} that $\Nm{\phi}=2\pi K$):
  \begin{align*}
    \frac{1}{\Nm{\phi}}     \cdot \int_{x+(c+p)}^\infty \absb{ \tau(x-s) } \,d\abs{\phi}(s)
    &\le
      \frac{c+p}{\Nm{\phi}} \cdot \int_{x+(c+p)}^\infty \,d\abs{\phi}
    \\
    &\le
      \frac{c+p}{\Nm{\phi}} \cdot \int_{c}^\infty \,d\abs{\phi}                                  &&\cmmt{$x\in[-p,+p]$}
    \\
    &=
      \frac{c+p}{2\pi K} \cdot 2K \sum_\sstack{ a\in\nfrac12+\ZZ \\ a \ge 2Kc } \frac{1}{\pi\, a^2}
    \\
    &\le
      \frac{c+p}{2\pi K}   \cdot \frac{2K}{\pi\, (2Kc-1)}
    \\
    &=
      \frac{1+p/c}{2\pi^2 K - \pi^2/c)}
    \\
    &\le
      \frac{1+p/c}{(2\pi^2 K - \pi^2 K)}                      &&\cmmt{$cK\ge 1$, Eqn.~\eqref{eq:overview:fold:quadratic:ck-ge-1}}
    \\
    &=
      \frac{1+p/c}{\pi^2 K}
    \\
    &=
      \frac{2}{\pi^2 K}.                                      &&\cmmt{$p\le c$}
  \end{align*}
  The case $c+p \le x-s$ is symmetrical and gives the same bound.

  For the case $-c-p < x-s < +c+p$ we upper bound as follows:
  \begin{multline*}
    \frac{1}{\Nm{\phi}}   \cdot \int \One_{\{-c-p<x-\place<c+p\}}(s) \absb{ \tau(x-s) } \,d\abs{\phi}(s)
    \le
      \frac{1}{\Nm{\phi}} \cdot \int_{x-(c+p)}^{x+c+p}                \abs{ x-s }       \,d\abs{\phi}(s)
    \\
    \le
        \frac{1}{\Nm{\phi}} \cdot \int_{x-(c+p)}^{x+c+p} \abs{x} \,d\abs{\phi}(s)
      + \frac{1}{\Nm{\phi}} \cdot \int_{x-(c+p)}^{x+c+p} \abs{s} \,d\abs{\phi}(s).
  \end{multline*}
  The first integral we upper-bound by $\abs{x}\cdot \int d\abs{\phi}$, so the first summand is upper bounded by~$\abs{x}$.
  For the second summand, invoking Theorem~\ref{thm:overview:opt} again, we continue
  \begin{multline*}
    \frac{1}{\Nm{\phi}} \cdot \int_{x-(c+p)}^{x+c+p} \abs{s} \,d\abs{\phi}(s)
    =
    \frac{1}{2\pi K} \cdot 2K\, \sum_\sstack{ a\in\nfrac12+\ZZ \\ x-(c+p)  \le a/2K \le x+c+p } \frac{a/2K}{\pi\,a^2}
    \\
    =
    \frac{1}{2\pi^2 K} \cdot \sum_\sstack{ a\in\nfrac12+\ZZ \\ x-(c+p)  \le a/2K \le x+(c+p) } \frac{1}{a}
    \qquad\lecmt{(*)}\qquad
    \frac{1}{2\pi^2 K} \cdot \sum_\sstack{ a\in\nfrac12+\ZZ \\ -c-2p  \le a/2K \le c+2p } \frac{1}{a}
    \\
    \le
    \frac{1}{2\pi^2 K} \cdot 2\cdot \bigl( \ln(2K(c+2p)) + 2  \bigr)
    =
    \frac{ \ln(K(c+2p)) + 2+\ln2 }{\pi^2 K}.
  \end{multline*}
  In the inequality marked with the asterisk, we have used the assumption $x\in[-p,+p]$.

  Adding everything up, we get, as the upper bound on the expected parameter value
  \begin{multline*}
    2 \cdot \frac{2}{\pi^2 K}
    + \abs{x}
    + \frac{ \ln(K(c+2p)) + 2+\ln2 }{\pi^2 K}
    =
    \abs{x} + \frac{ \ln(K(c+2p)) + 6+\ln2 }{ \pi^2 K },
  \end{multline*}
  as claimed.
\end{proof}

This concludes the proof of Prop.~\ref{prop:overview:fold:quadratic}.

\section{Sketches of proofs of non-existence results}\label{sec:Impossible}
\subsection{Impossibility of exponential concentration}\label{ssec:Impossible:expo-concentr}
\subsubsection{Background on exponential concentration}
We review some technicalities about exponential concentration.  The proof of the following lemma is purely technical and given
in Appendix~\ref{apx:math:expo-concentr} for the sake of completeness.

\begin{lemma}\label{lem:Impossible:expo-concentr:equiv}
  For complex measures~$\mu$ the following are equivalent:
  \begin{enumerate}[label=(\alph*)]
  \item $\mu$ is exponentially concentrated;
  \item There exists an $r>0$ with $\displaystyle \int e^{r\abs{x}} \,d\abs{\mu}(x) < \infty$.
  \end{enumerate}
\end{lemma}

The following is easily derived from Lemma~\ref{lem:Impossible:expo-concentr:equiv} by tracing the proof of Theorem~3 in
\cite{Stein-Shakarchi:book-complex:2010}.

\begin{lemma}\label{lem:overview:Impossible:expo-concentr=>holomo}
  Let $\mu$ be a complex measure on the real line.
  If $\mu$ is exponentially concentrated then $\hat\mu$ can be extended to a holomorphic function defined in a neighborhood of
  the real line.
\end{lemma}

\subsubsection{Proof of Theorem~\ref{thm:overview:Impossible:non-exist-expo-concentr}}
Theorem~\ref{thm:overview:Impossible:non-exist-expo-concentr} is proven with the same complex analysis argument as the result
for compact support in \S\ref{ssec:overview:Impossible:cpct}, but replacing Paley-Wiener with
Lemma~\ref{lem:overview:Impossible:expo-concentr=>holomo}.

Hence, the only challenging part of the proof is that we do not assume that~$\Xi$ contains an accumulation point, i.e., an
element $\xi\in\Xi$ that is in the closure of~$\Xi\setminus\{\xi\}$.

But with~$\Xi$ the set $\bar\Xi$ also satisfies the condition of the theorem.  This can be seen in two ways.
\textcal{(1)}~By first extending the proof of Lemma~\ref{lem:overview:linsyseq} to the case where the condition
``$\xi\in[-\nfrac12,+\nfrac12]$'' of equation~\eqref{eq:overview:linsyseq} is replaced with ``$\xi\in\Xi$'', and then noting
that both sides of equation~\eqref{eq:overview:linsyseq} are continuous functions of~$\xi$, so the equality extends to limit
points.
\textcal{(2)}~Di\-rect\-ly from~\eqref{eq:overview:deriv-half--measure}, with firstly $g := \cos(2\pi\xi\place)$, $\xi\in\Xi$,
and secondly $g := \sin(\pi\xi)$, $\xi\in\Xi$, by again noting that in all cases both sides of the equation are continuous
functions of~$\xi$.

\subsection{Shift rules with approximation error}\label{ssec:Impossible:apx-error}

\begin{proof}[Sketch of proof of Corollary~\ref{cor:overview:Impossible:non-exist-expo-concentr}.]
  For a proof by contradiction, assume that $C, r,\phi_*$ exist.

  We invoke Alaoglu's theorem: ``The dual unit ball is weak$^*$ compact.''  The space of complex measures is the dual of
  $\mathscr{C}_0$, the space of (complex valued) continuous functions vanishing in $\pm\infty$, and the total-variation norm is
  the operator norm on the dual (e.g., \cite[Theorem~6.19]{Rudin:real-complex-analysis:1987}).

  As $\{\phi_j\mid j\in\NN\}$ is norm bounded, by passing to a sub-sequence of $(\phi_j)_j$, there exists a complex
  measure~$\phi$ such that
  \begin{equation*}
    % \int h \,d\abs{\phi} &= \lim_{j\to\infty} \int h\,d\abs{\phi_j}       &&\text{for all $h\in\mathscr{C}_0$;}\\
    % \intertext{and hence also}
    \int h \,d\phi       = \lim_{j\to\infty} \int h\,d\phi_j             \qquad\text{for all $h\in\mathscr{C}_0$.}
  \end{equation*}

  The proof of the corollary is completed by showing that (a)~$\phi$ is exponentially concentrated, and (b)~$\phi$ is feasible
  for~$\GenericFun$.  The combination of (a) and~(b) contradicts Theorem~\ref{thm:overview:Impossible:non-exist-expo-concentr}.

  The arguments for (a,b) are entertaining exercises in calculus and measure theory, and we leave them to the bored reader.
\end{proof}

\section{Discussion and outlook}\label{sec:discussion}%----------------------------------------------------------------------
We have given a ``proper'' shift rule for analytic derivatives of ``perturbed-parametric'' quantum evolutions.  The support of
the measure is non-compact, and the tail, i.e., the fraction of the total variation that falls outside a given interval
$[-T,+T]$, $T\to\infty$, decays only linearly, as in a Cauchy-type probability distribution.

The question arises whether proper shift rules for analytic derivatives of perturbed-parametric quantum evolutions exist that
have lighter tails.  Given that we have also shown that the space of proper shift rules that give analytic derivatives has
infinite dimension, at first sight, that doesn't appear impossible.

However, we have also shown that proper shift rules (for perturbed-parametric quantum evolutions) with exponentially decaying
tails do not exist.  Indeed, the negative result holds for norm-bounded (i.e., worst-case variance bounded) families of proper
shift rules that approximate the derivative with an approximation error that tends to~$0$.

On the positive side, the best result we have is a $O(1/T^2)$ approximation error, for parameter values that are restricted to
$[-T,+T]$ (this, by nature, cannot be ``proper'' shift rule).

In future research, it might be interesting to either expand the non-existence results, or to give shift rules with faster
decaying tails.

\section*{Acknowledgments}%--------------------------------------------------------------------------------------------------
The author would like to thank the anonymous referees for their careful reading of the manuscript and valuable suggestions for
improving it.  The author also thanks L.~Savanna for discussions about the draft paper.  This research was partly funded by the
Estonian Center of Excellence in Computer Science (EXCITE), and by Estonian Research Agency (Eesti Teadusagentuur, ETIS) through
grant PRG946.

\appendix%===================================================================================================================
\section*{APPENDIX}
%\section{The $\hbar$ dictionary}\label{apx:hbar-dict}
\section{Miscellaneous math}\label{apx:math}
\subsection{Sums of reciprocals of squares of odd integers}\label{apx:math:sum-square-odd}
The following is a direct consequence of the \emph{Basel Problem}, $\sum_{n=1}^\infty \frac{1}{n^2} = \pi^2/6$.

\begin{lemma}\label{lem:math:sum-square-odd}
  $\displaystyle \sum_{j=0}^\infty \frac{1}{ (2j+1)^2 }   = \pi^2/8$
\end{lemma}
\begin{proof}
  In the following equations, the first and last equations are the Basel Problem:
  \begin{equation*}
    \pi^2/6
    =
    \sum_{n=1}^\infty \frac{1}{n^2}
    =
    \sum_{n=1}^\infty \frac{1}{(2n+1)^2}
    +
    \sum_{n=1}^\infty \frac{1}{(2n)^2}
    =
    \sum_{n=1}^\infty \frac{1}{(2n+1)^2}
    +
    \frac{1}{4} \cdot \pi^2/6.
  \end{equation*}
  Collecting the non-sum terms on the LHS gives the statement of the lemma.
\end{proof}

%\section{Supplementary source code}\label{apx:source-code}
\subsection{Simultaneous Diophantine approximation}\label{apx:math:diophant}
Recall Dirichlet's simultaneous diophantine approximation theorem.

\begin{theorem}[SDA]\label{thm:SDA}
  For all positive integers $d,m$ and real numbers $\xi_1,\dots,\xi_d$, there exists a positive integer $q\ge m$ and integers
  $k_1,\dots,k_d$ such that
  \begin{equation*}
    \Abs{ \xi_j - \frac{k_j}{q} }   \le \frac{1}{ q \, m^{1/d} }
  \end{equation*}
\end{theorem}

We use it to prove the following fact.

\begin{lemma}\label{lem:apx:math:diophant:near-periodicity}
  Let
  \begin{equation*}
    f\colon x\mapsto \sum_{j=1}^d e^{2\pii\xi_j x} \, b_j
  \end{equation*}
  be a bounded real-valued, complex-valued, or operator-valued function with finite Fourier spectrum, i.e., $\xi\in\RR^d$ and
  $b_1,\dots,b_d$ are real or complex numbers, or operators, respectively.

  If $\lim_{\abs{x}\to\infty} f(x) = 0$, then $f=0$.
\end{lemma}
\begin{proof}
  The lemma is trivial if the frequencies $\xi_1,\dots\xi_d$ have a common divisor, as then the function~$f$ is periodic.

  In the general case, let~$f$ be as described, and set
  \begin{equation*}
    c := \Nm{ f }_{\infty,[-1,0]} := \sup_{x\in[-1,0]} \Nm{f(x)},
  \end{equation*}
  where $\Nm{\centerdot}$ stands for any norm.

  Assume $f\ne 0$; this implies $c > 0$ (as~$f$ can be extended to a holomorphic function on the complex plane).
  We prove that for every $T>0$ we have $\Nm{ f }_{\infty,\lt[T,\infty\rt[} \ge c/2$.  This implies that
  $\lim_{\abs{x}\to\infty} f(x) =0$ does not hold.

  For given $T>0$, let~$m\ge T$ be a integer with $m^{\nfrac1d} \ge 8\pi\Nm{b}_1 / c$ (where
  $\Nm{b}_1 := \sum_{j=1}^d \Nm{b_j}$).

  Now apply Theorem~\ref{thm:SDA} to the numbers $\xi_1,\dots,\xi_d$, and $m,d$ as just defined, and let $q$ be the denominator
  and $k_1,\dots,k_d$ the numerators in the Diophantine approximations.  We find that, for all $j=1,\dots,d$,
  \begin{equation}\label{eq:RNDyxsgfeob}
    2\pi \cdot \Abs{  \xi_j - \tfrac{k_j}{q}  } \cdot q \le \frac{2\pi}{m^{\nfrac1d}} \le \frac{c}{4\Nm{b}_1},
  \end{equation}
  by our choice of~$m$.  It follows that, for all $x\in[-1,0]$, we have
  \begin{align}
    \Nm{ f(x) - f(x+q) }
    &\le
      \sum_{j=1}^d \Abs{  e^{2\pii\xi_j x} - e^{2\pii\xi_j (x+q)}  }\cdot \Nm{b_j}           \notag
    \\
    &\le
      \max_{j=1,\dots,d} \Abs{  e^{2\pii\xi_j x} - e^{2\pii\xi_j (x+q)}  } \cdot \Nm{b}_1    \tag{$*$}\label{eq:RNDbg8fvw67}
  \end{align}
  For $j=1,\dots,d$, we continue
  \begin{multline}\tag{$*\!*$}\label{eq:RND89bc427h}
    \Abs{  e^{2\pii\xi_j x} - e^{2\pii\xi_j (x+q)}  }
    \\
    \le
    \Abs{  e^{2\pii\xi_j x} - e^{2\pii \frac{k_j}{q} x} } + \Abs{ e^{2\pii\xi_j (x+q)}  - e^{2\pii \frac{k_j}{q} (x+q)}}
    + \Abs{  e^{2\pii \frac{k_j}{q} x} - e^{2\pii \frac{k_j}{q} (x+q)}  }
  \end{multline}
  The third term on the RHS in~\eqref{eq:RND89bc427h} is~$0$.

  We treat the first two terms in~\eqref{eq:RND89bc427h} simultaneously by setting $s := x$ or $s:=x+q$, respectively, noting
  $\abs{s}\le q$.  We find, using~\eqref{eq:RNDyxsgfeob},
  \begin{equation*}
    \absB{  e^{2\pii\xi_j s} - e^{2\pii \frac{k_j}{q} s} }
    \le
    2\pi\cdot \Abs{ \xi_j - \tfrac{k_j}{q} }\cdot q
    \le
    c/4\Nm{b}_1.
  \end{equation*}
  Hence, we can upper-bound the RHS in~\eqref{eq:RND89bc427h} by $c/2\Nm{b}_1$.
  We continue~\eqref{eq:RNDbg8fvw67}:
  \begin{equation*}
    \Nm{ f(x) - f(x+q) }
    \le
    c/2\Nm{b}_1 \cdot \Nm{b}_1
    \le c/2,
  \end{equation*}
  and hence, for some $x\in[-1,0]$ we have $\Nm{f(x+q)} \ge c/2$.
\end{proof}

\begin{proposition}\label{prop:apx:math:diophant:direct-sum}
  The following sum of vector spaces is direct: \textcal{(1)} the vector space of all bounded (real, complex or operator-valued)
  functions with finite Fourier spectrum (each function can have its own Fourier spectrum, but it must be finite) and
  \textcal{(2)} the vector space of (real, complex or operator-valued, resp.) functions that vanish in~$\pm\infty$.

  In other words, if $f_1, f_0, g_1, g_0$ are (real, complex or operator-valued) functions which satisfy
  \begin{itemize}
  \item $f_1 + f_0 = g_1 + g_0$;
  \item $\lim_{\abs{x}\to\infty} f_0(x)=0$, $\lim_{\abs{x}\to\infty} g_0(x)=0$;
  \item Each of $f_1,g_1$ has a finite Fourier spectrum, both are bounded;
  \end{itemize}
  then $f_1=g_1$ and $f_0=g_0$ hold.
\end{proposition}
\begin{proof}
  The first statement a direct consequence of Lemma~\ref{lem:apx:math:diophant:near-periodicity}; the second statement follows
  by grouping terms.
\end{proof}

\subsubsection{Bounded functions with finite Fourier spectrum}\label{apx:math:diophant:finite-Fou-spec}
We have used the following well-known fact all the time.

\begin{proposition}
  Let~$f$ be a measurable (complex-valued) function.
  If~$f$ is bounded and has finite Fourier spectrum, then~$f$ is a (finite) linear combination of Fourier characters
  $e^{2\pii\xi\place}$, $\xi\in\RR$.
\end{proposition}

The tempered-distribution fact behind it is the following: If a tempered distribution has finite support~$\Xi\subset\RR$, then
$\tau$ is a linear combination of derivatives\footnote{%
  Defined traditionally as the $\alpha$th derivative of the Dirac distribution:
  $\dual{ \varphi }{ \partial^{\alpha}_a } = (-1)^\alpha \dual{ \partial^\alpha \varphi }{ \delta_a }$ for all Schwartz
  functions~$\varphi$.  } %
$\partial^{\alpha_\xi}_\xi$, $\xi\in\Xi$, with $\alpha \in (\ZZ_+)^\Xi$.  For $\tau := \hat f$, that makes $f$ have the form
\begin{equation*}
  f(x) = \sum_{\xi\in\Xi} x^{\alpha_\xi}\cdot e^{2\pii\xi x} \cdot b_\xi.
\end{equation*}
Elementary considerations about polynomials plus an argument similar to the one in the proof of
Lemma~\ref{lem:apx:math:diophant:near-periodicity} show that such a function is bounded only if $\alpha_\xi = 0$ for all
$\xi\in\Xi$.

\subsection{Exponential concentration}\label{apx:math:expo-concentr}
\begin{proof}[Proof of Lemma~\ref{lem:Impossible:expo-concentr:equiv}]
  Let $\mu := \abs{\nu}$ (or, simply, any finite positive measure).

  Firstly, suppose we have $C,r >0$ with $\nu(\RR\setminus[-T,+T]) \le C \cdot e^{-rT}$ for all $T\ge0$.  Recall from standard
  measure theory (e.g., \cite[Satz~23.8]{bauer:mass-u-int:1992}) that for all measurable functions $f\ge 0$ we have
  \begin{equation*}
    \int f \,d\nu = \int_0^\infty \nu(\{f\ge y\}) \,dy.
  \end{equation*}
  From this, with $f:=e^{r\abs{\place}}$, we find, with $r' := r/2$:
  \begin{multline*}
    \int e^{r'\abs{x}} \,d\nu
    =
    \int_0^\infty \nu(\{e^{r'\abs{\place}} \ge y\}) \,dy
    \\
    \le
    \Nm{\nu} + \int_1^\infty \nu(\{\abs{\place} \ge \ln(y)/r'\}) \,dy
    \le
    \Nm{\nu} + \int_1^\infty C y^{-2} \,dy
     < \infty.
  \end{multline*}

  Secondly, suppose we have $r>0$ such that $C := \int e^{r\abs{\place}} \,d\nu < \infty$.  We calculate
  \begin{multline*}
    \nu( \RR\setminus[-T,+T] )
    =
    \int \One_{\RR\setminus[-T,+T]} \,d\nu
    =
    e^{-rT} \int e^{rT} \One_{\RR\setminus[-T,+T]} \,d\nu
    \le
    e^{-rT} \int e^{r\abs{\place}}  \,d\nu
    \le
    C \cdot e^{-rT}.
  \end{multline*}
  This completes the proof of the lemma.
\end{proof}

\section{Finite-dimensional perturbation theory (for the theoretical computer scientist)}\label{apx:perturb-th}
We use the notations laid out in \S\ref{ssec:fourier:proof-decomp}; in addition, in this section, $e$ is just another letter
(and not Euler's number $\exp(1)$).

The mathematical cornerstone of finite-dimensional perturbation theory is \emph{Rellich's
  Theorem}~\cite[Theorem~1]{Rellich:perturb-book:1969}.  We give here a much simplified version adapted to our needs.

We employ the terminology of holomorphic functions on a complex disc: For $r>0$, we denote by $\DD_r$ the open disc of
radius~$r$ around~$0$ in~$\CC$.

\begin{theorem}[Rellich]\label{thm:rellich}
  Let $A,B$ be Hermitian operators on a Hilbert space~$\cH$ of dimension $d \in \NN$, and consider $z\mapsto C(z) := A + zB$;
  where $z\in\CC$.  Denote the (distinct) eigenvalues of $A$ by $\lambda^{(0)}_1,\dots,\lambda^{(0)}_e$, and by~$h_\ell$ the
  multiplicity of eigenvalue $\lambda^{(0)}_\ell$ of~$A$, $\ell=1,\dots,e$.

  There exists a radius $r>0$ such that the following holds.
  \begin{itemize}
  \item%
    For each $\ell=1,\dots,e$, $j=1,\dots,h_\ell$, there exists a holomorphic function
    $\lambda_{(\ell,j)}\colon \DD_{r} \to \CC$, such that
    \begin{enumerate}[label=(\alph*)]
    \item%
      For each $z\in \DD_{r}$, the eigenvalues of $C(z)$, listed with repetitions according to algebraic multiplicities, are
      $\lambda_{\ell,j}(z)$, $\ell=1,\dots,e$, $j=1,\dots,h_\ell$.
    \item For each $\ell,j$, we have $\lambda_{\ell,j}(0) = \lambda^{(0)}_\ell$.
    \end{enumerate}
  \item%
    For each $\ell=1,\dots,e$, $j=1,\dots,h_\ell$, there exists a holomorphic (vector-valued) function
    $w_{\ell,j}\colon \DD_{r} \to \cH$, such that
    \begin{enumerate}[resume*]
    \item For each $z\in \DD_{r}$, for each $\ell,j$, the vector $w_{\ell,j}(z)$ is eigenvector for the eigenvalue
      $\lambda_{\ell,j}(z)$ of $C(z)$, i.e., $C(z)w_{\ell,j}(z) = \lambda_{\ell,j}(z) w_{\ell,j}(z)$ (and $w_{\ell,j}(z)\ne 0$);
    \item\label{thm:rellich:ONB}%
      For each~$z\in \DD_{r}\cap\RR$, the vectors $w_{\ell,j}(z)$, where $\ell=1,\dots,e$, $j=1,\dots,h_\ell$, form an orthonormal
      basis of~$\cH$.
    \end{enumerate}
  \end{itemize}
\end{theorem}

Based on Theorem~\ref{thm:rellich} the usual undergrad quantum physics perturbation theory can be executed mathematically
rigorously in finite dimension; the result is shown in the following Corollary~\ref{cor:perturbation-theory} --- in matrix
notation (i.e., for the theoretical computer scientist).

\begin{corollary}\label{cor:perturbation-theory}
  Let $A,B$ be Hermitian operators on a Hilbert space~$\cH$ of finite dimension~$d\in\NN$.
  There exists an $r>0$ and an orthonormal basis of eigenvectors of~$A$, along with $d$-by-$d$ matrices $\tA,\tB$,
  $\tWcoeff{k}$, $k\in\ZZ_+$, and diagonal $d$-by-$d$ matrices $\tLambdacoeff{k}$, $k\in\ZZ_+$, such that the following holds.
  \begin{enumerate}[label=(\alph*)]
  \item%
    $\tA$ ($\tB$, resp.) is the matrix of the operator~$A$ ($B$, resp.) wrt the ONB;
  \item\label{cor:perturbation-theory:W-conv}%
    The matrix-valued power series $\tW(z) = \sum_{k=0}^\infty z^k \tWcoeff{k}$ has convergence ratio at least~$r$;
  \item\label{cor:perturbation-theory:W-unitary}%
    $\tW(z)$ is unitary for all $z\in\lt]-r,+r\rt[$;
  \item\label{cor:perturbation-theory:Lambda-conv}%
    The diagonal-matrix-valued power series $\tLambda(z) = \sum_{k=0}^\infty z^k \tLambdacoeff{k}$ has convergence ratio at
    least~$r$;
  \item\label{cor:perturbation-theory:diagonalization}%
    $\displaystyle \tA + z\tB = \tW(z)\tLambda(z)\tW(z)^\dag$ holds for all $z\in\lt]-r,+r\rt[$;
  \item\label{cor:perturbation-theory:W0}%
    $\displaystyle \tWcoeff{0} = \tW(0) = \tI$ (identity matrix);
  \item\label{cor:perturbation-theory:Lambda0}%
    $\displaystyle \tLambdacoeff{0} = \tLambda(0) = \tA$;
  \item\label{cor:perturbation-theory:Lambda1}%
    $\displaystyle \tLambdacoeff{1} = \Diag{\tB}$;
  \end{enumerate}
\end{corollary}

The proof of Corollary~\ref{cor:perturbation-theory} is well known; we sketch it.
The matrix-valued function $\tW$ is defined, for $\ell',\ell=1,\dots,e$, $j'=1,\dots,h_{\ell'}$, $j=1,\dots,h_\ell$, and all
small enough~$\abs{z}$, by
\begin{equation*}
  \tW_{ (\ell',j'), (\ell,j) }(z) := \cip{ w_{\ell,j}(0) }{ w_{\ell',j'}(z) };
\end{equation*}
where $\cip{.}{.}$ is the inner product in the Hilbert space~$\cH$, and the $w_*(\centerdot)$ are from Rellich's
Theorem~\ref{thm:rellich}.  In other words, the columns of~$\tW(z)$ are the orthonormal eigenstates at~$z$, wrt the basis
at~$0$.  The diagonal-matrix valued fuction~$\tLambda$ has on its diagonal the eigenvalue-valued functions $\lambda$ from
Rellich's theorem, in the corresponding ordering.  Now, just expand the power series in~$z$ on both sides of the equation
\begin{equation*}
(\tA + z\tB)\tW(z) = \tW(z)\tLambda(z)
\end{equation*}
and collect terms.  Hence, there is no need to repeat the proof here.

\section{Fourier-analytic characterization of derivative-com\-put\-ing measures, ge\-ne\-ral case}\label{apx:Half-deriv:linsyseq-lemma}
In \S\ref{ssec:overview:𝛍-𝛟:charac}, we have proven one direction of Lemma~\ref{lem:overview:linsyseq}, we now prove the
remaining one in full generality: That~\eqref{eq:overview:linsyseq} implies~\eqref{eq:overview:deriv-half--measure}.

Apart from standard calculations with Fourier(-Stieltjes) transforms and tempered distributions (most of which we are writing
out in this section), the proof is based on the technique in the Paley-Wiener-Schwartz theorem that is contained in the
following fact.

\begin{lemma}[Paley-Wiener-Schwartz \cite{Rudin:FA:1991,Reed-Simon:MathPhys-2-Fourier:1975}]\label{lem:Paley-Wiener-Schwartz}
  Let~$\tau$ be a tempered distribution with compact support, and~$\psi$ a Schwartz function with compact support such that
  $\psi=1$ on $\Supp\tau$.  The tempered distribution $\check\tau$ (inverse Fourier transform) is given by integrating against
  the following function\footnote{Note that $\psi \cdot e^{2\pii\place x}$ is a Schwartz function with compact support, for
    every~$x\in\RR$.}
  \begin{equation*}
    x \mapsto \dual{  \psi \cdot e^{2\pii\place x}  }{ \tau   },
  \end{equation*}
  which has at most polynomial growth and is analytic.
\end{lemma}
In this section we arbitrarily fix a Schwartz function~$\psi$ with compact support satisfying
\begin{equation*}
  \begin{split}
    0 \le \psi \le 1 &\quad\text{everywhere, }                          \\
    \psi =  1        &\quad\text{on $[-\nfrac12,+\nfrac12]$, and }      \\
    \psi =  0        &\quad\text{outside of $[-1,+1]$.}
  \end{split}
\end{equation*}

Let us pull out some lemmas for easier readability.

\begin{lemma}\label{lem:apx:Half-deriv:linsyseq-lemma:FIT}
  Let~$f$ be a smooth function of at most polynomial growth which has compact Fourier spectrum.  For all $x\in\RR$ we have
  \begin{equation*}
    f(x) = \dual{  \psi \cdot e^{2\pii\place x}  }{ \hat f   }.
  \end{equation*}
\end{lemma}
\begin{proof}
  From Lemma~\ref{lem:Paley-Wiener-Schwartz} and the Fourier Inversion Theorem for tempered distributions, we know that the RHS
  defines a function of~$x$, integration against which defines a tempered distribution that is equal to the tempered
  distribution defined by integrating against~$f$.  As both~$f$ and (again by Lemma~\ref{lem:Paley-Wiener-Schwartz}) the
  function defined by the RHS are continuous, pointwise equality follows.
\end{proof}

\begin{lemma}\label{lem:apx:Half-deriv:linsyseq-lemma:deriv-and-moment}
  If $f$ is a smooth function of at most polynomial growth, such that $\hat f$ has compact support, then~$f'$ has at most
  polynomial growth.  Moreover,
  \begin{enumerate}[label=(\alph*)]
  \item\label{lem:apx:Half-deriv:linsyseq-lemma:deriv-and-moment:deriv}%
    $\displaystyle%
    f'(x)           = \dual{  \psi \cdot e^{2\pii\place x}  }{  \widehat{f'}  }
    $ %
    \quad for all $x\in\RR$, and
  \item\label{lem:apx:Half-deriv:linsyseq-lemma:deriv-and-moment:moment}%
    $\displaystyle%
    \widehat{f'} = (2\pii\place) \, \hat f.
    $ %
  \end{enumerate}
\end{lemma}
\begin{proof}
  Let~$f$ as stated, and~$\sigma$ be the tempered distribution ``integrating against~$f$''.  The tempered distribution
  $\partial\hat\sigma$ has support contained in $\Supp\hat\sigma$.  Indeed, if $\varphi$ is a Schwartz function with support in
  the complement of $\Supp\hat\sigma$, when so is $\partial\varphi$, so that
  \begin{equation*}
    \dual{ \varphi }{ \partial\hat\sigma }
    =
    - \dual{ \partial\varphi }{ \hat\sigma }
    = 0.
  \end{equation*}
  So Lemma~\ref{lem:Paley-Wiener-Schwartz} is applicable: There exists a continuous function~$g$ of at most polynomial growth
  such that $\partial\sigma$ is given by integrating against~$g$.  But for all compactly supported Schwartz functions~$\varphi$
  we have
  \begin{equation*}
    \int\varphi f'
    =
    -\int \partial\varphi f
    =
    -\dual{\partial\varphi}{\sigma}
    =
    \dual{\varphi}{\partial\sigma}.
    =
    \int\varphi g,
  \end{equation*}
  so that $f'=g$ holds pointwise.  This proves that~$f'$ is of at most polynomial growth, and we can apply
  Lemma~\ref{lem:apx:Half-deriv:linsyseq-lemma:FIT} to $f'$, which gives
  us~\eqref{lem:apx:Half-deriv:linsyseq-lemma:deriv-and-moment:deriv}.

  With~\eqref{lem:apx:Half-deriv:linsyseq-lemma:deriv-and-moment:deriv}, the
  equation~\eqref{lem:apx:Half-deriv:linsyseq-lemma:deriv-and-moment:moment} is purely a relation between tempered distributions and
  follows from $(\partial\sigma)^\Hat = (2\pii\place)\, \hat\sigma$.
\end{proof}

\begin{lemma}\label{lem:apx:Half-deriv:linsyseq-lemma:exchange}
  Let~$\mu$ be a finite measure.  For every Schwartz-function~$\varphi$, we have

  \begin{enumerate}[label=(\alph*)]
  \item\label{lem:apx:Half-deriv:linsyseq-lemma:exchange:Bd}%
    For all $x\in\RR$,
    \begin{equation*}
      \absb{   \dual{  \varphi  }{  \psi e^{-2\pii\place x}  }   }
      \le
      \Nm{\varphi}_\infty\cdot \Nm{\psi}_1,
    \end{equation*}
    and hence $x\mapsto  \dual{  \varphi  }{  \psi e^{-2\pii\place x}  }$ is $\mu$-integrable;
  \item\label{lem:apx:Half-deriv:linsyseq-lemma:exchange:X}%
    \(\displaystyle
    \dual{  \varphi  }{  \psi\hat\mu  }
    =
    \int \dual{  \varphi  }{  \psi e^{-2\pii\place x}  } \; d\mu(x).
    \)
  \end{enumerate}
\end{lemma}
\begin{proof}
  For~\ref{lem:apx:Half-deriv:linsyseq-lemma:exchange:Bd}, we simply calculate
  \begin{align*}
    \absb{   \dual{  \varphi  }{  \psi e^{-2\pii\place x}  }   }
    &=
      \Abs{   \int  \varphi(\xi)\psi(\xi) e^{-2\pii\xi x} \,d\xi   }
    \\
    &\le
      \int  \abs{\varphi(\xi)}\,\abs{\psi(\xi)} \, \abs{ e^{-2\pii\xi x} } \,d\xi
    \\
    &\le
      \Nm{\varphi}_\infty \int  \abs{\psi(\xi)} \,d\xi,
  \end{align*}
  which gives the bound.  For integrability, as~$\mu$ is finite, in addition to the boundedness of the integrand that we have
  just proven, we only require measurability, which follows from the usual facts about parameterized integrals (e.g.,
  \cite{Rudin:real-complex-analysis:1987}).

  For~\ref{lem:apx:Half-deriv:linsyseq-lemma:exchange:X}, we simply calculate
  \begin{align*}
    \int \dual{  \varphi  }{  \psi e^{-2\pii\place x}  } \; d\mu(x)
    &=
      \int\int  \varphi(\xi) \,  \psi(\xi)\, e^{-2\pii\xi x}  \;d\xi \, d\mu(x)
    \\
    &=
      \int\int  \varphi(\xi) \,  \psi(\xi)\, e^{-2\pii\xi x}  \;d\mu(x) \, d\xi &&\cmmt{Fubini with \ref{lem:apx:Half-deriv:linsyseq-lemma:exchange:Bd}}
    \\
    &=\int  \varphi(\xi) \; \psi(\xi) \cdot \int e^{-2\pii\xi x}  \,d\mu(x) \; d\xi
    \\
    &=
    \dual{  \varphi  }{  \psi\hat\mu  },
  \end{align*}
  and the proof of the lemma is completed.
\end{proof}

Now we address the task of this subsection.

\begin{proof}[Proof of sufficiency in Lemma~\ref{lem:overview:linsyseq}]
  Let $\mu$ be a finite measure, and $f\in\Fun_\nfrac12$.  We have to prove that if~$\mu$ satisfies the
  condition~\eqref{eq:overview:linsyseq} in Lemma~\ref{lem:overview:linsyseq}, then $f,\mu$ satisfy the equation
  in~\eqref{eq:overview:deriv-half--measure}: $f'(\nfrac12) = \int f \,d\mu$.

  Define a sequence of Schwartz functions $f_j := \psi(\place/2j)\cdot f$, $j\in\NN$, and note that, for all Schwartz
  functions~$\varphi$, we have
  \begin{equation*}
    \absb{ \dual{ \varphi }{ f - f_j } }
    \le
    \int_{\RR\setminus[-j,+j]} \abs{f} \cdot \abs{\varphi} \xrightarrow{j\to\infty} 0,
  \end{equation*}
  i.e., $f_j \xrightarrow{j\to\infty} f$ as tempered distributions (i.e., in the $\sigma(\Schwartz',\Schwartz)$ locally-convex
  topology), and hence, by continuity of the Fourier transform, $\widehat{f_j} \xrightarrow{j\to\infty} \hat f$ as tempered
  distributions.

  By Lemma~\ref{lem:apx:Half-deriv:linsyseq-lemma:exchange}\ref{lem:apx:Half-deriv:linsyseq-lemma:exchange:X}, we have, for all~$j$,
  \begin{equation}\label{eq:RND2icfng7o}\tag{$*$}
    \dual{  \widehat{f_j}  }{  \psi\hat\mu  }
    =
    \int \dual{  \widehat{f_j}  }{  \psi e^{-2\pii\place x}  } \; d\mu(x).
  \end{equation}
  We take the limit $j\to\infty$ on both sides of~\eqref{eq:RND2icfng7o}.

  The LHS of~\eqref{eq:RND2icfng7o} goes to $\dual{ \psi\hat\mu }{ \hat f }$ as $j\to\infty$, because
  $f_j\xrightarrow{j\to\infty} f$ as tempered distributions, and
  $\dual{ \widehat{f_j} }{ \psi\hat\mu } = \dual{ \psi\hat\mu }{ \widehat{f_j} }$ (both sides are Schwartz functions).

  Taking the limit of the RHS of~\eqref{eq:RND2icfng7o}, we first note that the expression under the $\mu$-integral goes to
  $\dual{ \psi e^{-2\pii\place x} }{ \hat f }$ for every $x\in\RR$ (same argument as for the LHS).  We can take the limit under
  the $\mu$-integral by the Dominated Convergence Theorem, as, by
  Lemma~\ref{lem:apx:Half-deriv:linsyseq-lemma:exchange}\ref{lem:apx:Half-deriv:linsyseq-lemma:exchange:Bd}, we have the
  $\abs{\mu}$-integrable upper bound
  \begin{equation*}
    \Nm{f}_\infty \cdot \Nm{\psi}_1
    \ge
    \Nm{f_j}_\infty \cdot \Nm{\psi}_1
    \ge
    \absb{  \dual{  \widehat{f_j}  }{  \psi e^{-2\pii\place x}  }   } \text{ (for all~$x$,$j$).}
  \end{equation*}

  Thus, after taking the limit $j\to\infty$ on both sides of~\eqref{eq:RND2icfng7o}, we find that
  \begin{equation}\label{eq:RNDcwfensdn}\tag{$*\!*$}
    \dual{  \psi\hat\mu  }{  \hat f  }
    =
    \int \dual{  \psi e^{-2\pii\place x}  }{  \hat f  } \; d\mu(x)
  \end{equation}
  holds.

  If~$\mu$ satisfies the condition~\eqref{eq:overview:linsyseq} in the statement of the lemma we are proving, we simply
  calculate:
  \begin{align*}
    \int f(x) \, d\mu(x)
    &=
      \int \dual{  \psi e^{2\pii\place x}  }{  \hat f  } \; d\mu(x)        &&\cmmt{by Lemma~\ref{lem:apx:Half-deriv:linsyseq-lemma:FIT}}
    \\
    &=
      \dual{  \psi \cdot\hat\mu(-\place)  }{  \hat f  }                    &&\cmmt{by~\eqref{eq:RNDcwfensdn}}
    \\
    &=
      \dual{  \psi \cdot 2\pii\place \cdot e^{i\pi\place}  }{  \hat f  }   &&\cmmt{condition on $\mu$, \eqref{eq:overview:linsyseq}}
    \\
    &=
      \dual{  \psi \cdot e^{i\pi\place}  }{  (2\pii\place)\,\hat f  }      &&
    \\
    &=
      \dual{  \psi \cdot e^{i\pi\place}  }{  \widehat{f'}  }               &&\cmmt{Lemma\ref{lem:apx:Half-deriv:linsyseq-lemma:deriv-and-moment}\eqref{lem:apx:Half-deriv:linsyseq-lemma:deriv-and-moment:moment}}
    \\
    &=
      \dual{  \psi \cdot e^{2\pii\place\cdot\nfrac12}  }{  \widehat{f'}  } &&
    \\
    &=
      f'(\nfrac12).                                                        &&\cmmt{Lemma\ref{lem:apx:Half-deriv:linsyseq-lemma:deriv-and-moment}\eqref{lem:apx:Half-deriv:linsyseq-lemma:deriv-and-moment:deriv}}
  \end{align*}
  This completes the proof of Lemma~\ref{lem:overview:linsyseq}.
\end{proof}

\section{Additional graphs from numerical simulations}\label{apx:numsim}
\begin{figure}[htbp]
  \centering
  \subfloat[]{%
    \begin{minipage}[b]{0.45\textwidth}\centering%
      \includegraphics[width=1.1\linewidth]{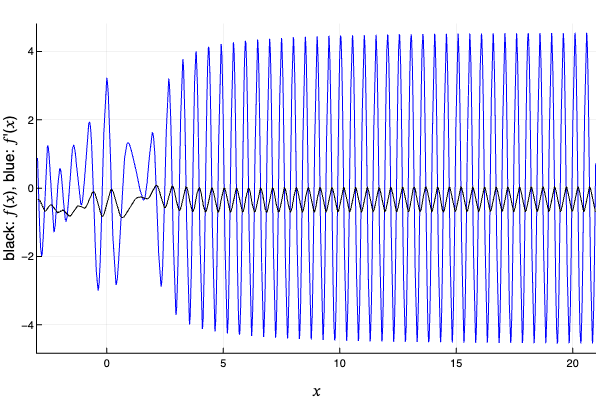}
    \end{minipage}%
  }
  \hfill
  \subfloat[]{%
    \begin{minipage}[b]{0.45\textwidth}\centering%
      \includegraphics[width=1.1\linewidth]{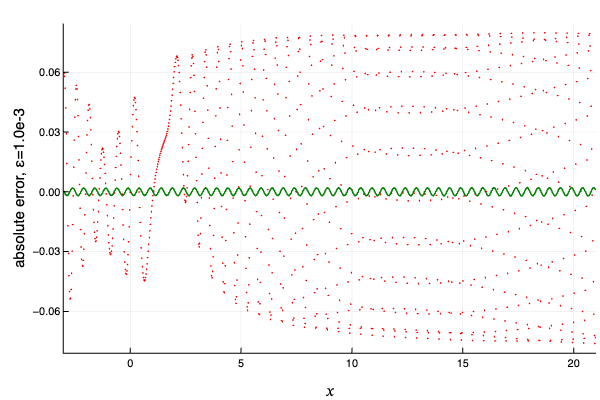}
    \end{minipage}%
  }
  \caption[]{\textbf{Second example of approximations provided by this paper's STNySR and BC's ASPSR.  }  %
    \small%
    Data from a randomly created (and selected because it looks pretty) expectation-value function~$f$ is presented (with
    matrices of size $4\times 4$).
    Graph (a) shows \textcolor{darkgray}{$f$ (black)} and \textcolor{blue}{$f'$ (blue)}, and (b) shows the absolute errors of
    the \textcolor{red}{ASPSR (red)} and the \textcolor{green}{STNySR (green)}, with $\eps=10^{-3}$.  The shown interval is well
    within the cut-off interval $[-125,+125]$ of STNySR; in this situation, STNySR typically gives a better approximation than
    ASPSR.}\label{fig:bc-ny-comparison:abserr-nice-win}
\end{figure}

\begin{figure}[htbp]
  \centering
  \subfloat[]{%
    \begin{minipage}[b]{0.45\textwidth}\centering%
      \includegraphics[width=1.1\linewidth]{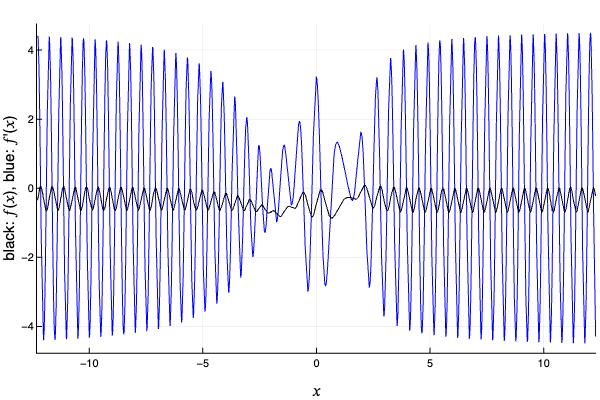}
    \end{minipage}%
  }
  % \hfill
  % \subfloat{%
  %   \begin{minipage}[b]{0.45\textwidth}\centering%
  %     %\includegraphics[width=1.1\linewidth]{abserr-nice-win-_-f+df.png}
  %   \end{minipage}%
  % }
  \\
  \subfloat[]{%
    \begin{minipage}[b]{0.45\textwidth}\centering%
      \includegraphics[width=1.1\linewidth]{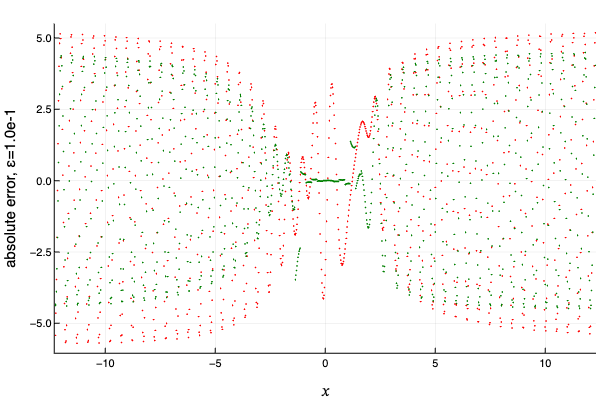}
    \end{minipage}%
  }
  \hfill
  \subfloat[]{%
    \begin{minipage}[b]{0.45\textwidth}\centering%
      \includegraphics[width=1.1\linewidth]{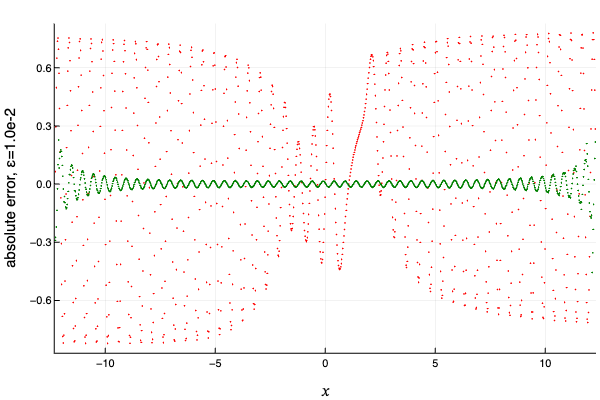}
    \end{minipage}%
  }
  \\
  \subfloat[]{%
    \begin{minipage}[b]{0.45\textwidth}\centering%
      \includegraphics[width=1.1\linewidth]{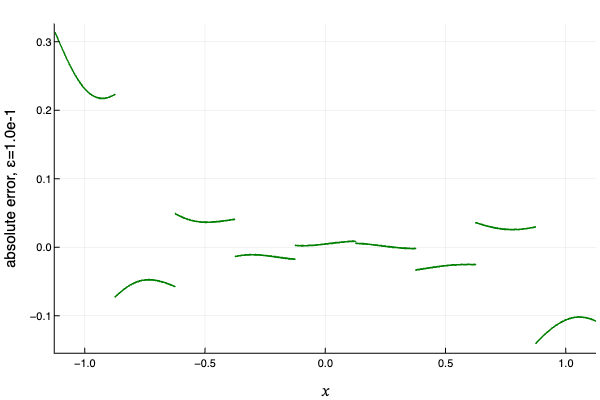}
    \end{minipage}%
  }
  \hfill
  \subfloat[]{%
    \begin{minipage}[b]{0.45\textwidth}\centering%
      \includegraphics[width=1.1\linewidth]{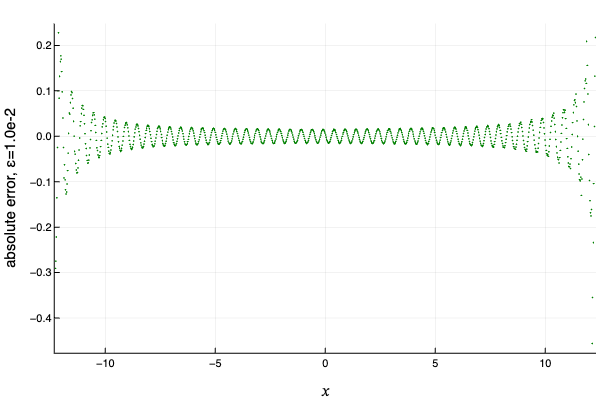}
    \end{minipage}%
  }
  \caption[]{\textbf{Second example of approximations provided by this paper's STNySR and BC's ASPSR.  }  %
    \small%
    Data from the same expectation-value function~$f$ from Fig.~\ref{fig:bc-ny-comparison:abserr-nice-win}.
    Graph (a) shows \textcolor{darkgray}{$f$ (black)} and \textcolor{blue}{$f'$ (blue)}, on a larger interval than in
    Fig.~\ref{fig:bc-ny-comparison:abserr-nice-win}.
    In (b,c,d,e) absolute errors of the \textcolor{red}{ASPSR (red)} and the \textcolor{green}{STNySR (green)} are plotted, on
    different parameter intervals.
    In (b), (d), we took $\eps=10^{-1}$ (leading to the cut-off $T=1.25$ for the STNySR) and in (c), (e) we took $\eps=10^{-2}$
    (leading to the cut-off $T=12.5$).
    \\
    For $\eps=0.1$, ASPSR is noise-only, while STNySR delivers what may be called an approximation to the derivative within the
    interval $[-1,+1]\subset [-T,+T]$.
    \\
    For $\eps=0.01$, STNySR provides a clearly superior approximation to the derivative, at almost every point.  Graph~(e) shows
    the effect that the error in STNySR blows up near the end-points of the cut-off interval
    $[-12.5,+12.5]$.}\label{fig:bc-ny-comparison:abserr-nice}
\end{figure}

\begin{figure}[htbp]
  \centering
  \subfloat[]{%
    \begin{minipage}[b]{0.45\textwidth}\centering%
      \includegraphics[width=1.1\linewidth]{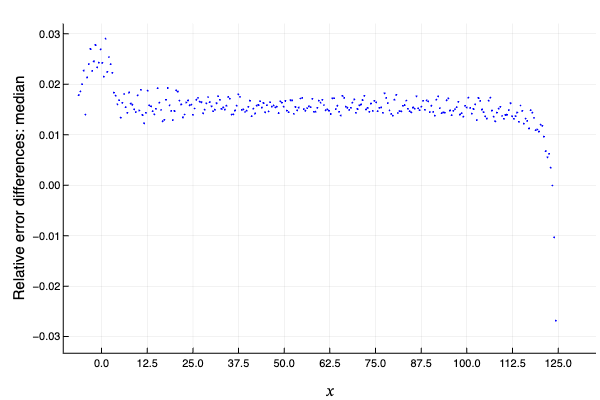}
    \end{minipage}%
  }
  \hfill
  \subfloat[]{%
    \begin{minipage}[b]{0.45\textwidth}\centering%
      \includegraphics[width=1.1\linewidth]{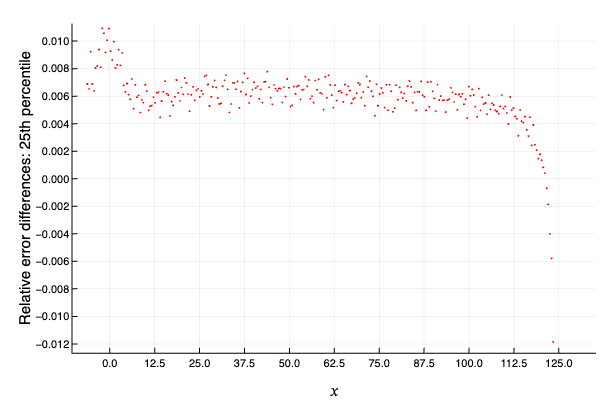}
    \end{minipage}%
  }
  \\
  \subfloat[]{%
    \begin{minipage}[b]{0.45\textwidth}\centering%
      \includegraphics[width=1.1\linewidth]{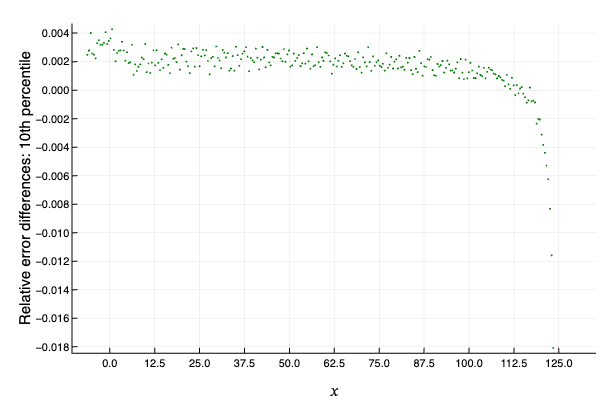}
    \end{minipage}%
  }
  \hfill
  \subfloat[]{%
    \begin{minipage}[b]{0.45\textwidth}\centering%
      \includegraphics[width=1.1\linewidth]{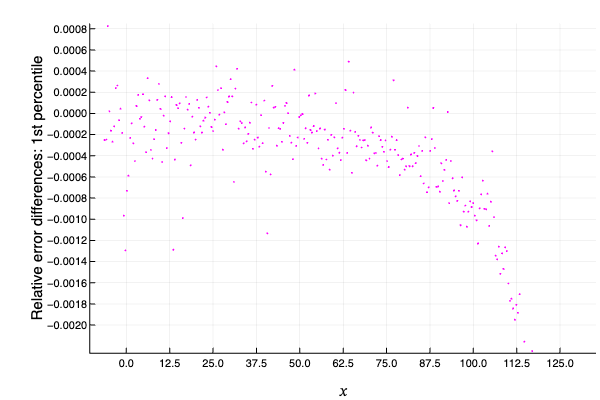}
    \end{minipage}%
  }
  \caption[]{\textbf{Another comparison of relative errors between STNySR and BC's ASPSR.  }  %
    \small%
    Plots of 200 random instances of expectation-value functions~$f$ chosen randomly as described in the text (with $8\times 8$
    matrices), considered at 300 parameter values, $x$.  For each combination of parameter value~$x$ and expectation-value
    function~$f$, the difference between the relative error of the ASPSR-approximation (with $\eps=10^{-3}$) of $f'(x)$ minus
    the relative error of the STNySR-approximation of $f'(x)$ (with parameter cut-off $1/8\eps=125$) is computed.  For each
    parameter value~$x$, the \textcolor{blue}{median (blue)}, \textcolor{red}{25th percentile (red)}, \textcolor{green}{10th
      percentile (green)}, \textcolor{magenta}{1st percentile (magenta)} of the 200 random instances are plotted, in different
    scales.  For 90\% of the instances, STNySR beats ASPSR by a clear margin over the all but the outside 10\% of the
    parameter region.  }\label{fig:bc-ny-comparison:relerr}
\end{figure}

We give some more graphs of numerical simulations.
Figs.\ \ref{fig:bc-ny-comparison:abserr-nice-win} and~\ref{fig:bc-ny-comparison:abserr-nice} are of the same type as
Fig.~\ref{fig:bc-ny-comparison:abserrB}, in that each shows data based on a single expectation-value function.
Fig.~\ref{fig:bc-ny-comparison:relerr} is of the same type as Fig.~\ref{fig:bc-ny-comparison:relerr2}, in that they show
statistics for a collection of 200 random instances.

\clearpage
\bibliographystyle{quantum}

\input{nyquist.camera.bbl}
%%%%%%%%%%%%%%%%%%%%%%%%%%%%%%%%%%%%%%%%%%%%%%%%%%%%%%%%%%%%%%%%%%%%%%%%%%%%%%%%%%%%%%%%%%%%%%%%%%%%%%%%%%%%%%%%
\end{document}

%% file: defsN.amsart.tex
\theoremstyle{plain}
\newtheorem{theorem}{Theorem}
\newtheorem*{theorem*}{Theorem}

\newtheorem{proposition}[theorem]{Proposition}
\newtheorem*{proposition*}{Proposition}

\newtheorem{corollary}[theorem]{Corollary}
\newtheorem*{corollary*}{Corollary}

\newtheorem{lemma}[theorem]{Lemma}
\newtheorem*{lemma*}{Lemma}

\newtheorem*{observation*}{Observation}

\newtheorem{remark}[theorem]{Remark}
\newtheorem*{remark*}{Remark}

\newtheorem*{remarks*}{Remarks}

\theoremstyle{definition}
\newtheorem{definition}[theorem]{Definition}
\newtheorem*{definition*}{Definition}

\newtheorem*{notation*}{Notation}

\newtheorem*{example*}{Example}

\newtheorem*{exercise*}{Exercise}

\theoremstyle{remark}

\newtheorem*{claim*}{Claim}

\newtheorem*{conjecture*}{Conjecture}

\newtheorem*{question*}{Question}

\newtheorem*{questions*}{Questions}

\newtheorem*{problem*}{Problem}

\newtheorem*{problems*}{Problems}

\newtheorem*{openproblem*}{Open Problem}

\newcommand{\subclass}[1]{}

\newcommand{\enumTi}[1]{\renewcommand{\theenumi}{#1}}

\newcommand{\alphenumi}{\enumTi{\alph{enumi}}}

\newcommand{\romenumi}{\enumTi{\roman{enumi}}}

%% file: defs.common.tex
\newlength{\hspaceforlengthglumpf}

\newcommand{\cmmt}[1]{\text{\footnotesize[#1]}}

\newcommand{\eqcmt}[1]{\mathrel{\mathop=\limits_{#1}}}
\newcommand{\lecmt}[1]{\mathrel{\mathop\le\limits_{#1}}}

\newcommand{\musteq}{\stackrel{!}{=}}

\DeclareMathOperator{\tr}{tr}

\DeclareMathOperator{\Diag}{Diag}

\DeclareMathOperator{\Supp}{Supp}
\DeclareMathOperator{\Spec}{Spec}

\newcommand{\One}{\mathbf{1}}

\newcommand{\lt}{\left}
\newcommand{\rt}{\right}

\renewcommand{\Hat}{\wedge}

\newcommand{\sstack}[1]{{\substack{#1}}}

\newcommand{\pii}{{\pi i}}

\newcommand{\nfrac}[2]{{\nicefrac{#1}{#2}}}

\newcommand{\CC}{\mathbb{C}}
\newcommand{\DD}{\mathbb{D}}

\newcommand{\LL}{\mathbb{L}}

\newcommand{\NN}{\mathbb{N}}

\newcommand{\RR}{\mathbb{R}}

\newcommand{\ZZ}{\mathbb{Z}}

\DeclareMathOperator*{\Exp}{\mathbb{E}}

\DeclareMathOperator*{\Var}{\mathbb{V}\mspace{-2mu}ar}

\newcommand{\eps}{\varepsilon}

\newcommand{\ceil}[1]{\lceil{#1}\rceil}

\newcommand{\abs}[1]{{\lvert{#1}\rvert}}                        \newcommand{\Nm}[1]{{\lVert{#1}\rVert}}
\newcommand{\Abs}[1]{{\lt\lvert{#1}\rt\rvert}}                  
\newcommand{\absb}[1]{{\bigl\lvert{#1}\bigr\rvert}}             \newcommand{\Nmb}[1]{{\bigl\lVert{#1}\bigr\rVert}}
\newcommand{\absB}[1]{{\Bigl\lvert{#1}\Bigr\rvert}}

\newcommand{\cip}[2]{({#1}\mid{#2})}

\newcommand{\dual}[2]{\langle{#1}:{#2}\rangle}

\newcommand{\ket}[1]{{\lvert  #1 \rangle}}

\newlength{\algotabbingwidth}